\documentclass[aip,twocolumn,reprint,superscriptaddress]{revtex4-1}
\usepackage{amsmath}
\usepackage{amssymb}
\usepackage[latin9]{inputenc}
\usepackage{graphicx}
\usepackage{hyperref}
\usepackage{physics}

\usepackage{graphicx}
\usepackage[color= blue!20]{todonotes}
\usepackage[normalem]{ulem}
\usepackage{color}
\usepackage{ulem}
\usepackage[symbol]{footmisc}

\usepackage{xcolor}
\hypersetup{
	colorlinks,
	linkcolor={blue!50!black},
	citecolor={blue!50!black}
}
\DeclareMathOperator{\atan2}{atan2}

\begin{document}
	
\addtocontents{toc}{\protect\setcounter{tocdepth}{0}}

\onecolumngrid

\noindent\textbf{\textsf{\Large Parametrically enhanced interactions and non-trivial bath dynamics\\ in a photon-pressure Kerr amplifier}}

\normalsize
\vspace{.3cm}
\noindent\textsf{  I.~C.~Rodrigues$^{1}$, G.~A.~Steele$^{1}$, and D.~Bothner$^{1,2}$,}

\vspace{.2cm}
\noindent\textit{$^1$Kavli Institute of Nanoscience, Delft University of Technology, PO Box 5046, 2600 GA Delft, The Netherlands\\$^2$Physikalisches Institut, Center for Quantum Science (CQ) and LISA$^+$, Universit\"at T\"ubingen, Auf der Morgenstelle 14, 72076 T\"ubingen, Germany}

\vspace{.5cm}

\date{\today}

{\addtolength{\leftskip}{10 mm}
\addtolength{\rightskip}{10 mm}

Photon-pressure coupling between two superconducting circuits is a promising platform for investigating radiation-pressure coupling in novel parameter regimes and for the development of radio-frequency (RF) quantum photonics and quantum-limited RF sensing.
So far, the intrinsic Josephson nonlinearity of photon-pressure coupled circuits has not been considered a potential resource for enhanced devices or novel experimental schemes.
Here, we implement photon-pressure coupling between a RF circuit and a microwave cavity containing a superconducting quantum interference device (SQUID) which can be operated as a Josephson parametric amplifier (JPA).
We demonstrate a Kerr-based enhancement of the photon-pressure single-photon coupling rate and an increase of the cooperativity by one order of magnitude in the amplifier regime.
In addition, we characterize the upconverted and Kerr-amplified residual thermal fluctuations of the RF circuit, and observe that the intracavity amplification reduces the measurement imprecision. 
Finally, we demonstrate that RF mode sideband-cooling is surprisingly not limited to the effective amplifier mode temperature arising from quantum noise amplification, which we explain by non-trivial bath dynamics due to a two-stage amplification process.
Our results demonstrate how Kerr nonlinearities and in particular Josephson parametric amplification can be utilized as resource for enhanced photon-pressure systems and Kerr cavity optomechanics.
}
\vspace{.5cm}

\twocolumngrid
\noindent\textbf{\textsf{\small INTRODUCTION}}
\vspace{1mm}

Photon-pressure and radiation-pressure coupled oscillators, where the amplitude of one oscillator modulates the resonance frequency of the second, have enabled a large variety of groundbreaking experiments in the recent decades.
In cavity optomechanics \cite{Aspelmeyer14}, this type of coupling has been used for unprecendented precision in the detection and control of mechanical displacement \cite{Teufel09, Anetsberger10, Teufel11, Chan11, Wollman15, Mason19}, to generate entanglement between two mechanical oscillators \cite{Riedinger18, OckeloenKorppi18}, to realize non-reciprocal signal processing \cite{Bernier17, Barzanjeh17, Fang17}, parametric microwave amplification \cite{Massel11, Cohen20, Bothner20}, frequency conversion \cite{Andrews14, OckeloenKorppi16, Forsch19} and the generation of entangled radiation \cite{Barzanjeh19}, to name just a few of the highlights.
More recently, the implementation of photon-pressure coupling between two superconducting circuits has attracted a lot of attention \cite{Johansson14, Kim15, Hardal17, Eichler18}.
Strikingly, within a short period of time the strong-coupling regime, the quantum-coherent regime, and sideband-cooling of a hot radio-frequency circuit into its quantum ground-state have been achieved \cite{Bothner21, Rodrigues21}.
These recent results open the door for quantum-limited photon-pressure microwave technologies, radio-frequency quantum photonics, quantum-enhanced dark matter axion detection at low-energy scales \cite{Chaudhuri19, Backes21} and for new approaches in circuit-based quantum information processing in terms of fault-tolerant bosonic codes \cite{Weigand20}. 
Photon-pressure coupled circuits utilize a superconducting quantum interference device (SQUID) as key coupling element, similar to flux-mediated optomechanics \cite{Shevchuk17, Rodrigues19, Zoepfl20, Schmidt20a, Bera21}, and therefore these platforms naturally come with Kerr cavities due to the Josephson nonlinearity of the SQUID inductance.
Most experimental and theoretical works on optomechanical and photon-pressure systems have considered only the case of photon-pressure-coupled linear oscillators but lately there has been growing interest in Kerr-like nonlinearities in photon-pressure interacting systems \cite{Nation08, Kumar10, Mikkelsen17, Asjad19, Gan19, Qiu19, Lau20, Bothner22}.
Kerr nonlinearities in superconducting circuits are already extremely useful resources for cat-state quantum computation \cite{Leghtas15}, for quantum-limited signal processing and detection by means of stand-alone Josephson parametric amplifiers, circulators and converters \cite{CastellanosBeltran08, Bergeal10, Pogorzalek17, Macklin15, Sliwa15}, and for Josephson metamaterials \cite{Martinez19, Winkel20, Planat20}.
Adding these exciting functionalities to photon-pressure coupled and optomechanical systems constitutes therefore a highly promising approach for enhanced quantum sensing devices and novel photon control schemes.
\begin{figure*}
	\centerline{\includegraphics[trim = {1cm, 1.5cm, 1cm, 1.5cm}, clip=True,width=0.98\textwidth]{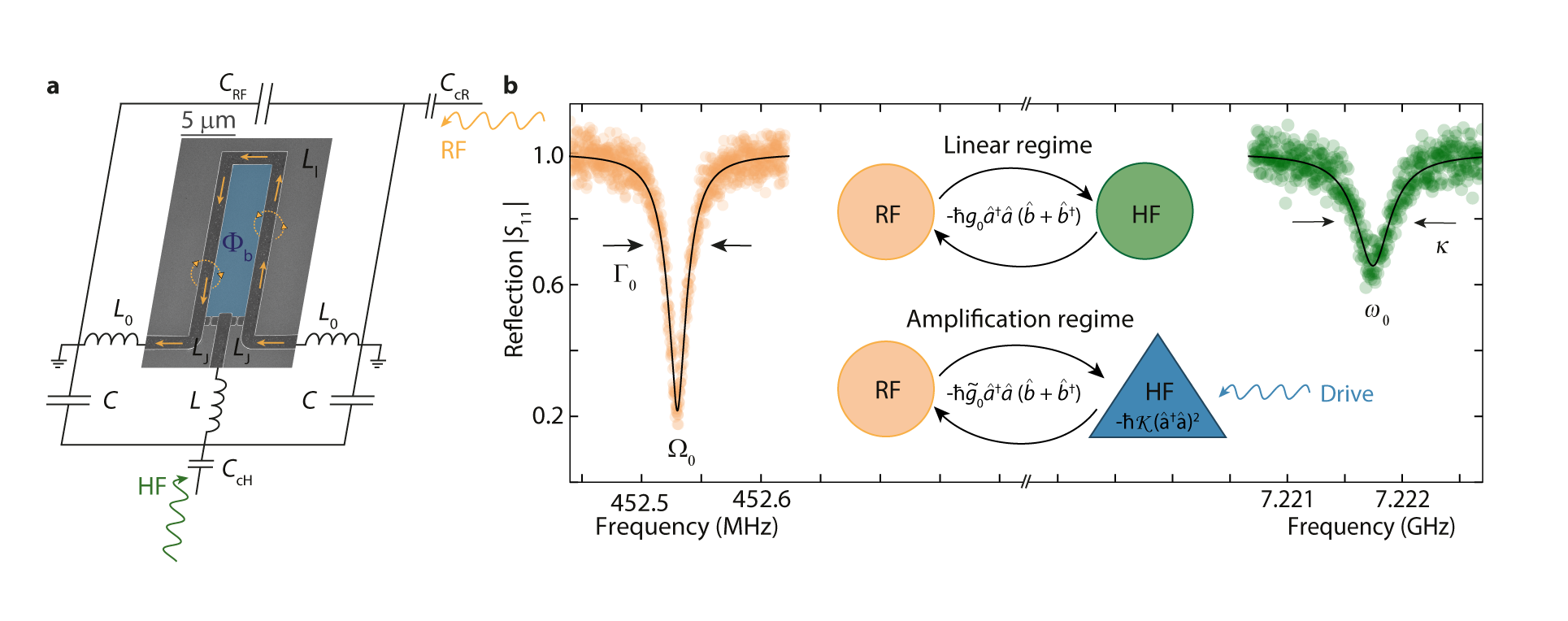}}
	\caption{\textsf{\textbf{Photon-pressure coupling between a radio-frequency LC circuit and a parametric amplifier SQUID cavity.} \textbf{a} Circuit schematic with an embedded scanning electron microscopy image of the superconducting quantum interference device. The high-frequency (HF) mode consists of the linear inductors $L$, $L_0$ the capacitor $C$ and the Josephson inductances $L_\mathrm{J}$. The RF mode consist of the capacitor $C_\mathrm{RF}$, and the linear inductors $L_0$ and $L_\mathrm{l}$. Each mode is capacitively coupled to an individual feedline for driving and readout by means of a coupling capacitor $C_\mathrm{cR}$ and $C_\mathrm{cH}$. The SQUID in the center of the circuit is biased with an external coil to a magnetic flux $\Phi_\mathrm{b}$ and any current from the RF mode flowing through the SQUID, indicated as yellow arrows, will add additional fluctuating flux $\Phi_\mathrm{RF}$. Both, the bias flux and the RF flux will change the inductance of the Josephson junctions in the SQUID $L_\mathrm{J}(\Phi)$. \textbf{b} shows the reflection response of the RF and HF mode in their corresponding frequency ranges and measured via their individual feedlines, respectively. The RF mode displays a resonance frequency $\Omega_0 = 2\pi\cdot 452.53\,$MHz and a total linewidth $\Gamma_0 = 2\pi\cdot 45\,$kHz. For the HF mode, we get $\omega_0 = 2\pi\cdot 7.2218\,$GHz and the linewidth $\kappa = 2\pi\cdot 400\,$kHz. Both, resonance frequency and linewidth depend on the flux bias and here $\Phi_\mathrm{b}/\Phi_0 = 0.48$. Inset shows schematically the two photon-pressure operation modes implemented in this paper. In the \textit{linear regime}, the RF circuit is coupled via photon-pressure to a linear HF cavity, in the \textit{amplification regime}, the RF mode is coupled to a Kerr parametric amplifier. The amplification regime is activated by a near-resonant strong HF cavity drive. The single-photon coupling rates are given by $g_0$ and $\tilde{g}_0 = g_0 + \frac{g_\mathcal{K}}{2}\hat{a}^\dagger \hat{a}$, respectively.}}
	\label{fig:Fig1}
\end{figure*}

Here we report photon-pressure coupling between a superconducting radio-frequency (RF) circuit and a strongly driven superconducting Kerr cavity, operated as a parametric amplifier.
As well-known from previous work\cite{Drummond80, Khandekar15, Huber20, Bothner22, FaniSani21}, by strongly driving the high-frequency SQUID cavity of our system, we can activate a four-wave mixing process and obtain an effective signal-idler double-mode cavity, here reaching up to $\sim 12\,$dB of intracavity gain.
Furthermore, by using an additional pump tone applied to the red sideband of the signal-mode resonance, we simultaneously switch on the photon-pressure coupling between this quasi-mode and the RF circuit.
We observe that the strong parametric drive enhances the single-photon coupling rate between the circuits, which in combination with further enhancement effects eventually leads to a more than tenfold increment in effective cooperativity.
Using the device as a radio-frequency thermal noise upconverter, we find that the output noise is accordingly amplified by the intrinsic Josephson amplification, which is potentially interesting for enhanced detection of weak radio-frequency signals.
Finally, we observe that sideband-cooling of the RF mode is not limited to the effective photon occupation of the quantum-heated amplifier mode and that the cooling tone is increasing the population imbalance between the two modes instead of reducing it.
Our results using a driven Kerr cavity disclose physical phenomena that have not yet been observed or described in standard radiation-pressure systems, and which are potentially useful for sensing of weak RF signals, microwave signal processing and Kerr optomechanical systems.
\vspace{2mm}

\noindent\textbf{\textsf{\small RESULTS}}
\vspace{1mm}

\noindent\textbf{\textsf{\small Device and photon-pressure coupling}}
\vspace{0mm}

Our device combines a superconducting radio-frequency LC circuit with a superconducting microwave SQUID cavity in a galvanic coupling architecture \cite{Rodrigues21}, cf. Fig.~\ref{fig:Fig1}\textbf{a}.
The circuit resonance frequencies are $\omega_0 = 2\pi \cdot 7.222\,$GHz for the high-frequency (HF) mode and $\Omega_0 = 2\pi \cdot 452.5\,$MHz for the RF mode, respectively, cf. Fig.~\ref{fig:Fig1}\textbf{b}.
At the heart of the device is a nanobridge-based SQUID, which translates the magnetic flux connected to oscillating currents in the RF inductor into resonance frequency modulations of the HF circuit.
To first order and without taking into account the nonlinearity of the Josephson nanobridges, the Hamiltonian of the undriven system is given by
\begin{equation}
\hat{H}_\mathrm{lin} = \hbar\omega_0\hat{a}^\dagger\hat{a} + \hbar\Omega_0\hat{b}^\dagger\hat{b} + \hbar g_0\hat{a}^\dagger\hat{a}\left(\hat{b} + \hat{b}^\dagger\right)
\end{equation}
where the photon-pressure single-photon coupling rate 
\begin{equation}
g_0 = \frac{\partial\omega_0}{\partial\Phi}\Phi_\mathrm{zpf}
\end{equation}
is given by the flux responsivity of the HF mode resonance frequency $\partial\omega_0/\partial\Phi$ and the effective zero-point RF flux $\Phi_\mathrm{zpf} \approx 635\,\mu\Phi_0$ coupling into the SQUID loop.
Note that here the annihilation (creation) operators $\hat{a},\hat{b}~(\hat{a}^\dagger,\hat{b}^\dagger)$ refer to a change in photon excitations of the HF and RF circuit, respectively, and that the RF induced flux $\hat{\Phi} = \Phi_\mathrm{zpf}\left(\hat{b} + \hat{b}^\dagger\right)$ threading the SQUID is analogous to the displacement of a mechanical resonator in an optomechanical system.
When the Kerr nonlinearity of the Josephson junctions is taken into account, the Hamiltonian $\hat{H} = \hat{H}_\mathrm{lin} + \hat{H}_\mathrm{Kerr}$ is extended with the Kerr terms \cite{Mikkelsen17}
\begin{equation}
\hat{H}_\mathrm{Kerr} = \frac{\hbar \mathcal{K}}{2}\left(  \hat{a}^\dagger \hat{a}\right)^2 + \frac{\hbar g_\mathcal{K}}{2}\left(  \hat{a}^\dagger \hat{a}\right)^2\left(\hat{b} + \hat{b}^\dagger \right)
\end{equation}
where the Kerr-related photon-pressure coupling constant is given by
\begin{equation}
g_\mathcal{K} = \frac{\partial \mathcal{K}}{\partial\Phi}\Phi_\mathrm{zpf}.
\end{equation}
Here, we omitted the nonlinearity of the RF circuit as it is extremly small with $\mathcal{K}_\mathrm{RF}\sim - 2\pi\cdot 1\,$Hz.
For the high-frequency circuit, the Kerr constant $\mathcal{K} = -\frac{e^2}{2\hbar C_\mathrm{HF}}  \frac{L_\mathrm{J}^3}{L_\mathrm{HF}^3}$ depends on the bias flux via the inductance ratio and is on the order of $\mathcal{K} \sim -2\pi\cdot 5\,$kHz.
The interaction part of the Hamiltonian is therefore given by
\begin{equation}
\hat{H}_\mathrm{int} = \hbar g_0\hat{a}^\dagger\hat{a}\left(\hat{b} + \hat{b}^\dagger\right) + \frac{\hbar g_\mathcal{K}}{2}\left(  \hat{a}^\dagger \hat{a}\right)^2\left(\hat{b} + \hat{b}^\dagger \right).
\end{equation}
In the following section we will investigate the linearized dynamics of this system under strong near-resonant driving and for the case of a combination of near-resonant driving and additional photon-pressure red-sideband pumping.

\begin{figure}[h!]
	\centerline{\includegraphics[trim = {0.5cm, 0.8cm, 0.0cm, 0.9cm}, clip=True,scale=0.51]{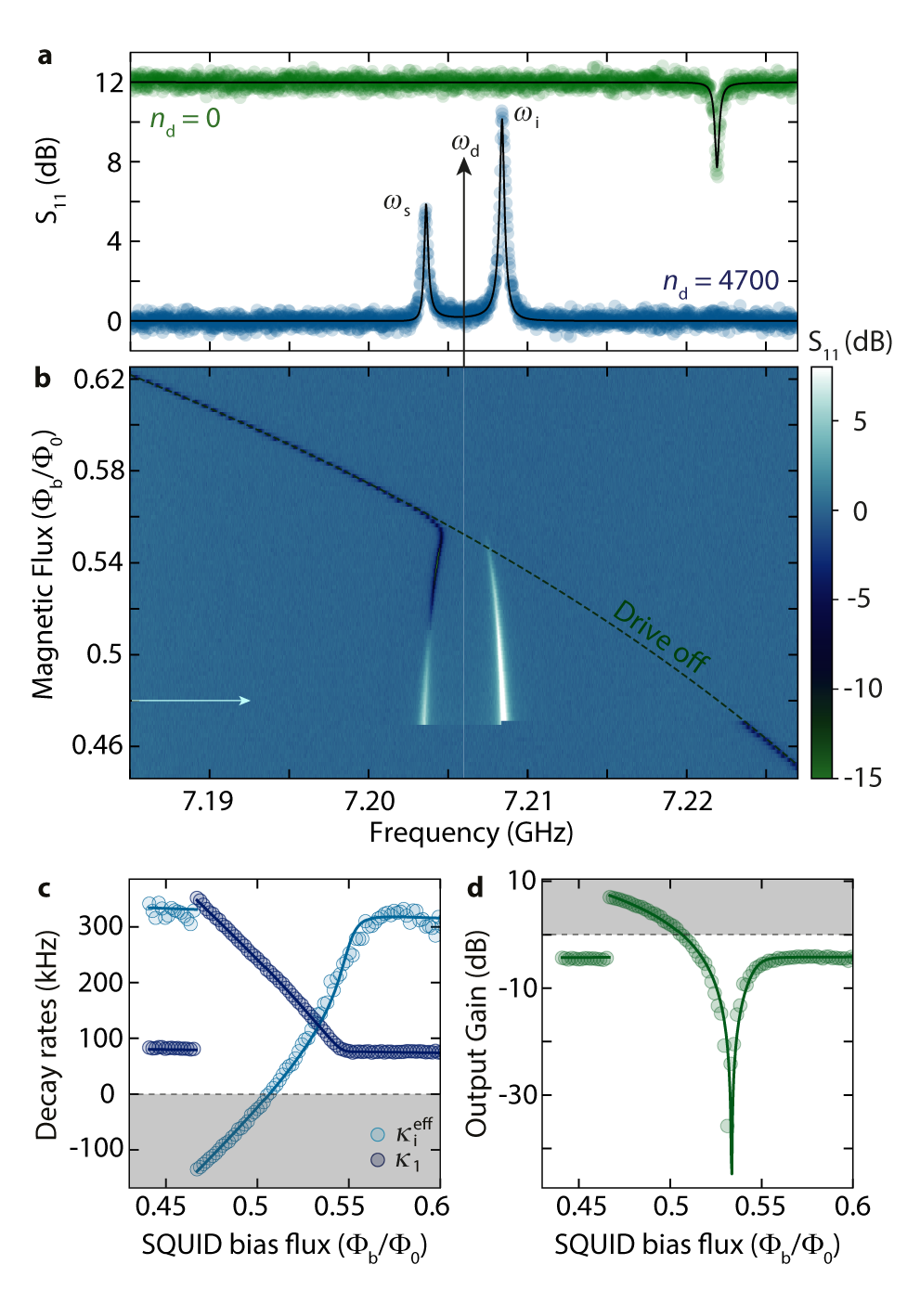}}
	\caption{\textsf{\textbf{Observation of parametric gain in a strongly driven photon-pressure Kerr cavity.} \textbf{a} Probe-tone reflection off the HF SQUID cavity $S_{11}$ with (blue) and without (green) a strong drive at $\omega_\mathrm{d}$. The undriven response labeled $n_\textrm{d} = 0$ is offset by $+12\,$dB for clarity. While the undriven cavity displays a single absorption resonance at $\omega_0$, the driven state ($n_\textrm{d} = 4700$) exhibits a double-resonance with output gain in both quasi-modes. The signal- and idler-mode resonance frequencies are $\omega_\mathrm{s}$ and $\omega_\mathrm{i}$, respectively. Circles are data, lines are fits. \textbf{b} Color-coded reflection $S_{11}$ in presence of the strong drive at $\omega_\mathrm{d}$ during a resonance frequency upsweep, displaying the continuous emergence of the double-mode response of the linescan shown in  \textbf{a} (position indicated by horizontal arrow). The undriven resonance frequency $\omega_0(\Phi_\mathrm{b})$ is indicated as dashed line. From a fit to the signal mode of each line with a single linear cavity response, we obtain the apparent external and internal decay rates $\kappa_1$ and $\kappa_\mathrm{i}^\mathrm{eff} = \kappa - \kappa_1$, plotted as circles in \textbf{c}. Lines show theoretical calculations based on the full driven Kerr model (see Supplementary Note 5 and 7), including drive-saturating two-level systems. The intracavity Josephson gain at the amplifier signal resonance is given by $\mathcal{G}_\mathrm{s} = \kappa_1/\kappa_\mathrm{e}$, and the gray shaded area indicates the regime of $\kappa_\mathrm{i}^\mathrm{eff}<0$, i.e., of output gain $>1$. In \textbf{d}, we show the corresponding output gain at $\omega = \omega_\mathrm{s}$, indicating that the effective coupling between the cavity and its readout feedline can be continuously changed using the drive tone, reaching for instance the regime of critically coupled at the point where the output gain is lowest (i.e. $\kappa_\mathrm{i}^\mathrm{eff} = \kappa_1$). Circles are extracted from data, line is the theoretical prediction.}}
	\label{fig:Fig2}
\end{figure}

\vspace{2mm}
\noindent\textbf{\textsf{\small Kerr amplifier quasi-modes}}
\vspace{0mm}

For a strong near-resonant drive, the dynamics of the HF cavity with respect to a small additional probe field is captured by that of current-pumped Josephson parametric amplifier (JPA).
In contrast to usual JPA experiments, however, we operate the amplifier in the high-amplitude state far beyond its bifurcation point and work with a small linewidth cavity $\kappa\sim 2\pi\cdot 250\,$kHz in the undercoupled regime.
We prepare the SQUID cavity in this state by using a fixed-frequency drive tone at $\omega_\mathrm{d}$ and by moving the HF cavity resonance frequency $\omega_0$ from lower to higher frequencies through $\omega_\mathrm{d}$ by means of the SQUID flux bias $\Phi_\mathrm{b}$, cf. Fig.~\ref{fig:Fig2}.
The drive-induced modification of the cavity susceptibility leads to several effects regarding the cavity response to an additional probe tone.
First, the resonance frequency of the driven mode $\omega_\textrm{s}$ deviates considerably from the undriven case (dashed line in Fig.~\ref{fig:Fig2}\textbf{b}) and even tends to shift to lower frequencies with decreasing flux.
The reason behind this is the nonlinear frequency shift due to an increasing intracavity drive photon number, which is compensating the flux shift \cite{Bothner22}.
Secondly, we observe that the intracavity Josephson gain turns the resonance absorption dip into a net gain peak, translating a clear change in the effective coupling between the cavity and its feedline.
And finally, as theoretically and experimentally explored in previous systems\cite{Drummond80, Khandekar15, Huber20, FaniSani21, Bothner22}, we also observe the emergence of a second peak in the spectral response due to a phenomenon one can describe as "idler resonance". Here, the probe tone image frequency, i.e. the frequency of the idler photons generated by nonlinear mixing from the drive and the probe, becomes resonant with the cavity mode \cite{FaniSani21}. In this regime we observe output field gain at the idler resonance and the cavity exhibits an internal feedback locking mechanism that has been used to stabilize the cavity against external flux noise in a related system \cite{Bothner22}.
In this experimental situation, the photon-pressure coupling can be neglected to first order and the linearized probe-tone response is given by
\begin{equation}
S_{11}(\Omega) = 1 - \kappa_\mathrm{e}\chi_\mathrm{g}(\Omega)
\end{equation}
with the driven susceptibility 
\begin{equation}
\chi_\mathrm{g}(\Omega) = \frac{\chi_\mathrm{p}(\Omega)}{1 - \mathcal{K}^2 n_\mathrm{d}^2\chi_\mathrm{p}(\Omega)\chi_\mathrm{p}^*(-\Omega)}.
\end{equation}
Here, $\kappa_\mathrm{e} \sim 2\pi\cdot 80\,$kHz is the external coupling rate, $n_\mathrm{d}$ is the intracavity drive photon number, $\Omega$ is the probe-tone frequency with respect to $\omega_\mathrm{d}$ and $\chi_\mathrm{p}^{-1} = \kappa/2 + i\left(\Delta_\mathrm{d} + 2\mathcal{K}n_\mathrm{d} + \Omega \right)$ with $\Delta_\mathrm{d} = \omega_\mathrm{d}-\omega_0$.

\begin{figure*}
	\centerline{\includegraphics[trim = {0.0cm, 9.5cm, 0.0cm, 3.0cm}, clip=True, width=0.99\textwidth]{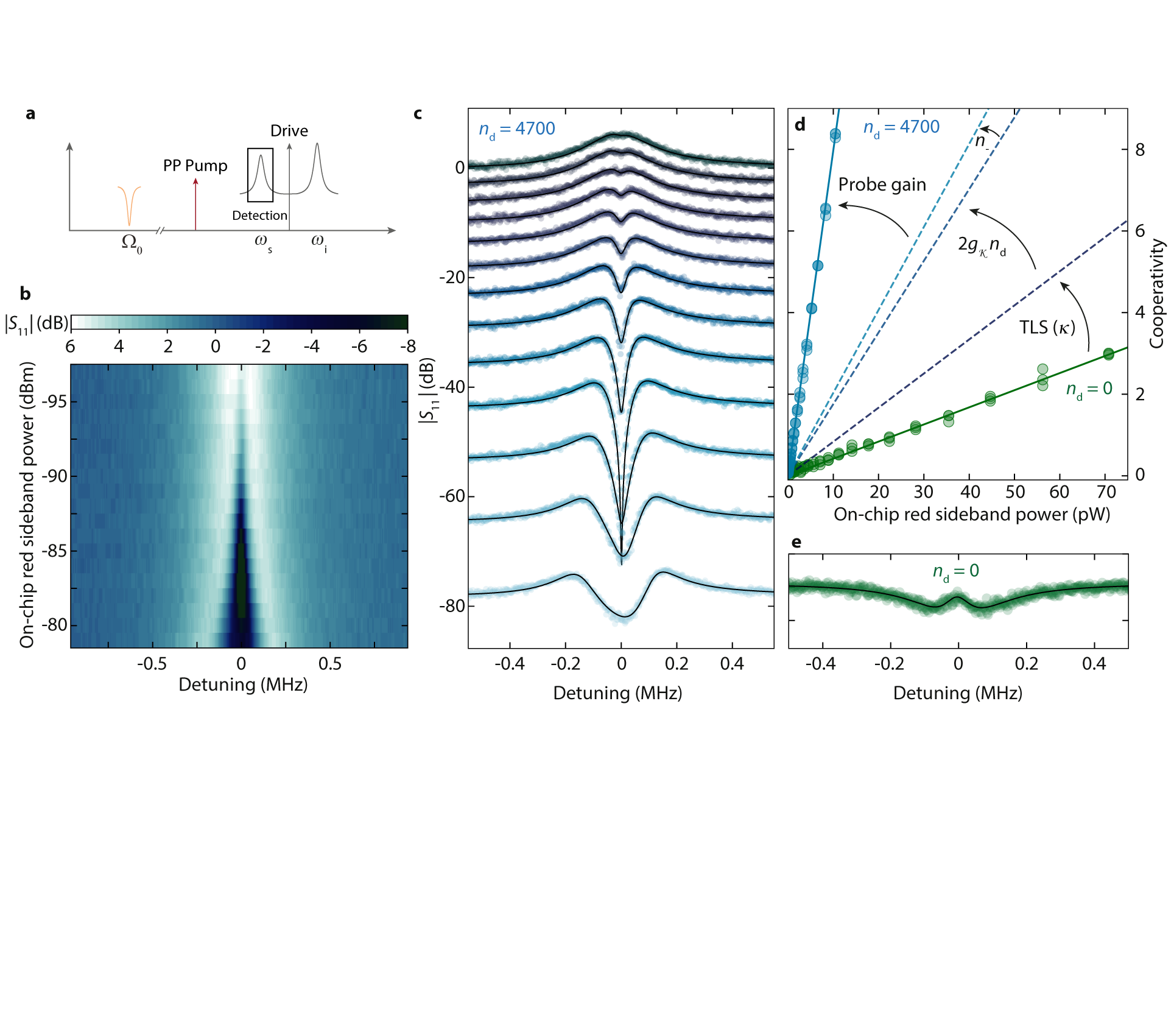}}
	\caption{\textsf{\textbf{Parametrically enhanced photon-pressure interaction by internal Kerr amplification.} \textbf{a} Experimental protocol for photon-pressure coupling in the amplifier high-gain regime. By means of a strong drive tone at $\omega_\mathrm{d}$, the HF mode is prepared in the the regime where both quasi-modes show output gain and where $n_\textrm{d} = 4700$. An additional photon-pressure pump tone (PP Pump) is applied at the red sideband of the signal resonance $\omega_\mathrm{p} \approx \omega_\mathrm{s} - \Omega_0 $. A third, weak probe tone around $\omega \approx \omega_\mathrm{s}$ detects the device reflection response $S_{11}$. \textbf{b} Color-coded HF reflection $S_{11}$ vs detuning from the signal resonance frequency $\omega_\mathrm{s}$ and photon-pressure pump power. Individual linescans are shown in panel \textbf{c}. Circles are data, lines are the result of theoretical calculations using the full Kerr model cf. Supplementary Note 8. The slight peak-height asymmetry for the largest photon-pressure pump powers originates from frequency-dependent Josephson gain. From fits to the data with the effective conventional-mode model discussed in the context of Fig.~\ref{fig:Fig2}, we obtain the effective cooperativity $\mathcal{C}_\mathrm{eff} = 4 g_\mathrm{eff}^2/(\kappa \Gamma_0) = 4\mathcal{G}_\mathrm{s}n_-\left(g_0 + 2g_\mathcal{K}n_\mathrm{d}\right)^2/(\kappa\Gamma_0)$ for each power. The result is shown in panel \textbf{d} blue circles in direct comparison with the cooperativity obtained without parametric drive (i.e. $n_\textrm{d} = 0$) and Josephson gain (i.e. $\mathcal{G}_\mathrm{s}=1$), respectively (green circles). The cooperativity of the parametrically driven HF cavity for $n_\textrm{d} = 4700$ is enhanced by more than one order of magnitude, which can be explained by a combination of several drive-induced and Kerr-related enhancement effects as indicated by the arrows. 'TLS ($\kappa$)' indicates a reduction of the cavity linewidth by two-level-system saturation, $2g_\mathcal{K}n_\mathrm{d}$ refers to an enhanced single photon coupling rate by modulation of the Kerr constant, $n_-$ refers to parametric amplification of the photon-pressure sideband pump and 'Probe gain' to amplification of the probe field with gain $\mathcal{G}_\mathrm{s} = 4$, details can be found in the main text. For a direct comparison at the photon-pressure pump power of highest cooperativity in \textbf{c}, the corresponding response in the absence of drive photons ($n_\textrm{d} = 0$) is shown in panel \textbf{e}, revealing only a small transparency window at the bottom of the undriven HF resonance. Detuning in \textbf{e} is with respect to $\omega_0$.}}
	\label{fig:Fig3}
\end{figure*}

To obtain some more intuitive insight, the reflection can also be approximated by a combination of two Kerr-modified conventional modes
\begin{equation}
S_{11}(\Omega) = 1 - \frac{\kappa_1}{\frac{\kappa}{2} + i\left(\Omega - \Omega_\mathrm{s} \right)}- \frac{\kappa_2}{\frac{\kappa}{2} + i\left(\Omega - \Omega_\mathrm{i} \right)}
\end{equation}
using the signal and idler mode resonance frequencies $\Omega_\mathrm{s, i}$, the apparent external linewidths $\kappa_1 = \mathcal{G}_\mathrm{s}\kappa_\mathrm{e}$, $\kappa_2 = \mathcal{G}_\mathrm{i}\kappa_\mathrm{e}$ and the intracavity Josephson gain
\begin{equation}
\mathcal{G}_\mathrm{s, i} = \frac{\Omega_\mathrm{s, i} - \Delta_\mathrm{d} + 2\mathcal{K}n_\mathrm{d}}{2\Omega_\mathrm{s, i}}
\end{equation}
at the signal and idler mode resonance frequencies.
The maximum intracavity Josephson gain for the signal mode is then given by $\mathcal{G}_\mathrm{s} = \kappa_1/\kappa_e \sim 12\,$dB, cf. Fig.~\ref{fig:Fig2}\textbf{c}, leading to the observed output gain of $\mathcal{G}_\mathrm{out} \sim 6\,$dB.
Both, Josephson gain and effective linewidths, are well captured by the theoretical model, cf. Figs.~\ref{fig:Fig2}\textbf{c} and \textbf{d}, if we take the effect of saturating two-level systems into account \cite{Capelle20} which reduces the total mode linewidth with drive photon number $n_\mathrm{d}$ (for more details see the Supplementary Material).
Note that when treating the signal resonance as a usual mode, the increasing drive photon number and Josephson gain, respectively, also induce a transition from an undercoupled to a critically coupled to an overcoupled cavity and finally to a cavity with a negative internal linewidth displaying a net output gain, cf. Fig.~\ref{fig:Fig2}\textbf{c},\textbf{d}. This can be described by a change in the magnitude of the ratio between the effective external and internal linewidths of the driven system $|\kappa_1/\kappa_\textrm{i}^\textrm{eff}|$ which goes from $<1$ (undercoupled) to $>1$ (overcoupled). Note that the total decay rate $\kappa$ is not affected by the amplification process. At last, this drive-tunable external coupling is not only advantageous to realize non-degenerate parametric amplifiers \cite{Bergeal10} but also for the engineering of tunable microwave attenuators \cite{Pogorzalek17}.
\vspace{2mm}

\noindent\textbf{\textsf{\small Parametrically enhanced interaction in a Kerr amplifier}}
\vspace{0mm}

Once the HF SQUID cavity is prepared in the parametric amplifier state with $n_\textrm{d} = 4700$, we activate the photon-pressure coupling to the RF circuit by an additional pump tone on the red sideband of the signal mode, i.e., at $\omega_\mathrm{p} = \omega_\mathrm{s} - \Omega_0$, cf. Fig.~\ref{fig:Fig3}\textbf{a}.
To characterize the interaction between the driven and pumped HF mode and the RF circuit, we then detect the device response around $\omega_\mathrm{s}$ with a weak third probe tone.
For low sideband-pump powers we observe a small dip inside the signal mode resonance peak, cf. Fig.~\ref{fig:Fig3}\textbf{b}, \textbf{c}, which gets wider and deeper with increasing pump power.
The appearance of this window indicates photon-pressure induced absorption \cite{Hocke12}, an effect originating in coherent driving of the RF mode by the pump-probe-beating and a corresponding interference between the original probe tone and an RF induced pump tone sideband.
The width of the window in the small power regime is therefore given by the effective RF mode damping rate $\Gamma_\mathrm{eff} = \Gamma_0 + \Gamma_\mathrm{pp}$, where $\Gamma_0 = 2\pi\cdot 45\,$kHz is the intrinsic RF circuit linewidth and $\Gamma_\mathrm{pp}$ is the  photon-pressure dynamical backaction damping\cite{Bothner21, Rodrigues21}.
For the largest powers, the absorption window gets shallower again and the HF response is at the onset of normal-mode splitting, as we are approaching the photon-pressure strong-coupling regime \cite{Bothner21}.
For each pump power, the effective cooperativity $\mathcal{C}_\mathrm{eff} = 4g_\mathrm{eff}^2/(\kappa\Gamma_0)$ can be determined with the RF mode linewidth $\Gamma_0 \approx 2\pi\cdot 45\,$kHz by fitting the reflection response using an effective linear mode model as discussed in the context of Fig.~\ref{fig:Fig2}, for details see Supplementary Note 8.
As result we find that the photon-pressure cooperativity with the HF cavity signal mode is more than one order of magnitude enhanced compared to the equivalent experiment with the undriven cavity, cf. Fig.~\ref{fig:Fig3}\textbf{d}.
This enhancement originates from three main physical phenomena: (1) the saturation of two-level-systems by the drive tone, (2) the RF induced flux modulation of the Kerr non-linearity and (3) the intracavity amplification process arising from the presence of the strong drive.
For a better understanding of the different effects, we account for all the individual contributions and indicate them as dashed lines in Fig.~\ref{fig:Fig3}\textbf{d}.
In the following paragraphs we discuss them one by one.
First of all the linewidth of the driven HF mode $\kappa \approx 2\pi\cdot 225\,$kHz is reduced compared to the undriven mode ($\sim 400\,$kHz), most likely due to a saturation of two-level-systems by the drive tone.
From this, we get an increase in the cooperativity of $\sim 400/225 = 1.8$ and the contribution is labeled in Fig.~\ref{fig:Fig3}\textbf{d} with TLS ($\kappa$).
Such a TLS saturation effect, however, is not related directly to the presence of a Kerr nonlinearity in the device and hence could be viewed as rather trivial, in contrast to the other contributions.
For the more interesting enhancement factors, which all arise from the Kerr nonlinearity in combination with the strong driving, we consider the linearized version of the multi-tone driven interaction Hamiltonian.
Using $|\alpha_\mathrm{d}| \gg |\gamma_-|, \langle\hat{c}\rangle$ where $\alpha_\mathrm{d}$ is the drive intracavity field, $\gamma_-$ is the sideband-pump intracavity field and $\hat{c}$ is the intracavity probe field, the dominant contribution to the linearized interaction Hamiltonian is given by
\begin{equation}
\hat{H}_\mathrm{int} = \hbar \left[ g_0 + 2g_\mathcal{K}n_\mathrm{d} \right]\left(\gamma_-^*\hat{c} + \gamma_-\hat{c}^\dagger \right)\left( \hat{b} + \hat{b}^\dagger \right)
\end{equation}
showing the usual multi-photon enhancement by the pump amplitude $\gamma_-$ and an additional enhancement of the linearized single-photon coupling-rate $\tilde{g}_0 = g_0 + 2g_\mathcal{K}n_\mathrm{d}$ by the modulation of the Kerr constant $g_\mathcal{K}$.
In our experiment $|g_\mathcal{K}| \sim 2\pi\cdot 6\,$Hz, $|g_0| \sim 2\pi\cdot 120\,$kHz, i.e., $|g_\mathcal{K}| \ll |g_0|$.
The Kerr-contribution to the single-photon coupling rate in the data of Fig.~\ref{fig:Fig3}, however, is additionally enhanced by the large drive-photon number $n_\mathrm{d} \sim 4700$ and therefore contributes significantly to the total coupling rate, in fact it increases the effective cooperativity by a factor $\sim 1.9$.
We label this enhancement contribution in Fig.~\ref{fig:Fig3}\textbf{d} with $2g_\mathcal{K}n_\mathrm{d}$.
The third and final contribution to the parametrically enhanced interaction is the parametric amplification of the intracavity fields by the drive, and both the sideband pump and the probe field are amplified with different gain.
Note that in Fig.~\ref{fig:Fig3}\textbf{d} the two parts (pump and probe) are shown individually.
The sideband pump field $\gamma_-$ is far detuned from the drive and therefore the parametric gain at this frequency is small; the pump photon number $n_- = |\gamma_-|^2$ is only increased by a factor $\sim 1.2$.
Nevertheless, it's a measurable contribution and in Fig.~\ref{fig:Fig3}\textbf{d} it is labeled with $n_-$.
The parametric amplification of the probe field $\hat{c}$ inside the driven HF resonance though is large, with an amplitude gain of $\mathcal{G}_\mathrm{s} = 4$.
Taking into account the deep sideband-resolved limit as well as red-sideband pumping and solving for the device response (see Supplementary Note 8), the effective multi-photon coupling rate is given by $g_\mathrm{eff} = \sqrt{\mathcal{G}_\mathrm{s}n_-}\tilde{g}_0$ with $\tilde{g}_0 = g_0 + 2g_\mathcal{K}n_\mathrm{d}$, i.e., also the cooperativity is enhanced by the resonance gain of the signal mode with $\mathcal{G}_\mathrm{s} \approx 4$.
Using the full linearized model for calculating the theoretical device response, we find excellent agreement with the data, cf. lines in Fig.~\ref{fig:Fig3}\textbf{c}.
Interestingly, the effect of the parametric drive is not only to  considerably enhance the linearized coupling rate and the cooperativity.
As with increasing gain the effective cavity resonance makes a continuous transition from an undercoupled to an overcoupled cavity, cf. Fig.~\ref{fig:Fig2}, also the shape of the photon-pressure induced RF resonance inside the cavity is strongly drive-dependent.
For vanishing or small parametric drives, i.e. when the HF cavity still exhibits a dip in the reflection spectrum, the RF signature on the cavity lineshape resembles the one of photon-pressure induced transparency, cf. Fig.~\ref{fig:Fig3}\textbf{e}. On the other hand, in the case of larger drives, i.e. when the cavity takes the shape of a peak whose resonance amplitude goes above the background, we get photon-pressure-induced absorption \cite{Hocke12}.
Therefore, we get a highly drive-tunable system response, potentially interesting for invertible narrowband filters and microwave signal control \cite{SafaviNaeini11, Zhou13}.
\vspace{2mm}
\noindent\textbf{\textsf{\small Enhanced radio-frequency upconversion}}
\vspace{0mm}
\begin{figure}[h!]
	\centerline{\includegraphics[trim = {0.1cm, 1.0cm, 0cm, 0cm}, clip=True, scale=0.53]{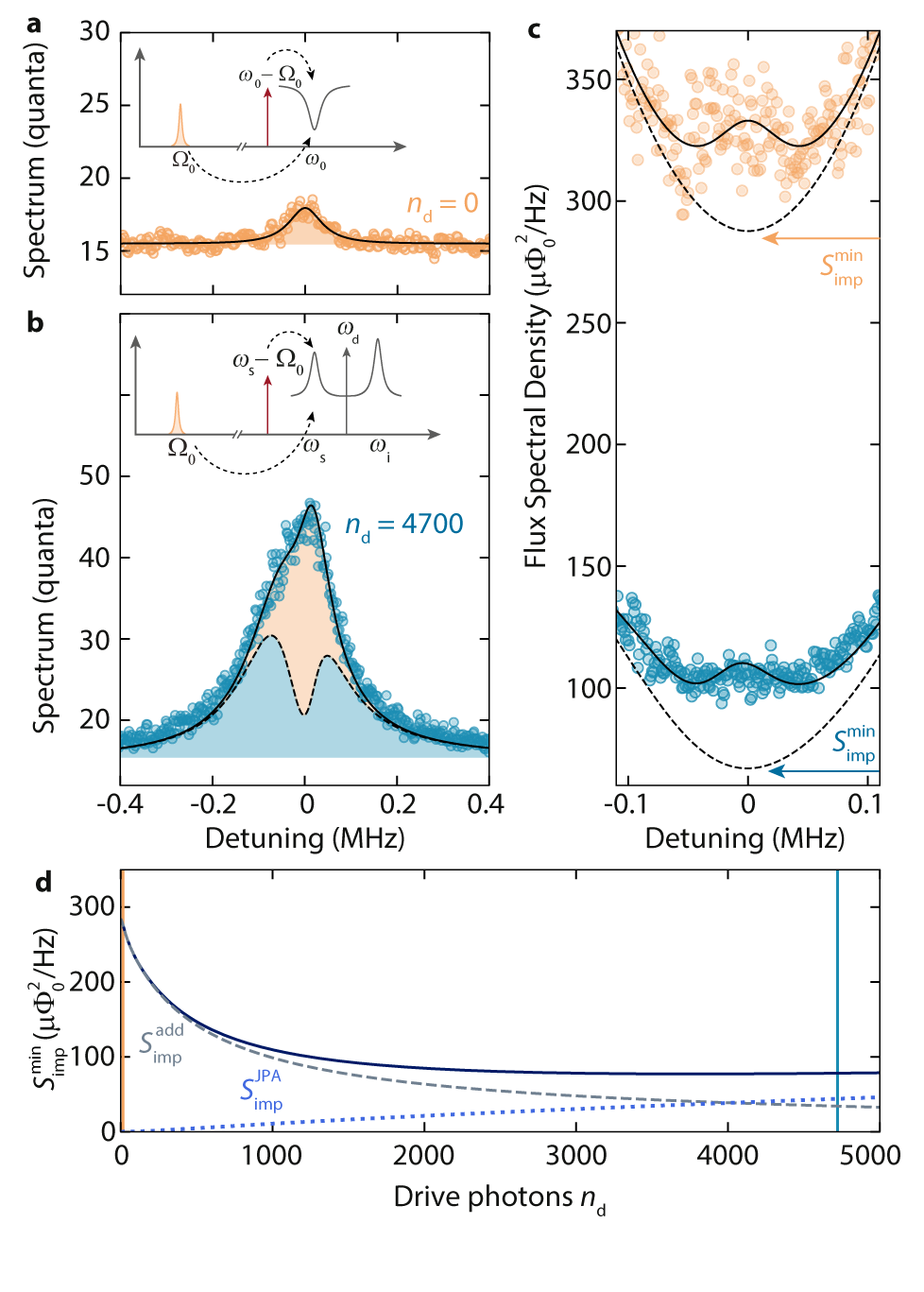}}
	\caption{\textsf{\textbf{Enhanced upconversion of radio-frequency thermal noise with a photon-pressure Kerr amplifier.} \textbf{a}, \textbf{b} Upconverted thermal noise of the RF mode, detected in the output spectrum of the SQUID cavity signal mode, for $n_\mathrm{d} = 0$ and $n_\mathrm{d} = 4700$, respectively. The noise contribution from the RF mode is shaded in orange, the blue-shaded area shows the amplifier output noise for $n_\mathrm{RF}^\mathrm{th} = 0$, which is basically amplified quantum noise with noise squashing due to $n_\mathrm{RF}^\mathrm{th} < \tilde{n}_\mathrm{th}^\mathrm{HF}$ with $\tilde{n}_\mathrm{th}^\mathrm{HF}$ being the effective signal mode occupation. The detuning is given with respect to the HF cavity signal mode resonance frequency and insets show a sketch of the experimental scheme. For both data $\mathcal{C}_\mathrm{eff} \sim 0.8$. The detected noise spectra in units of quanta can be converted to RF flux spectral densities, cf. text, as plotted in panel \textbf{c}. With internal amplification, the device RF flux sensitivity is enhanced by a factor of $\sim 3$, and exactly on signal mode resonance the imprecision noise by the amplifier chain is reduced by a factor of $\sim 3.2$. The paraboloid background shape of the imprecision noise (dashed line) originates from the HF cavity susceptibility and in particular from its small linewidth in our device. The theoretical curve for the minimum imprecision noise vs drive photon number is shown in panel \textbf{d}. Dashed and dotted lines show the individual contributions from the usual added noise (cryogenic HEMT and losses) and from the amplifier quantum noise, respectively. The two drive points from \textbf{a}-\textbf{c} are labeled with corresponding vertical lines.}}
	\label{fig:Fig4}
\end{figure}
Photon-pressure circuits are a highly promising platform for quantum-limited sensing of radio-frequency signals by upconversion and they are discussed in this context e.g. for dark matter axion detection\cite{Chaudhuri19, Backes21}.
The platform investigated here with a parametric amplifier being the RF upconverter itself might be a very interesting option towards an enhanced detection efficiency and similar approaches have also been discussed for other Josephson-based upconversion and detection schemes \cite{Eddins19, Schmidt20, Hatridge11}. 
To characterize the potential enhancement in RF flux sensitivity by the Josephson amplification in our setup, we detect the upconverted thermal fluctuations of the RF mode in the output field of the signal mode resonance with and without parametric gain, cf. Fig.~\ref{fig:Fig4}.
For this experiment, we work with a small photon-pressure cooperativity $\mathcal{C}_\mathrm{eff} \approx 0.8$ for both, the undriven and the amplification case to minimize the effects of dynamical backaction and mode hybridization, while still having a clearly detectable signal in the undriven case.
In a direct comparison between the detected output spectrum of both setups, we observe a significant intrinsic amplification of the upconverted RF noise in the amplification regime and in addition a significant background noise contribution from Josephson amplified HF cavity quantum noise, cf. Fig.~\ref{fig:Fig4}\textbf{a}, \textbf{b}.
For a quantification of the measurement imprecision of each configuration, the detected spectrum in units of quanta $S_{nn}(\Omega)$ is converted to RF flux spectral density using
\begin{eqnarray}
S_\Phi^\mathrm{tot}(\Omega) & = & \frac{2\Phi_\mathrm{zpf}^2}{\kappa_\mathrm{e}|\mathcal{G}(\Omega)|^2|\chi_\mathrm{s}|^2 n_- g_0^2}S_{nn}(\Omega) \\
& = & S_\Phi(\Omega) + S_\mathrm{imp}(\Omega)
\end{eqnarray}
with the Josephson gain $\mathcal{G} \neq 1$ being the main difference in the prefactor to the undriven case.
The imprecision noise takes the form
\begin{equation}
S_\mathrm{imp}(\Omega) = \frac{2\Phi_\mathrm{zpf}^2}{\kappa_\mathrm{e}|\mathcal{G}(\Omega)|^2|\chi_\mathrm{s}|^2 n_- g_0^2}\left[\frac{1}{2} + n_\mathrm{add} + n_\mathrm{JPA}(\Omega)\right],
\end{equation}
where $n_\mathrm{add} \approx 15$ is the effective noise added by the HEMT amplifier and the last term $n_\mathrm{JPA}(\Omega)$ is the imprecision noise contribution by the amplified quantum noise of the HF cavity.
In Fig.~\ref{fig:Fig4}\textbf{c} the effective RF flux spectral density is displayed, revealing a significant enhancement of the detection sensitivity in the amplifier state with $|\mathcal{G}| \sim 4$ at $\omega = \omega_\mathrm{s}$.
Due to the small ratio of linewidths in our device of $\kappa/\Gamma_0 \sim 5$ we see a strong frequency-dependence of the imprecision noise background, which could be easily compensated for in future implementations \cite{Hatridge11, LevensonFalk16} and which would be naturally reduced for a smaller linewidth RF mode.
The minimum imprecision noise at the signal mode resonance frequency, however, is still improved by a factor of $\sim3$.
To evaluate the minimum imprecision depending on the flux bias point and Josephson gain, we calculate the minimum at $\Omega = \Omega_\mathrm{s}$ for varying drive photon number and obtain
\begin{equation}
S_\mathrm{imp}^\mathrm{min} = \frac{\kappa^2\Phi_\mathrm{zpf}^2}{2\kappa_\mathrm{e}\mathcal{G}_\mathrm{s}^2 n_- g_0^2}\left[\frac{1}{2} + n_\mathrm{add} + n_\mathrm{JPA}(\Omega_\mathrm{s})\right]
\end{equation}
where $n_\mathrm{add} \approx 15$ and $n_\mathrm{JPA}(\Omega_\mathrm{s}) = 4\frac{\kappa_\mathrm{e}}{\kappa} \mathcal{G}_s\left(\mathcal{G}_\mathrm{s}-1\right)$.
The result is shown in Fig.~\ref{fig:Fig4}\textbf{d}.
Note that in the calculation of $S_\mathrm{imp}^\mathrm{min}$ and its individual contributions, shown in Fig.~\ref{fig:Fig4}\textbf{d}, we keep the effective cooperativity and the drive frequency constant but take into account the power- and flux-dependent parameters of our device, which is equivalent to having $\omega_0, \omega_\mathrm{s}, \kappa, \kappa_\mathrm{e}, \mathcal{K}, g_0$ and $n_-$ change with drive photon number $n_\mathrm{d}$ according to the flux sweep, cf. Fig.~\ref{fig:Fig2}.
Details on the expressions and their derivations can be found in the Supplementary Material.
As the drive photon number $n_\mathrm{d}$ is increased, the imprecision noise is significantly decreased by a factor of $3.4$, due to the additional gain provided by the intracavity amplification.
As the power is increased, however, eventually the imprecision noise becomes limited by the amplification of the quantum noise of the cavity.
One way to understand why the imprecision noise does not continue to improve for higher gain is that the quantum fluctuations of the cavity undergo amplification with gain $\mathcal{G}_\mathrm{s}^2$, while the intracavity fields from the photon-pressure coupling to the RF mode undergo only a net amplification of $\mathcal{G}_\mathrm{s}$: the second factor $\mathcal{G}_\mathrm{s}$ contributes instead to enhancing the cooperativity.
As the intracavity gain is increased and the amplified cavity input noise begins to dominate the amplification chain of the measurement, the imprecision noise for detecting the RF fields becomes worse again as it does not undergo the same amount of amplification as the cavity input fields.
\vspace{2mm}
\noindent\textbf{\textsf{\small Non-trivial bath dynamics}}
\vspace{0mm}
\begin{figure*}
	\centerline{\includegraphics[trim = {0.5cm, 11.3cm, 0.5cm, 1.0cm}, clip=True, width=0.95\textwidth]{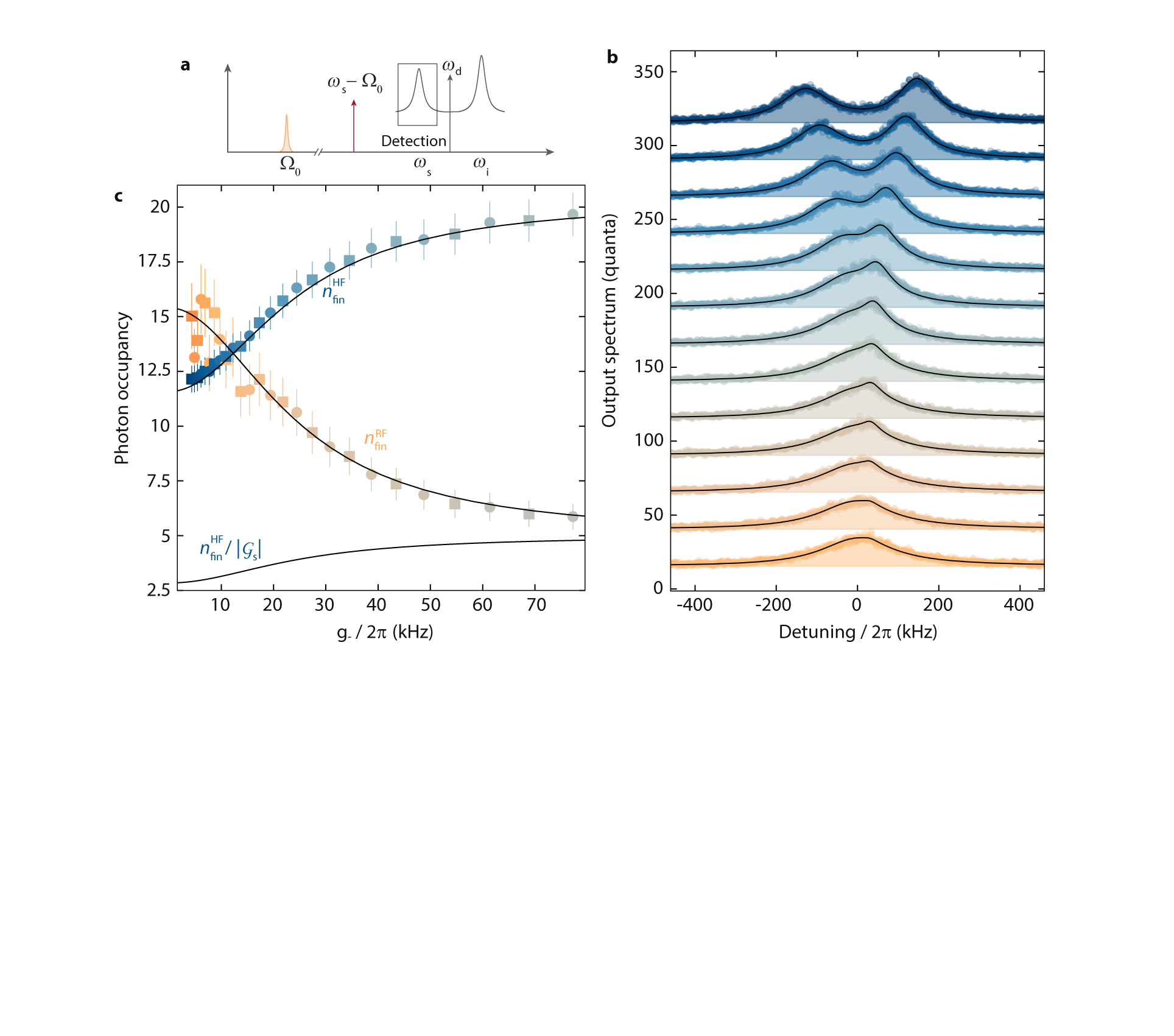}}
	\caption{\textsf{\textbf{Non-trivial bath dynamics in sideband-cooling with an amplified quantum bath.} \textbf{a} Schematic of the experiment. The amplifier signal resonance ($n_\textrm{d} = 4700$) is photon-pressure pumped on its red sideband with $\omega_\mathrm{p} = \omega_\mathrm{s}-\Omega_0$. The output power spectrum around $\omega \sim \omega_\mathrm{s}$ is detected using a spectrum analyzer. \textbf{b} Output spectra of the amplifier signal mode for increasing red-sideband pump power in units of quanta (bottom curve: lowest power, top curve: largest power). Circles are data, lines and shaded areas are fits with $n_\mathrm{th}^\mathrm{RF}$ and $\tilde{n}_\mathrm{th}^\mathrm{HF}$ as free parameters. Subsequent datasets and fit curves are offset by $+25$ each for clarity. For the lowest pump powers the output spectrum is dominated by the amplified HF cavity quantum noise, for medium powers an additional peak on top of the HF noise is emerging and for the highest powers, the two modes begin to hybridize and the output spectrum exhibits the onset of normal-mode splitting. From the fits to each pump power we determine the sideband-cooled final RF mode occupation and the resulting final HF mode occupation, which are plotted vs photon-pressure coupling rate $g_- = |\gamma_-|\tilde{g}_0$ in panel \textbf{c}. The residual RF mode occupation without photon-pressure pump is around $n_\mathrm{th}^\mathrm{RF}\sim 15$ and the sideband cooling reduces this thermal occupation to about $n_\mathrm{cool}^\mathrm{RF}\sim 6$ for the largest power used here. Strikingly, the effective HF mode occupation, arising from amplified quantum noise, is almost as high as the RF occupation for low pump powers and increases further with larger cooling of the RF mode, arriving at $n_\mathrm{fin}^\mathrm{HF} \approx 19 \gg n_\mathrm{fin}^\mathrm{RF}$. This cooling is considerably different from the usual sideband cooling with a hot HF mode, where the starting HF occupation is a fundamental limit for $n_\mathrm{lim}^\mathrm{RF}$ and indicates nonequilibrium heat flow from a cold to a hot reservoir. The effective HF photon number as seen by the RF mode is shown as line labeled with $n_\mathrm{fin}^\mathrm{HF}/|\mathcal{G}_\mathrm{s}|$ where $|\mathcal{G}_\mathrm{s}| \approx 4$.}}
	\label{fig:Fig5}
\end{figure*}
The asymmetric amplification, which is limiting the improvement of the imprecision noise with gain, leads to very unusual and non-trivial bath dynamics.
To reveal this effect, we discuss what happens in sideband-cooling with internal parametric gain in the HF cavity.
In the high gain regime, the effective temperature of the signal mode is in good approximation given by $T_\mathrm{eff} \approx \frac{\hbar\omega_\mathrm{s}}{k_\mathrm{B}}\tilde{n}_\mathrm{th}^\mathrm{HF}$ with the resonance frequency of the signal mode $\omega_\mathrm{s}$ and the effective mode occupation $\tilde{n}_\mathrm{th}^\mathrm{HF} = \frac{\mathcal{K}^2 n_\mathrm{d}^2}{|\kappa + 2i\Omega_\mathrm{s}|^2} = \mathcal{G}_\mathrm{s}\left(\mathcal{G}_\mathrm{s} -1 \right)$, arising from amplified quantum noise \cite{Clerk10}.
At the operation point for this experiment, we get $\tilde{n}_\mathrm{th}^\mathrm{HF} \approx 12$ and $T_\mathrm{eff} \approx 4.1\,$K.
To investigate experimentally how this large effective occupation impacts the RF mode in a sideband-cooling scheme, we prepare the HF cavity again in the amplifier state by a strong drive tone and pump the signal resonance with an additional red-detuned cooling tone, cf. Fig.~\ref{fig:Fig5}\textbf{a}.
From the output spectrum of the driven and pumped signal resonance, which contains the amplified upconverted RF fluctuation spectral density, the RF and HF mode occupations can then be extracted, cf. Fig.~\ref{fig:Fig5}\textbf{b}.
The RF mode occupation in equilibrium, i.e., without the cooling tone, is about $n_\mathrm{th}^\mathrm{RF} \sim 15$, i.e., considerably higher than complete thermalization with the mixing chamber at $T_\mathrm{b} = 15\,$mK would suggest.
Similar results have been observed before \cite{Bothner21, Rodrigues21} and are mainly explained by an imperfect radiation isolation between the sample and the cryogenic RF amplifier on the $3\,$K plate in our setup.
From comparison with the undriven case, where we get $n_\mathrm{th}^\mathrm{RF} \sim 13$, cf. Supplementary Material , we also find that the RF mode seems to be slightly heated by the parametric drive.
In any case, the naive expectation would be that considerable sideband-cooling of the RF mode will not be possible in this configuration as $n_\mathrm{th}^\mathrm{RF} \sim \tilde{n}_\mathrm{th}^\mathrm{HF}$.
The HF mode output spectra for varying power of the cooling tone, cf. Fig.~\ref{fig:Fig5}, in combination with a theoretical analysis, however, reveal a different and surprising scenario.
The theoretical model for the output power spectral density in units of quanta leads in good approximation to 
\begin{eqnarray}
S_{nn}(\omega) & = & \frac{1}{2} + n_\mathrm{add} + \kappa_1\kappa|\chi_\mathrm{s}^\mathrm{eff}|^2\frac{\tilde{n}_\mathrm{th}^\mathrm{HF}}{\mathcal{G}_\mathrm{s}} \nonumber \\
& & + \kappa_1g_\mathrm{eff}^2|\chi_+|^2|\chi_\mathrm{s}^\mathrm{eff}|^2\Gamma_0 n_\mathrm{th}^\mathrm{RF}
\end{eqnarray}
where $\chi_+^{-1} = \Gamma_0/2 + i(\Omega - \Omega_0)$, $\chi_\mathrm{s}^{-1} = \kappa/2 + i(\omega - \omega_\mathrm{s})$ and $\chi_\mathrm{s}^\mathrm{eff} = \chi_\mathrm{s}/(1 + g_\mathrm{eff}^2\chi_\mathrm{s}\chi_+)$.
Note that a step-by-step derivation is presented in Supplementary Note 8.
From the extracted occupation numbers, we find that the RF cooling factor increases with increasing power of the red-sideband tone and that the RF occupation gets significantly reduced to values far below $\tilde{n}_\mathrm{th}^\mathrm{HF}$.
The occupation of the HF mode simultaneously increases considerably beyond the original occupation of both modes, indicating that even in the strong-coupling regime where the RF and HF modes fully hybridize, the populations of the two bare modes are not in balance, in stark contrast to the phenomenology of photon-pressure cooling without parametric amplification.
For the largest cooling powers we report here, the device is already slightly above the threshold for normal-mode splitting and we obtain a final bare mode occupation of $\tilde{n}_\mathrm{th}^\mathrm{HF} \approx 19$ and $\tilde{n}_\mathrm{th}^\mathrm{RF} \approx 6$.
In fact, the final occupation of the RF mode can be expressed in a way that resembles the cooled occupation of linear sideband cooling
\begin{eqnarray}
n_\mathrm{fin}^\mathrm{RF} & = & \frac{\Gamma_0}{\kappa + \Gamma_0}\frac{4g_\mathrm{eff}^2 + \kappa(\kappa + \Gamma_0)}{4g_\mathrm{eff}^2 + \kappa\Gamma_0}n_\mathrm{th}^\mathrm{RF}\nonumber\\ & & + \frac{\kappa}{\kappa + \Gamma_0}\frac{4g_\mathrm{eff}^2 }{4g_\mathrm{eff}^2 + \kappa\Gamma_0}\frac{\tilde{n}_\mathrm{th}^\mathrm{HF}}{\mathcal{G}_\mathrm{s}}.
\end{eqnarray}
Here, it is not the effective thermal occupation of the signal mode that plays a role, but the effective occupation divided by the amplitude gain $\tilde{n}_\mathrm{th}^\mathrm{HF}/\mathcal{G}_\mathrm{s}$.
The analogue expression for the HF mode can be found in the Supplementary Material and there both occupations acquire an additional factor $\mathcal{G}_\mathrm{s}$.
This suggests that from the viewpoint of the HF mode both modes seem hotter by $\mathcal{G}_\mathrm{s}$.
A way to intuitively interpret this result is that due to the simultaneous parametric driving and photon-pressure coupling, it is not clear which effect happens first.
Are HF photons first transferred to the RF mode or first amplified?
The answer according to this interpretation would be that they acquire one amplitude gain factor before and one amplitude gain factor after the interaction.
\vspace{2mm}
\noindent\textbf{\textsf{\small DISCUSSION}}
\vspace{2mm}
In summary, we have presented a series of experiments based on photon-pressure coupled circuits, one of which could be operated as a parametric amplifier.
This operation mode leads to several interesting effects.
First, the amplifier regime leads to a large parametric enhancement of the linearized single-photon coupling rate and of the photon-pressure cooperativity between the two circuits, in total up to more than an order of magnitude compared to the gainless operation.
Part of this enhancement is originating from a photon-pressure modulation of the HF cavity Kerr nonlinearity, an effect described hitherto only in theoretical work.
Secondly, we demonstrated that the internal amplification also significantly reduces the imprecision noise of upconverted radio-frequency flux signals, which is a promising perspective for optimized RF sensing applications.
Finally, we found that parametric amplification within the photon-pressure coupled system allows for non-trivial sideband-cooling of the RF mode with a quantum-heated amplifier, where the effective, quantum-noise related temperature of the amplifier mode is not constituting the cooling limit for the RF mode.
Furthermore, we have shown that Kerr amplification in the photon-pressure cavity leads to unexpected bath dynamics that if further explored could potentially lead to interesting applications in quantum bath engineering.
Our experiments reveal that Kerr nonlinearities can be an extremely versatile and useful resource for engineering enhanced and novel photon-pressure based devices.
We believe the investigation of the possibilities has just begun, and a fruitful exchange of ideas and protocols with closely related platforms such as Kerr optomechanics will advance the exploration of nonlinearities in these systems further.
The Josephson-based Kerr nonlinearity has also already been demonstrated to allow for a variety of interesting microwave photon manipulation techniques such as cat state generation and stabilization, bosonic code quantum information processing, nonreciprocal photon transport or the implementation of superconducting qubits.
Integrating some of these possibilities into photon-pressure or Kerr optomechanical platforms might allow for elaborate quantum control of RF circuits and mechanical oscillators in the future.
\vspace{10mm}

\noindent\textbf{\textsf{\small References}}

%
\vspace{3mm}
\noindent\textbf{\textsf{\small Acknowledgements}}
This research was supported by the Netherlands Organisation for Scientific Research (NWO) in the Innovational Research Incentives Scheme -- VIDI, project 680-47-526.
This project has received funding from the European Research Council (ERC) under the European Union's Horizon 2020 research and innovation programme (grant agreement No 681476 - QOMD) and from the European Union's Horizon 2020 research and innovation programme under grant agreement No 732894 - HOT.
The authors thank A.~Nunnenkamp for valuable discussions.
\vspace{3mm}
\noindent\textbf{\textsf{\small Author contributions}}
All authors developed the concepts and ideas.
ICR and DB designed and fabricated the device, performed the measurements, analyzed the data and developed the theoretical treatment.
ICR and DB edited the manuscript with input from GAS.
All authors discussed the results and the manuscript.
\vspace{3mm}
\noindent\textbf{\textsf{\small Competing interest}}
The authors declare no competing interests.

\clearpage

\widetext

\noindent\textbf{\textsf{\Large  Supplementary Material for:\\\\ Parametrically enhanced interactions and non-trivial bath dynamics \\ in a photon-pressure Kerr amplifier}}

\normalsize
\vspace{.3cm}

\noindent\textsf{I.C.~Rodrigues, G.~A.~Steele, and D.~Bothner}

\vspace{.2cm}

\renewcommand{\theequation}{S\arabic{equation}}

\renewcommand{\bibnumfmt}[1]{[S#1]}
\renewcommand{\citenumfont}[1]{S#1}

\setcounter{figure}{0}
\setcounter{equation}{0}

\addtocontents{toc}{\protect\setcounter{tocdepth}{1}}

\tableofcontents

\newpage

\renewcommand{\figurename}{Supplementary Figure}

\section{Supplementary Note 1: Device fabrication}
\label{Section:Fab}
\begin{figure}[h]
	\centerline {\includegraphics[trim={0.5cm 6cm 0.5cm 0cm},clip=True,scale=0.75]{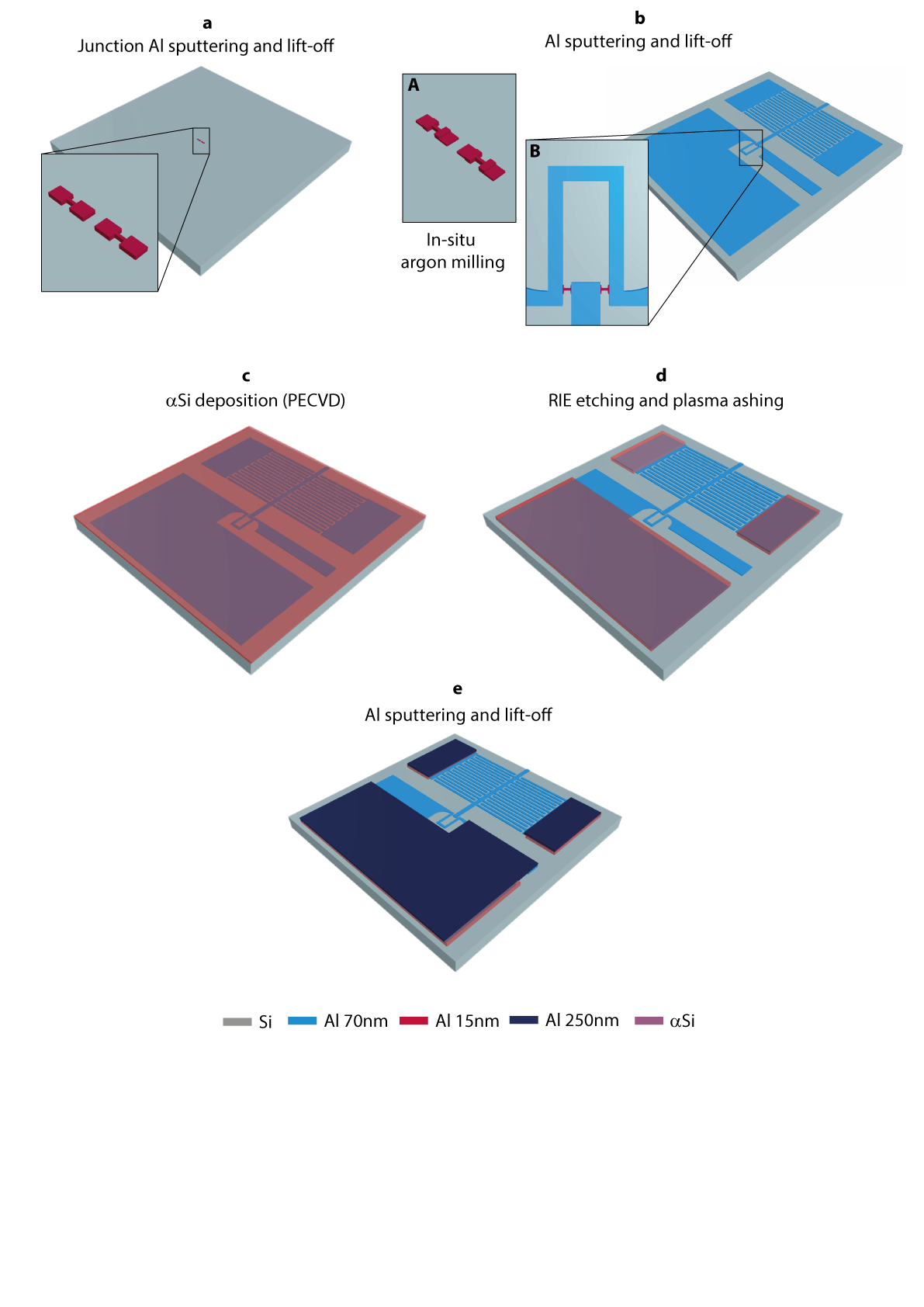}}
	\caption{\textsf{\textbf{Schematic device fabrication.} \textbf{a} shows the weak-link Josephson junctions with contact pads, patterned in the first fabrication step. \textbf{b} shows the patterned second Aluminum layer, forming the bottom of the RF parallel plate capacitor, the SQUID loop and the HF cavity. Inset \textbf{A} is showing the in-situ Argon-milled Josephson junctions prior to the deposition (the resist is not shown for better visibility of the milled structures). Inset \textbf{B}  shows a zoom-in of the 3D SQUID. \textbf{c} shows the sample after the deposition of $\alpha$Si. \textbf{d} shows the device after the subsequent SF$_6/$He reactive ion etching step, finished by an in-situ $\textrm{O}_{2}$ plasma ashing. \textbf{e} shows the final device after the deposition of the last Aluminum layer. }}
	\label{fig:Fab}
\end{figure}

\begin{itemize}
	\item \textbf{Step 0: Marker patterning.} Prior to the device fabrication, we performed the patterning of alignment markers on a full $4\,$inch Silicon wafer (intrinsic, high resisitivity, thickness $500\,$nm), required for the electron-beam lithography (EBL) alignment of the following fabrication steps.
	The structures were patterned using a CSAR62.13 resist mask and sputter deposition of $50\,$nm Molybdenum-Rhenium alloy.
	After undergoing a lift-off process, the only remaining structures on the wafer were the markers.
	The complete wafer was diced into $14\times14\,$mm$^2$ chips, which were used individually for the subsequent fabrication steps.
	The step was finalized by a series of several acetone and IPA rinses.
	\item \textbf{Step 1: Junctions patterning.} As first step in the fabrication, we pattern weak links which afterwards result in constriction type Josephson junctions (cJJs) between the arms of the SQUID.
	The weak link nanowires were patterned together with larger pads, cf. Supplementary Fig.~\ref{fig:Fab}\textbf{a}, which were used to achieve good electrical contact with the rest of the circuit, cf. Step~3.
	The nanowires are designed to be $\sim 50\,$nm wide and $\sim 100\,$nm long at this point of the fabrication, and each pad is $500\times500\,$nm$^2$ large.
	For this fabrication step, a CSAR62.09 was used as EBL resist and the development was done by dipping the exposed sample into Pentylacetate for $60\,$seconds, followed by a solution of MIBK:IPA (1:1) for $60\,$seconds, and finally rinsed in IPA, where MIBK is short for methyl isobutyl ketone and IPA for isopropyl alcohol. 
	The sample was subsequently loaded into a sputtering machine where a $15\,$nm layer of Aluminum was deposited.
	Finally, the chip was placed at the bottom of a beaker containing a small amount of Anisole and inserted into an ultrasonic bath for a few minutes where the sample underwent a lift-off process.
	The step was finalized by a series of several acetone and IPA rinses.
	\item \textbf{Step 2: Bottom RF capacitor plate and HF resonator patterning.} As second step in the fabrication, we pattern the bottom plate of the parallel plate capacitor, the inductor wire of the radio-frequency cavity (which also forms part of the  SQUID loop), the remaining part of the SQUID cavity (cf. Supplementary Fig.~\ref{fig:Fab}\textbf{b}) and the center conductor of the SQUID cavity feedline by means of EBL using CSAR62.13 as resist. 
	After the exposure, the sample was developed in the same way as in the first fabrication step and loaded into a sputtering machine.
	In the sputter system, we performed an argon milling step for two minutes and afterwards deposited $70\,$nm of Aluminum.
	The milling step, performed in-situ and prior to the deposition, very efficiently removes the oxide layer which was formed on top of the previously sputtered weak link pads, and therefore allows for good electrical contact between the two layers.
	After the deposition, the unpatterned area was lifted-off by means of an ultrasonic bath in room-temperature Anisole for a few minutes. 
	The step was finalized by a series of several acetone and IPA rinses.
	\item \textbf{Step 3: Amorphous silicon deposition.} The deposition of the dielectric layer of the parallel plate capacitor was done using a plasma-enhanced chemical vapor deposition (PECVD) process.
	To guarantee low dielectric losses in the material, the chamber underwent an RF cleaning process overnight and only afterwards the deposition of $\sim130\,$nm of amorphous silicon ($\alpha$Si) was performed.
	At this point of the fabrication, the whole sample is covered with dielectric, cf. Supplementary Fig.~\ref{fig:Fab}\textbf{c}.
	\item \textbf{Step 4: Reactive ion etch patterning of $\alpha$Si.} We spin-coat a double layer of resist (PMMA 950K A4 and ARN-7700-18) on top of the $\alpha$Si-covered sample, and expose the next pattern with EBL. 
	Prior to the development of the pattern, a post-bake of 2 minutes at $\sim115\,^{\circ}{\rm C}$ was required.
	Directly after, the sample was dipped into MF-321 developer for 2 minutes and 30 seconds, followed by H$_2$O for 30 seconds and lastly rinsed in IPA.
	To finish the third step of the fabrication, the developed sample underwent a SF$_6$/He reactive ion etching (RIE) to remove the amorphous Silicon.
	To conclude the etching step, we performed a O$_2$ plasma ashing in-situ with the RIE process to remove resist residues, the result is shown schematically in Supplementary Fig.~\ref{fig:Fab}\textbf{d}.
	\item \textbf{Step 5: Top capacitor plate and ground plane patterning.} As final step, the sample was again spin-coated with CSAR62.13 and the top plate of the RF capacitor as well as all ground plane and the low-frequency feedline were patterned with EBL.
	The resist development was done identical to the ones in the second and third steps.
	Afterwards, the sample was loaded into a sputtering machine where an argon milling process was performed in-situ for 2 minutes, in order to have good electrical contact between the top and bottom plates of the low-frequency capacitor, similar to what was done between the second and third fabrication steps.
	After the milling, a $250\,$nm thick layer of Aluminum was deposited and finally an ultrasonic lift-off procedure was performed.
	The step was finalized by a series of several acetone and IPA rinses.
	With this, the sample fabrication process was essentially completed, cf. Supplementary Fig.~\ref{fig:Fab}\textbf{e}.
	\item \textbf{Step 6: Dicing and mounting.} At the end of the fabrication, the sample was diced to a $10\times10\,$mm$^2$ size and mounted to a printed circuit board (PCB), wire-bonded to microwave feedlines and ground, and packaged into a radiation tight copper housing.
\end{itemize}

A schematic representation of the fabrication process can be seen in Supplementary Fig.~\ref{fig:Fab}, omitting the initial patterning of the electron beam markers and the sample mounting.

\section{Supplementary Note 2: Measurement setup}
\label{Section:Setup}

\begin{figure}[h]
	\centerline{\includegraphics[trim = {0cm, 4cm, 0.6cm, 1.5cm}, clip, scale=0.8]{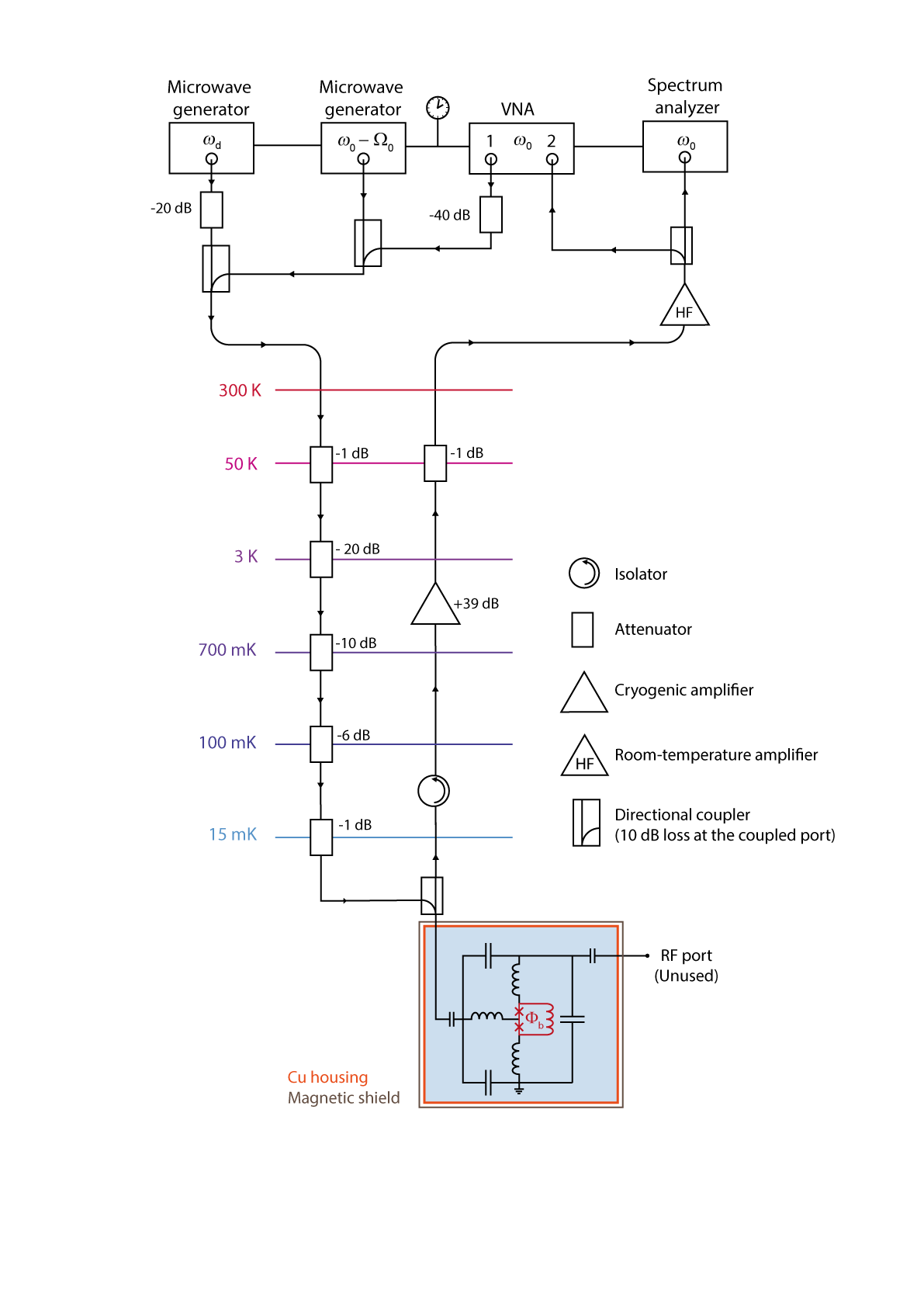}}
	\caption{\textsf{\textbf{Schematic of the measurement setup.} Detailed information is given in the text.}}
	\label{fig:Measurement Setup}
\end{figure}

The device used for the experiments reported in this paper was mounted on the bottom plate of a dilution refrigerator with a base temperature $T_\textrm{b} \approx 15\,$mK.
A schematic representation of the experimental measurement setup is shown in Supplementary Fig.~\ref{fig:Measurement Setup}.
The fabricated sample was glued and wire-bonded onto a printed circuit board (PCB) and afterwards mounted in a radiation tight copper housing.
In order to apply an out-of-plane magnetic field needed for flux biasing the SQUID cavity, a superconducting magnet was placed in very close proximity with the device inside the copper case.
At last, the PCB was connected to two coaxial lines by means of SMP connectors and the whole assembly was placed in a cryoperm magnetic shield.
The magnet was connected with DC wires, allowing for the field to be tuned with a DC current (not shown).
One of the coaxial lines was used as input/output for the high-frequency (HF) SQUID cavity and the other one as input/output access port to the radio-frequency (RF) circuit mode.
Since the RF line was never used during the course of this experiment, it was left disconnected in Supplementary Fig.~\ref{fig:Measurement Setup} to avoid any confusions.
In the real setup and experiment, however, it was connected to an RF directional coupler and a cryogenic RF amplifier (switched off), cf. also the Supplementary Material of Ref.~\cite{Rodrigues21S}.
As the HF cavity was measured in a reflection geometry, the input and output signals were split by means of directional coupler located between the $15$ and $100\,$mK plate.
In addition, the output signal was sent through an isolator and amplified by a cryogenic amplifier situated further in the output chain and the input line was heavily attenuated in order to balance the thermal radiation from the line to the base temperature of the individual fridge stages.
At room temperature, all the experiments were conducted with a single experimental setup.
This configuration was composed of two microwave generators, one responsible for providing a coherent drive tone to parametrically drive the HF cavity and the second one acting as a continuous red-sideband tone.
As these two tones were combined via a directional coupler, each of them could be switched on and off as desired, without requiring a physical change in the setup.  
Furthermore, an additional weak probe signal was also combined with the other existing tones to measure the cavity response in presence of drive only or the combined drive and pump tones. 
Once outside of the fridge, the output signal was firstly amplified by a room-temperature low-noise microwave amplifer and then analyzed individually by a spectrum analyzer and a VNA.
During the detection of thermal noise with the signal analyzer, the VNA scan was stopped and the VNA output power was completely switched off.
For all experiments, the microwave sources and vector network analyzers (VNA) as well as the spectrum analyzer used a single reference clock of one of the devices.

\section{Supplementary Note 3: Flux dependence}
\label{Section:Flux Dependence}

\begin{figure}[h]
	\centerline{\includegraphics[trim = {0.6cm, 13.2cm, 0.3cm, 6cm}, clip, scale=0.8]{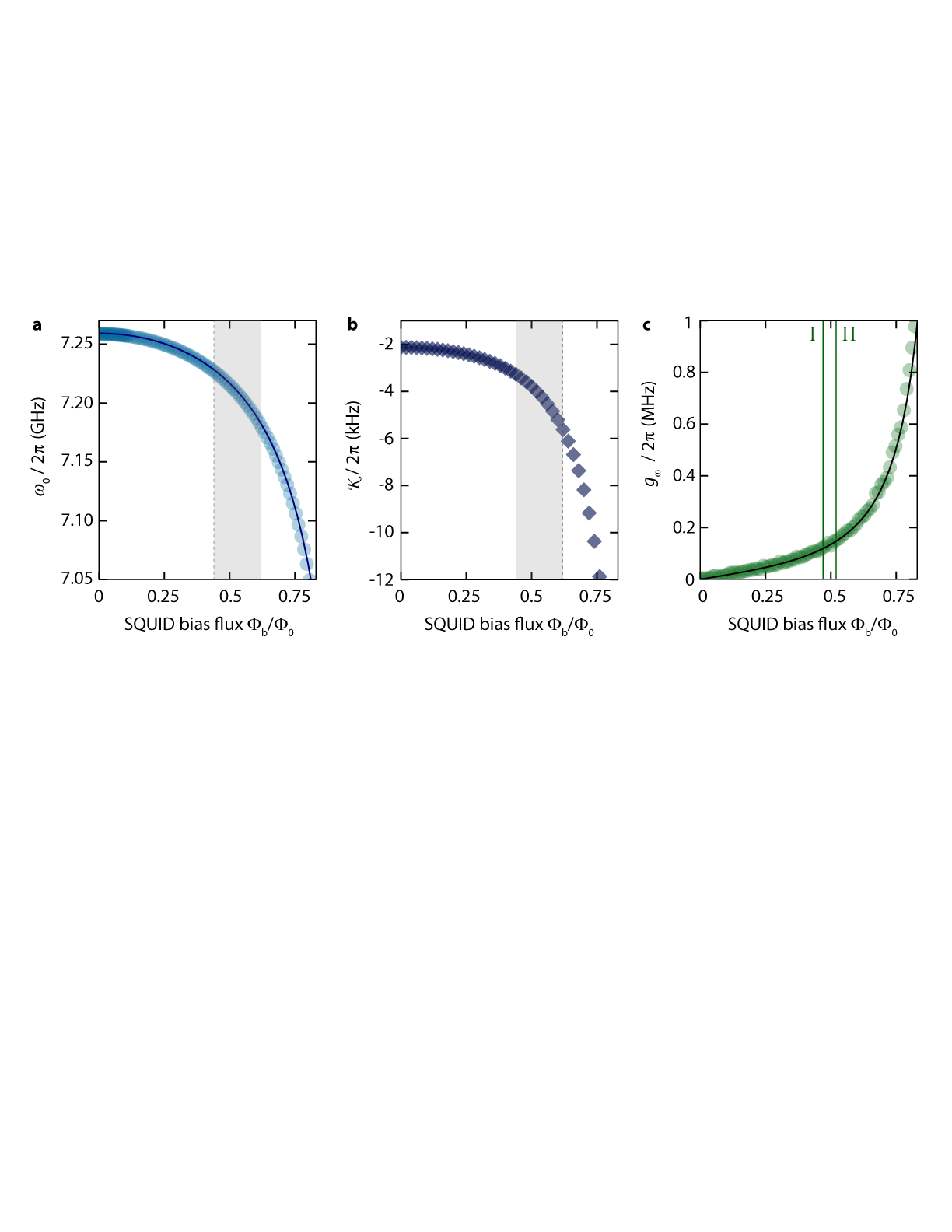}}
	\caption{\textsf{\textbf{Characterization of the SQUID cavity and photon-pressure coupling strength vs SQUID bias flux.} Panel \textbf{a} shows the experimentally measured SQUID cavity resonance frequency $\omega_0$ vs SQUID bias flux $\Phi_\textrm{b}/\Phi_0$. Circles are data, line is a fit curve using Eq.~(\ref{eqn:FluxDep}). Based on the circuit parameters we extract the cavity Kerr nonlinearity $\mathcal{K}$ via Eq.~(\ref{eqn:beta}), the result is shown vs bias flux in \textbf{b}. The gray regions in \textbf{a} and \textbf{b} indicate the flux operation range of main paper Fig.~2. Panel \textbf{c} shows the single-photon strength $g_0 = \frac{\partial\omega_0}{\partial\Phi}\Phi_\textrm{zpf}$ vs magnetic flux. The points are based on the experimentally measured values of $\frac{\partial\omega_0}{\partial\Phi}$ considering $\Phi_\textrm{zpf} \approx 635\,\mu\Phi_0$ and the line is calculated based on the fit curve to the flux arch. The two green vertical lines indicate the two different flux operation points of main paper Figs.~3-5 (point I, left line) and of point II (right line), respectively. }}
	\label{fig:stuffvsflux}
\end{figure}

The resonance frequency of a SQUID cavity with a symmetric SQUID can be described by \cite{Rodrigues21S}
\begin{equation}
\omega_0(\Phi_\mathrm{b}) = \frac{\omega_0(0)}{\sqrt{\Lambda + \frac{1-\Lambda}{\cos{\left(\pi\frac{\Phi}{\Phi_0}\right)}}}}
\label{eqn:FluxDep}
\end{equation}
where $\Phi$ corresponds to the total flux threading the SQUID loop and $\omega_0(0)$ is the sweetspot resonance frequency at an external bias flux $\Phi_\textrm{b} = 0$.
The parameter $\Lambda = (L_\mathrm{HF}-\frac{1}{2}L_\mathrm{j0})/L_\mathrm{HF}$ with the total high-frequency sweetspot inductance $L_\mathrm{HF}$ and the single junction Josephson inductance $L_{\mathrm{j}0}$ is a measure for the contribution of the Josephson inductance to the total inductance.
For zero bias current and the (geometric plus kinetic) loop inductance $L_\mathrm{loop}$ the total (normalized) flux threading the SQUID is given by
\begin{eqnarray}
\frac{\Phi}{\Phi_0} = \frac{\Phi_\mathrm{b}}{\Phi_0} + \frac{L_\mathrm{loop}J}{\Phi_0}
\end{eqnarray}
with the circulating current $J$.
The circulating current can also be expressed as
\begin{equation}
J = -I_\mathrm{c}\sin{\left(\pi\frac{\Phi}{\Phi_0}\right)}
\end{equation}
with the zero bias critical current of a single junction $I_\mathrm{c} = \frac{\Phi_0}{2\pi L_\mathrm{j0}}$.
Using the screening parameter $\beta_L = \frac{2L_\mathrm{loop}I_\mathrm{c}}{\Phi_0} = \frac{L_\mathrm{loop}}{\pi L_\mathrm{j0}}$ the relation for the total flux can be written as
\begin{eqnarray}
\frac{\Phi}{\Phi_0} = \frac{\Phi_\mathrm{b}}{\Phi_0} - \frac{\beta_L}{2}\sin{\left(\pi\frac{\Phi}{\Phi_0}\right)}.
\label{eqn:TotFlux}
\end{eqnarray}
Figure~1\textbf{d} of the main paper and Supplementary Fig.~\ref{fig:stuffvsflux}\textbf{a} show the experimentally determined SQUID cavity resonance frequency tuning with external magnetic flux $\Phi_\mathrm{b}$ and a fit curve using Eq.~(\ref{eqn:FluxDep}), where the relation between the applied external flux $\Phi_\mathrm{b}$ and the total flux in the SQUID $\Phi$ is given by Eq.~(\ref{eqn:TotFlux}).
As fit parameters we obtain $\beta_L =1.07$ and $\Lambda = 0.946$, i.e., the SQUID Josephson inductance contributes about $5.4\%$ to the total HF inductance.
Based on the sweet-spot resonance frequency of the SQUID cavity 
\begin{equation}
\omega_0(0) = \frac{1}{\sqrt{L_{\mathrm{HF}} C_{\textrm{HF}} }},
\end{equation} 
and on the capacitance of the SQUID cavity $C_{\textrm{HF}} \sim1.3\,$pF, we extract the total inductance of the high frequency mode to be $L_{\textrm{HF}} = 370\,$pH and with $\Lambda$ we get the sweet-spot inductance of a single junction $L_\mathrm{j0} = 40\,$pH.
As shown in Supplementary Figure \ref{fig:stuffvsflux}\textbf{b}, based on these parameters we calculate the flux dependent Kerr non-linearity $\mathcal{K}$ via
\begin{equation}
\mathcal{K} (\Phi_\mathrm{b}) = -\frac{e^2}{2\hbar C_{\textrm{HF}}} \left(\frac{L_\mathrm{j}(\Phi_\mathrm{b})}{L_{\textrm{HF}} - L_\mathrm{j0}/2 + L_\mathrm{j}(\Phi_\mathrm{b})}\right)^3.
\label{eqn:beta}
\end{equation} 
Furthermore, based on the flux responsivity of the cavity $\frac{\partial\omega_0}{\partial\Phi}$ we can estimate the single-photon coupling strength $g_0 = \frac{\partial\omega_0}{\partial\Phi}\Phi_\textrm{zpf}$, the result is shown in Supplementary Fig.~\ref{fig:stuffvsflux}\textbf{c}.
The estimation for the zero-point flux fluctuation value $\Phi_\textrm{zpf} \approx 635\,\mu\Phi_0$ used for the calculation of $g_0$ can be found in Supplementary Note 5 of Ref.~\cite{Rodrigues21S}.

\section{Supplementary Note 4: Data analysis and fitting}

\subsection{Background correction of network analysis data}
Due to impedance imperfections in both, the input and output lines, the ideal reflection response is modified by cable resonances and interferences within the setup.
Origin of these imperfections are all connectors, directional couplers, circulators, attenuators in the signal lines.
In addition, the cabling has a frequency-dependent attenuation.
To take all these modifications into account, we assume that the final reflection parameter $S_{11}^\mathrm{real}$ can be described by
\begin{equation}
S_{11}^\mathrm{real} = \left(a_0 + a_1\omega + a_2\omega^2  \right)\left[1 - f(\omega)e^{i\theta}\right]e^{i\left(\phi_0  + \phi_1\omega \right)}
\end{equation}
when the ideal reflection would be given by
\begin{equation}
S_{11}^\mathrm{ideal} = 1 - f(\omega).
\end{equation}
The real-valued numbers $a_0, a_1, a_2, \phi_0, \phi_1$ describe a frequency dependent modification of the background reflection, and the phase factor $\theta$ takes into account possible interferences such as parasitic signals bypassing the reflection from the device itself.
The function $f(\omega)$ depends on the exact experiment described but usually correponds to a function of the form $f(\omega) = \kappa_\mathrm{e}\chi(\omega)$ with the external decay rate $\kappa_\mathrm{e}$ and the system susceptibility $\chi(\omega)$.
Our standard fitting routine begins with removing the actual resonance signal from the reflection, leaving us with a gapped background reflection, which we fit using
\begin{equation}
S_{11}^\mathrm{bg} = \left(a_0 + a_1\omega + a_2\omega^2  \right)e^{i\left(\phi_0  + \phi_1\omega \right)}.
\end{equation}
Subsequently, we remove this background function from all measurement traces by complex division.
The resonance circle rotation angle $\theta$ can then be rotated off additionally, but is very small for our experiments, indicating that there is no significant reflection interference in our setup.
The result of both corrections is what we present as background-corrected data or reflection data in all figures, respectively.

\subsection{Two-level system losses}
\label{sec:TLS_theory}

In the experiments, we observe photon number dependent cavity linewidths, which we attribute to the presence and saturation of two-level systems (TLSs) in our device.
We model the power-dependent losses therefore with a TLS-induced decay rate according to\cite{Capelle20S}
\begin{equation}
\kappa_\mathrm{tls} = \kappa_1\left[ 1 - \frac{n_\mathrm{d}/n_\mathrm{cr}}{\sqrt{1 + n_\mathrm{d}/n_\mathrm{cr}}} \frac{1 + \sqrt{1 + n_\mathrm{d}/n_\mathrm{cr}}}{\left(\Delta_\mathrm{tls}/\Gamma_2\right)^2 + \left( 1 + \sqrt{1 + n_\mathrm{d}/n_\mathrm{cr}}  \right)^2}  \right]
\label{eqn:TLSloss}
\end{equation}
where $n_\mathrm{d}$ is the drive photon number inside the cavity, $n_\mathrm{cr}$ is the characteristic or critical photon number for TLS saturation, $\Delta_\mathrm{tls}$ is the detuning between the drive and the frequency of interest and $\Gamma_2$ is the effective mean TLS dephasing rate.
Note that we typically take an average value for $\Delta_\mathrm{tls}$, for instance the resonance frequency of the mode of interest.

\section{Supplementary Note 5: Theory of a driven Kerr cavity}
For many experiments presented here, namely main paper Fig.~2 and Supplementary Note~VI, the photon-pressure coupling can be neglected to first order in the device description and data analysis. 
The relevant experiments are conducted with a strong parametric drive only or with a parametric drive and a probe tone few MHz detuned from the drive.
Due to the parameter regime of the device, in particular due to the large sideband-resolution factor $\kappa/\Omega_0 \sim 1000$, the contribution from the photon-pressure to the HF cavity properties, does not play a significant role then.
Similarly, the dynamical backaction to the RF mode by the drive can be neglected.
For this reason, we will describe and analyze these experiments without a photon-pressure sideband pump with a bare Kerr cavity.
We note, that we also performed all the calculations including the photon-pressure term to confirm this statement quantitatively, but do not explicitly include the single-drive situation including photon-pressure terms in this Supplementary Material.

\subsection{Classical equation of motion}
We model the classical intracavity field $\alpha$ of the HF circuit without photon-pressure coupling using the equation of motion
\begin{equation}
\dot{\alpha} = \left[i(\omega_0 + \mathcal{K}|\alpha|^2) - \frac{\kappa}{2}\right]\alpha + i\sqrt{\kappa_\mathrm{e}}S_\mathrm{in}
\end{equation}
where $\omega_0$ is the bare cavity resonance frequency, $\mathcal{K}$ is the Kerr nonlinearity (frequency shift per photon), $\kappa$ is the bare total linewidth, $\kappa_\mathrm{e}$ is the external linewidth and $S_\mathrm{in}$ is the input field.
The intracavity field is normalized such that $|\alpha|^2 = n_\mathrm{c}$ corresponds to the intracavity photon number $n_\mathrm{c}$ and $|S_\mathrm{in}|^2$ to the input photon flux (photons per second).

\subsection{Single-tone solution}
With a single tone drive field $S_\mathrm{in} = S_\mathrm{d}e^{i(\omega_\mathrm{d}t + \phi_\mathrm{d})}$ and the Ansatz $\alpha_\mathrm{d}e^{i\omega_\mathrm{d}t}$, where $S_\mathrm{d}$ and $\alpha_\mathrm{d}$ are chosen to be real-valued, we get
\begin{equation}
\left[\frac{\kappa}{2}+i\Delta_\mathrm{d}\right]\alpha_\mathrm{d} - i\mathcal{K}\alpha_\mathrm{d}^3 = i\sqrt{\kappa_\mathrm{e}}S_\mathrm{d}e^{i\phi_\mathrm{d}}
\end{equation}
with $\Delta_\mathrm{d} = \omega_\mathrm{d} - \omega_0$ the detuning between the drive and the undriven cavity resonance frequency.
From this, by multiplication with its complex conjugate, we obtain a third order polynomial for the intracavity photon number $n_\mathrm{d} = \alpha_\mathrm{d}^2$, which is given by
\begin{equation}
\mathcal{K}^2n_\mathrm{d}^3 - 2\mathcal{K}\Delta_\mathrm{d} n_\mathrm{d}^2 + \left(\Delta_\mathrm{d}^2 + \frac{\kappa^2}{4}\right)n_\mathrm{d} - \kappa_\mathrm{e}S_\mathrm{d}^2 = 0.
\end{equation}
To obtain the full, complex intracavity field with respect to the drive field, we also need the phase $\phi_\mathrm{d}$, which is given by
\begin{equation}
\phi_\mathrm{d} = \atan2\left(-\frac{\kappa}{2}, \Delta_\mathrm{d} - \mathcal{K} n_\mathrm{d}\right).
\end{equation}
The intracavity field is then given by $\alpha = \sqrt{n_\mathrm{d}}e^{-i\phi_\mathrm{d}}$.
\subsection{Linearized two-tone solution}
If the Kerr cavity is driven by a strong drive field and a weaker second input field at frequency $\omega_\mathrm{p}$, we write for the total input field
\begin{equation}
S_\mathrm{in} = S_\mathrm{d}e^{i(\omega_\mathrm{d}t + \phi_\mathrm{d})} + S_\mathrm{p}e^{i\omega_\mathrm{p}t}
\end{equation}
and as Ansatz for the intracavity field we choose
\begin{equation}
\alpha = \alpha_\mathrm{d}e^{i\omega_\mathrm{d}t} + \gamma_-e^{i\omega_\mathrm{p}t} + \gamma_+e^{i(2\omega_\mathrm{d}-\omega_\mathrm{p})t}
\end{equation}
with complex-valued amplitudes $\gamma_-$ and $\gamma_+$.
(Note that this Ansatz already anticipates the well-known solution for a linearized two-tone response of the Kerr cavity.)
Inserting these into the equation of motion, going to the frame rotating with the signal $\omega_\mathrm{p}$, and linearizing the solution by dropping all terms not linear in $\gamma_-, \gamma_+$  yields
\begin{eqnarray}
& &\left[ \frac{\kappa}{2} +i(\Delta_\mathrm{d} - \mathcal{K} n_\mathrm{d})\right]\alpha_\mathrm{d}e^{i\Omega_\mathrm{dp}t} + \left[ \frac{\kappa}{2} +i(\Delta_\mathrm{p} - 2\mathcal{K} n_\mathrm{d})\right]\gamma_- + \left[ \frac{\kappa}{2} +i(\Delta_\mathrm{p} - 2\mathcal{K} n_\mathrm{d} + 2\Omega_\mathrm{dp})\right]\gamma_+e^{i2\Omega_\mathrm{dp}t}\nonumber\\
& & =  i\mathcal{K} n_\mathrm{d}\gamma_+^* + i\mathcal{K} n_\mathrm{d}\gamma_-^*e^{i2\Omega_\mathrm{dp}t} +i\sqrt{\kappa_\mathrm{e}}S_\mathrm{d}e^{i(\Omega_\mathrm{dp}t + \phi_\mathrm{d})} + i\sqrt{\kappa_\mathrm{e}}S_\mathrm{p}.
\end{eqnarray}
where $\Omega_\mathrm{dp} = \omega_\mathrm{d} - \omega_\mathrm{p}$ and $\Delta_\mathrm{p} = \omega_\mathrm{p} - \omega_0$.
We sort this equation by frequency components now and get three individual equations
\begin{eqnarray}
\left[\frac{\kappa}{2} +i(\Delta_\mathrm{d} - \mathcal{K} n_\mathrm{d})\right]\alpha_\mathrm{d} & = & i\sqrt{\kappa_\mathrm{e}}S_\mathrm{d}e^{i\phi_\mathrm{d}}\\
\left[\frac{\kappa}{2} +i(\Delta_\mathrm{p} - 2\mathcal{K} n_\mathrm{d})\right]\gamma_- - i\mathcal{K} n_\mathrm{d}\gamma_+^*& = & i\sqrt{\kappa_\mathrm{e}}S_\mathrm{p}\\
\left[\frac{\kappa}{2} +i(\Delta_\mathrm{p} - 2\mathcal{K} n_\mathrm{d} + 2\Omega_\mathrm{dp})\right]\gamma_+ - i\mathcal{K} n_\mathrm{d}\gamma_-^*& = & 0.
\end{eqnarray}
The first of these equations is now exactly the same as the one we obtained for single-tone driving.
With the procedure described in the previous section, the intracavity field $\alpha_\mathrm{d}$, the intracavity photon number $n_\mathrm{d}$ and the phase $\phi_\mathrm{d}$ can be determined.
Having solved for $n_\mathrm{d}$ allows then to solve also for $\gamma_-$ and $\gamma_+$.
We write the second and third equations as
\begin{eqnarray}
\frac{\gamma_-}{\chi_\mathrm{p}(0)} - i\mathcal{K} n_\mathrm{d}\gamma_+^*& = & i\sqrt{\kappa_\mathrm{e}}S_\mathrm{p}\\
\frac{\gamma_+}{\chi_\mathrm{p}(2\Omega_\mathrm{dp})} - i\mathcal{K} n_\mathrm{d}\gamma_-^* & = & 0
\end{eqnarray}
where we defined
\begin{equation}
\chi_\mathrm{p}(\Omega) = \frac{1}{\frac{\kappa}{2} + i\left(\Delta_\mathrm{p} -2\mathcal{K}n_\mathrm{d} + \Omega\right)}.
\end{equation}
We solve for $\gamma_+$ and get by complex conjugation
\begin{equation}
\gamma_+^* = -i\mathcal{K} n_\mathrm{d} \chi_\mathrm{p}^*(2\Omega_\mathrm{dp})\gamma_-
\end{equation}
Inserting this into the equation for $\gamma_-$ gives
\begin{eqnarray}
\gamma_- & = & i\frac{\chi_\mathrm{p}(0)}{1-\mathcal{K}^2 n_\mathrm{d}^2\chi_\mathrm{p}(0)\chi_\mathrm{p}^* (2\Omega_\mathrm{dp})}\sqrt{\kappa_\mathrm{e}}S_\mathrm{p}\\
& = & i\chi_\mathrm{g}(0) \sqrt{\kappa_\mathrm{e}} S_\mathrm{p}
\end{eqnarray}
where in the last step we defined
\begin{equation}
\chi_\mathrm{g}(\Omega) = \frac{\chi_\mathrm{p}(\Omega)}{1-\mathcal{K}^2 n_\mathrm{d}^2 \chi_\mathrm{p}(\Omega)\chi_\mathrm{p}^*(\Omega + 2\Omega_\mathrm{dp})}.
\end{equation}
To find the probe tone resonances of the driven Kerr oscillator, we solve for the complex frequencies $\tilde{\omega}_\mathrm{p}$, for which $\chi_\mathrm{g}^{-1} = 0$.
This is equivalent to
\begin{equation}
1-\mathcal{K}^2 n_\mathrm{d}^2\chi_\mathrm{p}(0)\chi_\mathrm{p}^*(2\tilde{\Omega}_\mathrm{dp}) = 0
\end{equation}
or
\begin{equation}
\left[\frac{\kappa}{2} + i\left(\tilde{\Delta}_\mathrm{p} - 2\mathcal{K} n_\mathrm{d}\right)\right]\cdot\left[\frac{\kappa}{2} - i\left(\tilde{\Delta}_\mathrm{p} - 2\mathcal{K} n_\mathrm{d} + 2\tilde{\Omega}_\mathrm{dp}\right)\right] - \mathcal{K}^2 n_\mathrm{d}^2 = 0.
\end{equation}
After multiplying out and sorting for terms with $\tilde{\omega}_\mathrm{p}$, we can write down the two complex solutions as
\begin{eqnarray}
\tilde{\omega}_{1/2} = \omega_\mathrm{d} + i\frac{\kappa}{2} \pm \sqrt{\left(\Delta_\mathrm{d} - \mathcal{K} n_\mathrm{d}\right)\left(\Delta_\mathrm{d} - 3\mathcal{K} n_\mathrm{d}\right)}.
\end{eqnarray}
Later, we will identify the low-frequency solution with the signal quasi-mode and the high-frequency solution with the idler quasi-mode.
Therefore, we will also denote the real part of these two solutions as
\begin{eqnarray}
\omega_\mathrm{s} & = & \omega_\mathrm{d} - \sqrt{\left(\Delta_\mathrm{d} - \mathcal{K} n_\mathrm{d}\right)\left(\Delta_\mathrm{d} - 3\mathcal{K} n_\mathrm{d}\right)} \\
\omega_\mathrm{i} & = & \omega_\mathrm{d} + \sqrt{\left(\Delta_\mathrm{d} - \mathcal{K} n_\mathrm{d}\right)\left(\Delta_\mathrm{d} - 3\mathcal{K} n_\mathrm{d}\right)}
\end{eqnarray}
in the regime $\Delta_\mathrm{d}-\mathcal{K}n_\mathrm{d} > 0$ and $\Delta_\mathrm{d}-3\mathcal{K}n_\mathrm{d} > 0$ and vice versa.
The regime of an imaginary square root, i.e., two modes with identical frequencies but different linewidths, is not relevant for our experiments.

\subsection{Probe tone response}

The reflection off the driven cavity for a single input probe tone is given by
\begin{eqnarray}
S_{11} & = & 1 + i\sqrt{\kappa_\mathrm{e}}\frac{\gamma_-}{S_\mathrm{p}} \\
& = & 1- \kappa_\mathrm{e}\chi_\mathrm{g}
\label{eqn:FM_Kerr_S11}
\end{eqnarray}

\subsection{Quantum description with input noise}

We denote the quantum operator for the HF intracavity fluctuation fields  $\hat{c}$.
The linearized Heisenberg-Langevin equation of motion is then given by
\begin{eqnarray}
\dot{\hat{c}} = \left[i\left(\omega_0 - \omega_\mathrm{d} + 2\mathcal{K}n_\mathrm{d}\right)  - \frac{\kappa}{2}\right]\hat{c} + i\mathcal{K}n_\mathrm{d}\hat{c}^\dagger + \sqrt{\kappa_\mathrm{e}}\hat{\xi}_\mathrm{e} + \sqrt{\kappa_\mathrm{i}}\hat{\xi}_\mathrm{i} 
\end{eqnarray}
with the noise input terms $\hat{\xi}$ and the subscripts 'e' and 'i' describing external and internal baths.
Note also, that we are treating the system here in the frame rotating with the drive.
By Fourier transform we obtain
\begin{eqnarray}
\frac{\hat{c}(\Omega)}{\chi_\mathrm{p}} - i\mathcal{K}n_\mathrm{d}\hat{c}^\dagger(-\Omega) & = & \sqrt{\kappa_\mathrm{e}}\hat{\xi}_\mathrm{e}(\Omega) + \sqrt{\kappa_\mathrm{i}}\hat{\xi}_\mathrm{i}(\Omega) \\
\frac{\hat{c}^\dagger(-\Omega)}{\overline{\chi}_\mathrm{p}} + i\mathcal{K}n_\mathrm{d}\hat{c}(\Omega) & = & \sqrt{\kappa_\mathrm{e}}\hat{\xi}_\mathrm{e}^\dagger(-\Omega) + \sqrt{\kappa_\mathrm{i}}\hat{\xi}_\mathrm{i}^\dagger(-\Omega)
\end{eqnarray}
where
\begin{equation}
\chi_\mathrm{p} = \frac{1}{\frac{\kappa}{2} + i\left(\Delta_\mathrm{d} - 2\mathcal{K}n_\mathrm{d} + \Omega  \right)}, ~~~ 	\overline{\chi}_\mathrm{p} = \frac{1}{\frac{\kappa}{2} - i\left(\Delta_\mathrm{d} - 2\mathcal{K}n_\mathrm{d} - \Omega  \right)}
\end{equation}
and $\Omega = \omega - \omega_\mathrm{d}$.
These equations can be combined to obtain
\begin{equation}
\frac{\hat{c}(\Omega)}{\chi_\mathrm{g}} = \sqrt{\kappa_\mathrm{e}}\hat{\xi}_\mathrm{e}(\Omega) + \sqrt{\kappa_\mathrm{i}}\hat{\xi}_\mathrm{i}(\Omega) + i\mathcal{K}n_\mathrm{d}\overline{\chi}_\mathrm{p}\sqrt{\kappa_\mathrm{e}}\hat{\xi}_\mathrm{e}^\dagger(-\Omega) + i\mathcal{K}n_\mathrm{d}\overline{\chi}_\mathrm{p}\sqrt{\kappa_\mathrm{i}}\hat{\xi}_\mathrm{i}^\dagger(-\Omega)
\end{equation}
with again
\begin{equation}
\chi_\mathrm{g} = \frac{\chi_\mathrm{p}}{1 - \mathcal{K}^2 n_\mathrm{d}^2\chi_\mathrm{p}\overline{\chi}_\mathrm{p}}
\end{equation}
For the output field, we get
\begin{equation}
\hat{c}_\mathrm{out}(\Omega) =  \left(1 - \kappa_\mathrm{e}\chi_\mathrm{g}\right)\hat{\xi}_\mathrm{e}(\Omega) - \sqrt{\kappa_\mathrm{i}\kappa_\mathrm{e}}\chi_\mathrm{g}\hat{\xi}_\mathrm{i}(\Omega)  - i\mathcal{K}n_\mathrm{d}\overline{\chi}_\mathrm{p}\chi_\mathrm{g}\kappa_\mathrm{e}\hat{\xi}_\mathrm{e}^\dagger(-\Omega) -i\mathcal{K}n_\mathrm{d}\overline{\chi}_\mathrm{p}\chi_\mathrm{g}\sqrt{\kappa_\mathrm{i}\kappa_\mathrm{e}}\hat{\xi}_\mathrm{i}^\dagger(-\Omega).
\end{equation}
The positive frequency contribution to the power spectral density (PSD) is then given by
\begin{equation}
S_{n+} =  \left|1 - \kappa_\mathrm{e}\chi_\mathrm{g}\right|^2n_\mathrm{e}^\mathrm{HF} + \kappa_\mathrm{i}\kappa_\mathrm{e}|\chi_\mathrm{g}|^2n_\mathrm{i}^\mathrm{HF}  + \mathcal{K}^2n_\mathrm{d}^2|\overline{\chi}_\mathrm{p}|^2|\chi_\mathrm{g}|^2\kappa_\mathrm{e}^2\left(n_\mathrm{e}^\mathrm{HF} + 1 \right) +\mathcal{K}^2n_\mathrm{d}^2|\overline{\chi}_\mathrm{p}|^2|\chi_\mathrm{g}|^2\kappa_\mathrm{i}\kappa_\mathrm{e}\left(n_\mathrm{i}^\mathrm{HF} + 1 \right)
\end{equation}
where we used the relations $\langle\hat{\xi}_x^\dagger(\Omega_1)\hat{\xi}_x(\Omega_2)\rangle = n_x^\mathrm{HF}\delta(\Omega_2 - \Omega_1)$ and $\langle\hat{\xi}_x(\Omega_1)\hat{\xi}_x^\dagger(\Omega_2)\rangle = \left(n_x^\mathrm{HF} + 1\right)\delta(\Omega_2 - \Omega_1)$.
The negative frequency contribution is given by
\begin{equation}
S_{n-} =  \left|1 - \kappa_\mathrm{e}\chi_\mathrm{g}\right|^2\left(n_\mathrm{e}^\mathrm{HF}+1\right) + \kappa_\mathrm{i}\kappa_\mathrm{e}|\chi_\mathrm{g}|^2\left(n_\mathrm{i}^\mathrm{HF} + 1\right)  + \mathcal{K}^2n_\mathrm{d}^2|\overline{\chi}_\mathrm{p}|^2|\chi_\mathrm{g}|^2\kappa_\mathrm{e}^2n_\mathrm{e}^\mathrm{HF} +\mathcal{K}^2n_\mathrm{d}^2|\overline{\chi}_\mathrm{p}|^2|\chi_\mathrm{g}|^2\kappa_\mathrm{i}\kappa_\mathrm{e}n_\mathrm{i}^\mathrm{HF}.
\end{equation}
The total symmetric PSD then is
\begin{eqnarray}
S_{nn} & = &  \left|1 - \kappa_\mathrm{e}\chi_\mathrm{g}\right|^2\left(n_\mathrm{e}^\mathrm{HF} + \frac{1}{2} \right) + \kappa_\mathrm{i}\kappa_\mathrm{e}|\chi_\mathrm{g}|^2 \left(n_\mathrm{i}^\mathrm{HF} + \frac{1}{2} \right)  + \mathcal{K}^2n_\mathrm{d}^2|\overline{\chi}_\mathrm{p}|^2|\chi_\mathrm{g}|^2\kappa_\mathrm{e}^2\left(n_\mathrm{e}^\mathrm{HF} + \frac{1}{2} \right) +\mathcal{K}^2n_\mathrm{d}^2|\overline{\chi}_\mathrm{p}|^2|\chi_\mathrm{g}|^2\kappa_\mathrm{i}\kappa_\mathrm{e}\left(n_\mathrm{i}^\mathrm{HF} + \frac{1}{2} \right) \nonumber \\
& = & \left[1 - \kappa_\mathrm{e}\kappa|\chi_\mathrm{g}|^2\left( 1 - \mathcal{K}^2 n_\mathrm{d}^2|\overline{\chi}_\mathrm{p}|^2\right)  + \kappa_\mathrm{e}^2|\chi_\mathrm{g}|^2 \right]\left(n_\mathrm{e}^\mathrm{HF} + \frac{1}{2} \right)+ \kappa_\mathrm{i}\kappa_\mathrm{e}|\chi_\mathrm{g}|^2 \left(n_\mathrm{i}^\mathrm{HF} + \frac{1}{2} \right) \nonumber \\
& & + \mathcal{K}^2n_\mathrm{d}^2|\overline{\chi}_\mathrm{p}|^2|\chi_\mathrm{g}|^2\kappa_\mathrm{e}^2\left(n_\mathrm{e}^\mathrm{HF} + \frac{1}{2} \right) +\mathcal{K}^2n_\mathrm{d}^2|\overline{\chi}_\mathrm{p}|^2|\chi_\mathrm{g}|^2\kappa_\mathrm{i}\kappa_\mathrm{e}\left(n_\mathrm{i}^\mathrm{HF} + \frac{1}{2} \right).
\end{eqnarray}
We can simplify this to
\begin{equation}
S_{nn} = \frac{1}{2} + n_\mathrm{e}^\mathrm{HF} + \kappa_\mathrm{i}\kappa_\mathrm{e}|\chi_\mathrm{g}|^2\left(n_\mathrm{i}^\mathrm{HF} - n_\mathrm{e}^\mathrm{HF}\right) + \kappa_\mathrm{e}^2\mathcal{K}^2n_\mathrm{d}^2|\overline{\chi}_\mathrm{p}|^2|\chi_\mathrm{g}|^2\left(2n_\mathrm{e}^\mathrm{HF} + 1 \right) + \kappa_\mathrm{i}\kappa_\mathrm{e}\mathcal{K}^2n_\mathrm{d}^2|\overline{\chi}_\mathrm{p}|^2|\chi_\mathrm{g}|^2\left(n_\mathrm{i}^\mathrm{HF} + n_\mathrm{e}^\mathrm{HF} + 1 \right)
\end{equation}
or for negligible thermal occupation $n_\mathrm{e}^\mathrm{HF}, n_\mathrm{i}^\mathrm{HF} \ll 0.5$
\begin{equation}
S_{nn}^\mathrm{q} = \frac{1}{2} + \kappa_\mathrm{e}\kappa\mathcal{K}^2 n_\mathrm{d}^2|\overline{\chi}_\mathrm{p}|^2|\chi_\mathrm{g}|^2.
\end{equation}
Taking into account the effective number of photons added by the detection chain $n_\mathrm{add}$, we get as total power spectral density in units of quanta
\begin{equation}
S_{nn}^\mathrm{q, tot} = \frac{1}{2} + n_\mathrm{add} + \kappa_\mathrm{e}\kappa\mathcal{K}^2 n_\mathrm{d}^2|\overline{\chi}_\mathrm{p}|^2|\chi_\mathrm{g}|^2.
\end{equation}
If the thermal occuption cannot be dropped, the full equation for the power spectral density is given by
\begin{equation}
S_{nn}^\mathrm{tot} = \frac{1}{2} + n_\mathrm{add} + n_\mathrm{e}^\mathrm{HF} + \kappa_\mathrm{i}\kappa_\mathrm{e}|\chi_\mathrm{g}|^2\left(n_\mathrm{i}^\mathrm{HF} - n_\mathrm{e}^\mathrm{HF}\right) + \kappa_\mathrm{e}^2\mathcal{K}^2n_\mathrm{d}^2|\overline{\chi}_\mathrm{p}|^2|\chi_\mathrm{g}|^2\left(2n_\mathrm{e}^\mathrm{HF} + 1 \right) + \kappa_\mathrm{i}\kappa_\mathrm{e}\mathcal{K}^2n_\mathrm{d}^2|\overline{\chi}_\mathrm{p}|^2|\chi_\mathrm{g}|^2\left(n_\mathrm{i}^\mathrm{HF} + n_\mathrm{e}^\mathrm{HF} + 1 \right)
\label{eqn:PSD_Kerr}
\end{equation}
where $n_\mathrm{add}$ is the (input-referenced) noise added by the cryogenic amplifier in the detection chain in units of quanta.

\section{Supplementary Note 6: Driven Kerr cavity - Measurements of $S_{11}$ and output noise}
In this section we present measurements of the linearized probe tone response of the strongly driven Kerr cavity as well as measurements of the (amplified) output noise.
We note that the experiments presented in this section were performed during a later cooldown of the dilution refrigerator than all the photon-pressure experiments presented below and in the main paper.
In between the cooldowns, the device had to be positioned differently on the mixing chamber and connected slightly differently to the input/output cabling.
For these reasons, the characteristic parameters of the device itself such as nonlinearity and linewidths as well as characteristics of the output line such as the number of added noise quanta, can deviate slightly from the parameters during the earlier cooldown, i.e., from the parameters in the main paper.
\subsection{Driven cavity response for a bias flux upsweep}
\begin{figure}[h!]
	\centerline{\includegraphics[trim = {0.6cm, 11cm, 1cm, 5cm}, clip, scale = 0.6]{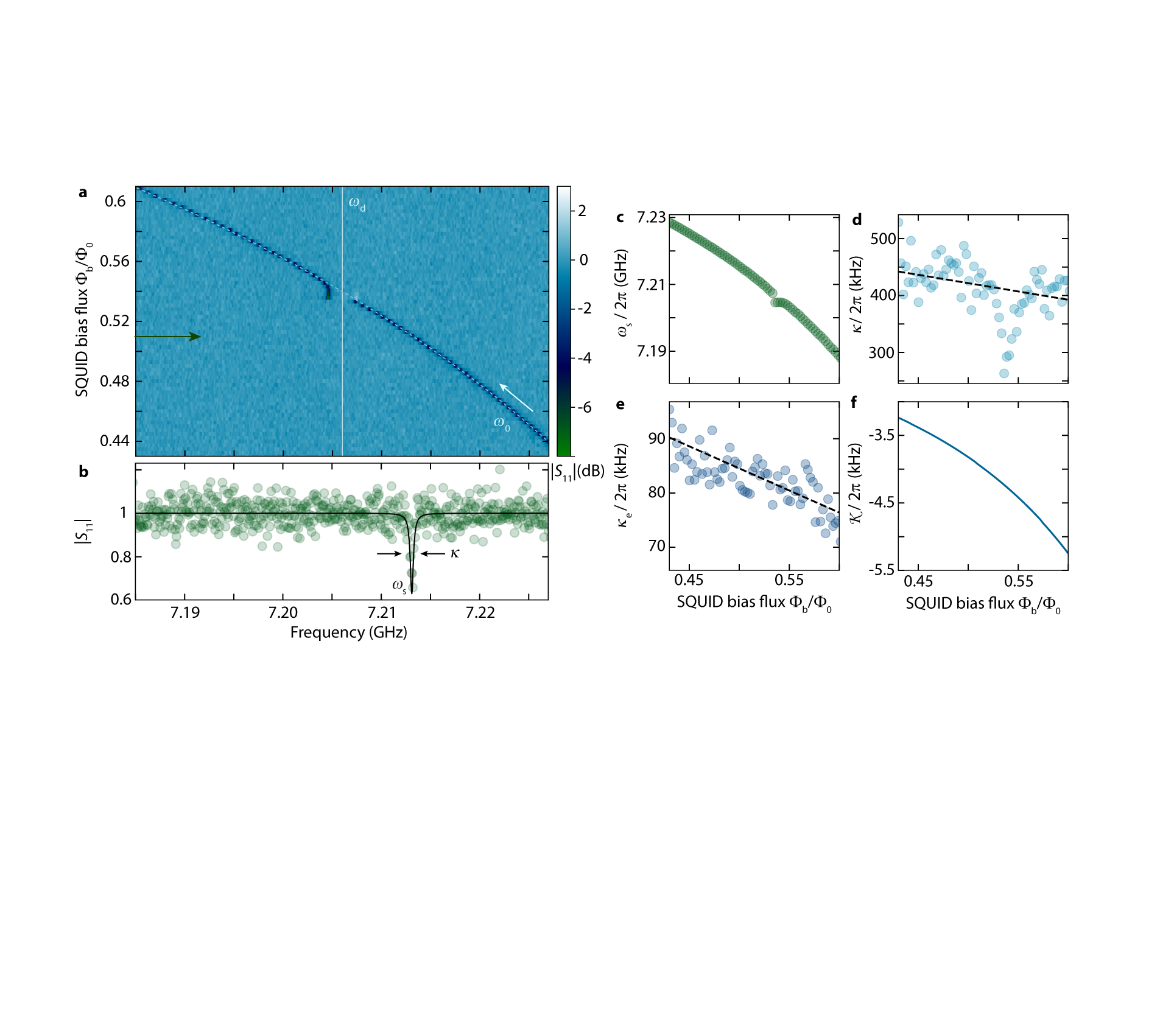}}
	\caption{\textsf{\textbf{SQUID cavity characterization in presence of a parametric drive during a flux upsweep.} \textbf{a} shows color-coded the background-corrected probe tone reflection $|S_{11}|$ of the SQUID cavity while its resonance frequency is swept through a parametric drive at $\omega_\mathrm{d}$, the drive frequency is indicated by a vertical white line. The HF cavity resonance frequency $\omega_0(\Phi_\mathrm{b})$ in absence of the drive is shown as white dashed line and the white arrow indicates the sweep direction. For small detunings between $\omega_\mathrm{d}$ and $\omega_0$ the probe tone resonance shows significant deviations from the undriven case. We fit each of the linescans using Eq.~(\ref{eqn:FM_Kerr_S11}), an example is shown in panel \textbf{b}. Green circles are data points, black line is a fit. Green arrow in \textbf{a} indicates the position of the linescan shown in \textbf{b}. As fit parameters, we obtain the total linewidth $\kappa$, the external linewidth $\kappa_\mathrm{e}$ and the intracavity drive photon number $n_\mathrm{d}$, determining the resonance frequency of the driven mode $\omega_\mathrm{s}$. In panels \textbf{c}, \textbf{d}, \textbf{e} we show $\omega_\mathrm{s}$, $\kappa$, $\kappa_\mathrm{e}$, respectively, and the Kerr nonlinearity $\mathcal{K}$ used as input for the fit for completeness in panel \textbf{f}. We observe that both, the total linewidth $\kappa$ and the external linewidth $\kappa_\mathrm{e}$, slightly decrease with increasing flux. The dashed lines show linear fits to the flux-dependence of the linewidth. The total linewidth in addition exhibits a significant reduction in the flux region where cavity and drive are near-resonant. We attribute this to saturation of two-level systems by the drive.}}
	\label{fig:UpSweep}
\end{figure}

As first experiment, we present the linearized reflection response of the driven HF cavity during a sweep of its resonance frequency from high to low frequencies, hereby crossing through the frequency of a strong drive signal at $\omega_\mathrm{d}$, cf. Supplementary Fig.~\ref{fig:UpSweep}.
The cavity resonance frequency is swept by increasing the bias magnetic flux through the SQUID loop using the external magnet coil.
When the undriven cavity resonance frequency $\omega_0(\Phi_\mathrm{b})$ is sufficiently far detuned from the drive at $\omega_\mathrm{d}$, the cavity response does not deviate from the undriven case.
This holds for both, positive and negative detunings.

Once the cavity comes close in frequency to the drive, however, a sudden jump of the HF cavity to the other side of the drive is observed.
This origin of this jump lies in the bifurcation of the driven Kerr oscillator, where the intracavity photon number from the drive at $\omega_\mathrm{d}$ abruptly increases, abruptly shifting the photon-number dependent frequency of the probe tone resonance.
Sweeping the flux further at first does not considerable change the probe tone resonance frequency as the additional shift due to the change in flux is compensated by a reduction of the intracavity drive photons.
Once the bare cavity frequency $\omega_0$ is detuned from the drive again by several linewidths, the drive-induced Kerr shift becomes negligible and the response follows the undriven case again.
For each flux value, we fit the background-corrected reflection using Eq.~(\ref{eqn:FM_Kerr_S11}) with total linewidth $\kappa$, external linewidth $\kappa_\mathrm{e}$ and drive photon number $n_\mathrm{d}$ as fit parameter, cf. Supplementary Fig.~\ref{fig:UpSweep}\textbf{b}-\textbf{e}.
From the analysis of the flux-dependence presented in Supplementary Note~\ref{Section:Flux Dependence}, we use the Kerr nonlinearity $\mathcal{K}$ and the bare resonance frequency $\omega_0$ as given input parameters for each flux point.
The linewidths depend slightly on flux with a dependence that -- in the regime of negligible $n_\mathrm{d}$ -- can be captured by a linear approximation, cf. dashed lines in Supplementary Fig.~\ref{fig:UpSweep}\textbf{d} and \textbf{e}.
The configuration of sweeping the frequency from larger to smaller values with a fixed drive frequency is equivalent to sweeping the drive from lower to higher frequencies with a fixed-frequency Kerr cavity.
Such a drive sweep is the most common way to drive a Kerr oscillator to the bifurcation regime and to observe the characteristic tilted lineshape of the resonance with a sudden jump when the power is above the critical power for bifurcation.
For a negative Kerr nonlinearity, this procedure means that the cavity will always be in the low-amplitude state and this is also what we effectively do here in the flux upsweep protocol shown and analyzed in Supplementary Fig.~\ref{fig:UpSweep}.

\subsection{Driven cavity response for a bias flux downsweep}

When the flux is swept from higher to lower values instead of the opposite direction as in the previous section, the cavity is operated in the high-amplitude state far beyond the bifurcation point.
This corresponds to the frequency downsweep in the standard configuration where the drive is swept and the cavity frequency is fixed.
In Supplementary Fig.~\ref{fig:DownSweep} we show and discuss the results of this flux-downsweep approach.

\begin{figure}[h!]
	\centerline{\includegraphics[trim = {0cm, 8.5cm, 1.5cm, 5cm}, clip, scale=0.6]{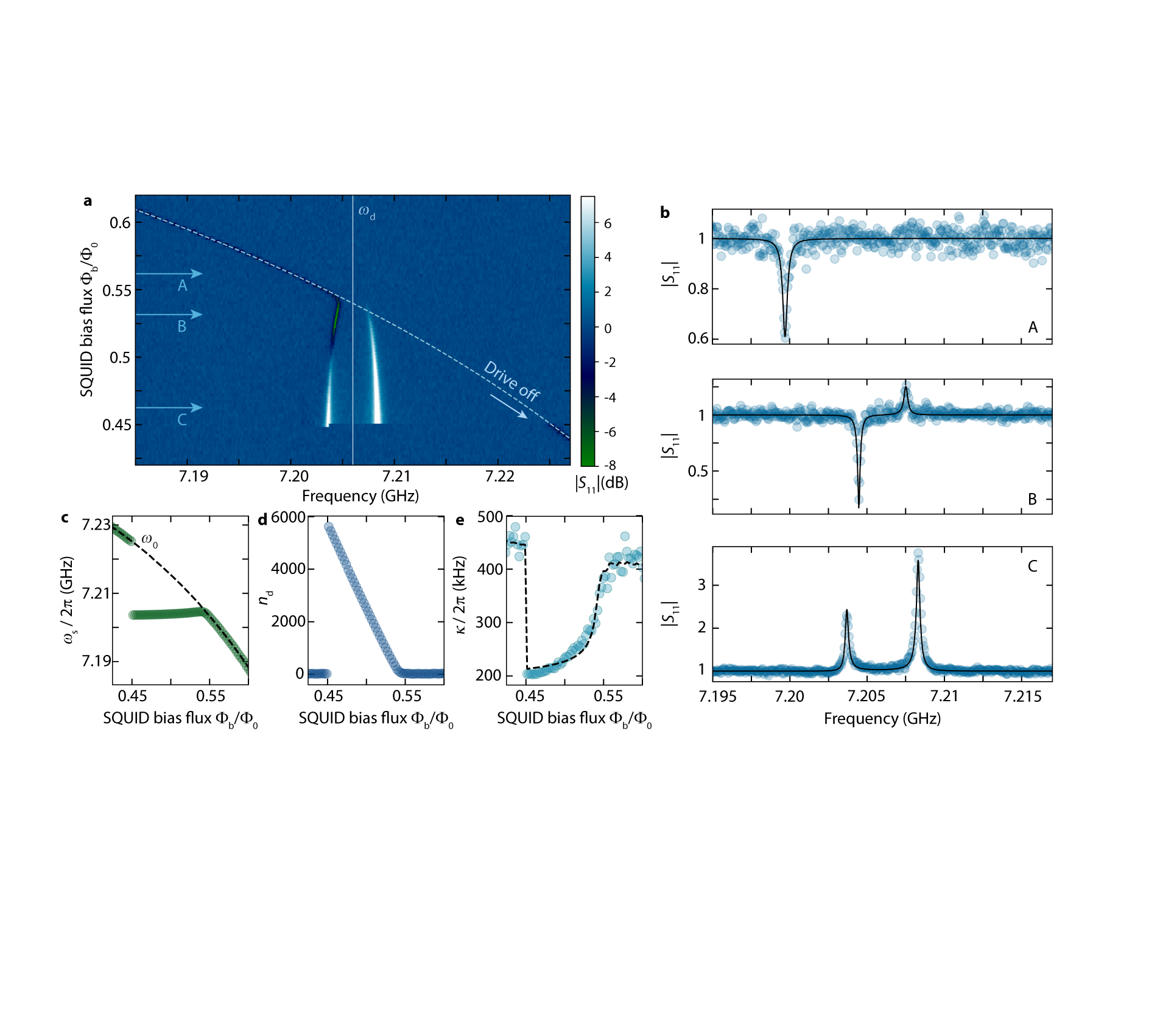}}
	\caption{\textsf{\textbf{SQUID cavity characterization in presence of a parametric drive during a flux downsweep.} \textbf{a} shows color-coded the background-corrected probe tone reflection $|S_{11}|$ of the SQUID cavity while its resonance frequency is swept through a strong drive at $\omega_\mathrm{d}$, the drive frequency is indicated by a vertical white line. The HF cavity resonance frequency $\omega_0(\Phi_\mathrm{b})$ in absence of the drive is shown as white dashed line and the white arrow indicates the sweep direction. For a large range of detunings between $\omega_\mathrm{d}$ and $\omega_0$ the probe tone resonance frequency shows significant deviations from the undriven case. The inversion from a resonance dip to a resonance peak and the appearance of a mirror mode peak in this regime are signatures of Josephson parametric amplification and degenerate four-wave mixing. We fit each of the linescans using Eq.~(\ref{eqn:FM_Kerr_S11}), three representative examples are shown in panel \textbf{b}. Top curve shows the signal resonance far detuned from the drive tone, middle panel shows the regime of small amplification with a small idler resonance peak and bottom shows the response in the state of large amplification, where both quasi-modes show output gain. Blue circles are data points, black lines are fits. Light blue arrows in \textbf{a} labelled with "A", "B" and "C" indicate the position of the linescans shown in \textbf{b}. As fit parameters, we obtain the total linewidth $\kappa$ and the intracavity drive photon number $n_\mathrm{d}$, which determines the resonance frequency of the driven signal mode $\omega_\mathrm{s}$. In panels \textbf{c}, \textbf{d} and \textbf{e} we show $\omega_\mathrm{s}$, $n_\mathrm{d}$, $\kappa$, respectively. The Kerr nonlinearity $\mathcal{K}(\Phi_\mathrm{b})$, the bare resonance frequency $\omega_0(\Phi_\mathrm{b})$ and the external linewidth $\kappa_\mathrm{e}(\Phi_\mathrm{b})$ are used as fixed input for the fit as determined from the flux dependence without drive. The intracavity drive photon number $n_\mathrm{d}$ in panel \textbf{d} resembles the characteristic triangular shape of a Kerr oscillator driven beyond bifurcation. The total linewidth plotted in panel \textbf{d} shows a strong dependence on flux and drive photon number, respectively. We model the linewidth with the equations for saturating two-level systems\cite{Capelle20S} and find reasonable agreement with the data, the fit is shown as black dashed line in \textbf{e}.}}
	\label{fig:DownSweep}
\end{figure}

When the probe tone resonance approaches the drive tone frequency $\omega_\mathrm{d}$, it stops shifting upwards with flux and instead remains nearly constant over a broad range of fluxes, indicating that the drive induced Kerr-shift is compensating the flux shift of the undriven cavity.
The probe tone cavity susceptibility, however, is strongly modified by the presence of the drive.
With increasing drive photon number, the original absorption dip first gets deeper indicating reduced internal losses (and an undercoupled cavity), then turns around and gets less deep again (overcoupled regime) until finally even turns into a resonance peak, indicating net Josephson gain for the input field.
This behaviour is fully explained by Josephson parametric amplification of the intracavity probe field with increasing drive photon number.
A second signature for the Josephson amplification regime is the appearance of a mirror mode on the opposite side of the drive tone.
The Josephson amplification process corresponds also to degenerate four-wave mixing and therefore the idler of the probe tone can also be resonant with the cavity, which explains the appearance of the mirror mode.
These two modes have been discussed and observed also in the context of optical cavities\cite{Drummond80S, Khandekar15S} and mechanical oscillators with a Kerr nonlinearity\cite{Huber20S} and are sometimes also called quasi-modes.
We will call them signal and idler mode or signal and idler resonance, respectively, and describe their resonance frequencies with $\omega_\mathrm{s}$ and $\omega_\mathrm{i}$.
Note that in the regime of sufficiently large drive photon number $n_\mathrm{d}$, both quasi-modes show output gain with up to $8\,$dB for the signal resonance and $11\,$dB for the idler resonance.
\begin{figure}[h!]
	\centerline{\includegraphics[trim = {0cm, 6.5cm, 0cm, 2.5cm}, clip, scale=0.55]{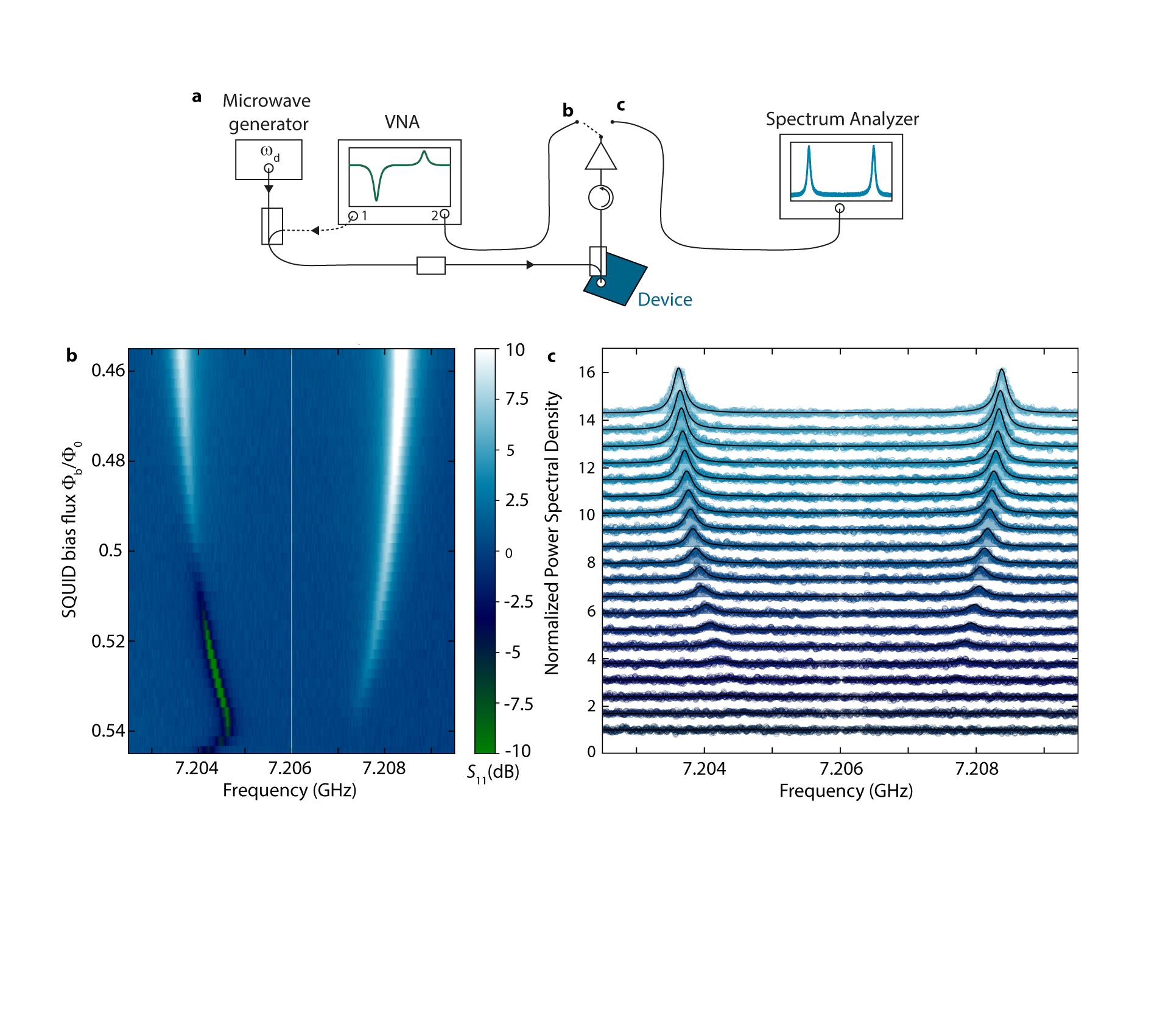}}
	\caption{\textsf{\textbf{Observation of Josephson-amplified quantum noise beyond cavity bifurcation.} Panel \textbf{a} shows a schematic of the experiment presented here. The device input (HF cavity port) is connected simultaneously to a microwave generator, providing the fixed-frequency and fixed-power drive at $\omega_\mathrm{d}$, and to port 1 (output) of a vector network analyzer (VNA). The device output field is connected to a cryogenic amplifier through a circulator and the amplified output field can be routed to both, port 2 (input) of the network analyzer or a spectrum analyzer. Whenever the output field is detected with the spectrum analyzer, the input field from the VNA is switched off, so the only signal going to the device during a measurement of the spectrum is the generator drive tone. \textbf{b} shows the VNA reflection $|S_{11}|$ color-coded vs a downsweep of the bias flux. It is equivalent to a high-resolution zoom-in of Supplementary Fig.~\ref{fig:DownSweep}\textbf{a} with inverted flux axis. The white vertical line in the center indicates the position of the drive at $\omega_\mathrm{d}$. Two (quasi)-modes symmetric to the drive are visible. The signal resonance (lower-frequency mode) develops with decreasing flux from a dip in $|S_{11}|$ to a peak, the origin for this transition is Josephson parametric amplification of the probe field. For the same reason and due to four-wave mixing, an idler resonance (higher-frequency mode) is appearing, forming an output gain peak in $|S_{11}|$ for all fluxes. In \textbf{c} the corresponding output noise power spectral density, measured with the spectrum analyzer, is shown. The spectrum shows two nearly identical noise peaks at the frequencies of signal resoannce $\omega_\mathrm{s}$ and idler resonance $\omega_\mathrm{i}$. The spectra are normalized to the white noise background and subsequent curves are offset by $+0.7$ for better visibility with the bottom curve having no offset. The drive tone signal in the center has been removed from the data. Bottom curve corresponds to $\Phi_\mathrm{b}/\Phi_0 = 0.544$ and top curve to $\Phi_\mathrm{b}/\Phi_0 = 0.456$. Circles are data points and black lines/shaded areas are theoretical curves based on Eq.~(\ref{eqn:PSD_Kerr}). Both, symmetry of the peaks and their amplitude with respect to the background, indicate that we observe parametrically amplified quantum noise here.}}
	\label{fig:KerrNoise}
\end{figure}

From the fits to each probe tone reflection curve $S_{11}$ using Eq.~(\ref{eqn:FM_Kerr_S11}), cf. Supplementary Fig.~\ref{fig:DownSweep}, we extract the total linewidth $\kappa$ and the intracavity drive photon number $n_\mathrm{d}$.
As fixed input parameters for the fit function, we use $\mathcal{K}(\Phi_\mathrm{b})$, $\omega_0(\Phi_\mathrm{b})$ and $\kappa_\mathrm{e}(\Phi_\mathrm{b})$ as obtained from independent measurements described above.
The results for $\omega_\mathrm{s}$, $n_\mathrm{d}$ and $\kappa$ are shown in Supplementary Fig.~\ref{fig:DownSweep}\textbf{c}-\textbf{e}.
While signal resonance frequency and intracavity drive photon number show no signatures of unusual behaviour, the total linewidth $\kappa$ shows a strong dependence on drive photon number, respectively.
We attribute the significant reduction of $\kappa$ from $\sim2\pi\cdot 430\,$kHz in the undriven case to $\sim2\pi\cdot 220\,$kHz in the driven case to saturation of two-level systems (TLSs) in the HF circuit.
Taking a simple model for the TLSs into account in the form of Eq.~(\ref{eqn:TLSloss}) we find acceptable agreement between data and theoretical expectations, cf. data and dashed line in Supplementary Fig.~\ref{fig:DownSweep}\textbf{d}.
There is still a slight disagreement between the data and the theoretical fit though, which we think might be originating either in a distribution of TLSs with different saturation powers and dephasing times or in additional sources of linewidth broadening in the undriven cavity, as e.g.~external or internal flux noise.
With the current modelling of TLSs, we obtain as fit parameters an average TLS dephasing rate $\Gamma_2 = 2\pi\cdot 2.5\,$MHz and a saturation intracavity photon number $n_\mathrm{cr} \approx 75$, cf. Sec.~\ref{sec:TLS_theory} and Ref.~\cite{Capelle20S} for the model.

\subsection{Driven cavity response and detection of parametrically amplified quantum noise}

The regime of parametric amplification discussed in the previous section is the most relevant for this work.
Therefore we will discuss this regime in more detail now, including a measurement of the cavity noise output power spectral density.
In Supplementary Fig.~\ref{fig:KerrNoise}\textbf{a} the experimental protocol is visualized.
In essence, we perform a similar measurement as in the previous section, i.e., we sweep the cavity bias flux from larger to smaller values, herewith shifting the bare cavity resonance frequency from smaller values to larger values.
We focus here on the most interesting regime, which is the regime where the cavity probe tone response is strongly modified by the presence of a drive tone at $\omega_\mathrm{d}$.
For each bias flux value, we first take a probe tone response trace $S_{11}$ using a VNA and afterwards measure the output power spectrum with a spectrum analyzer and with the VNA output power completely switched off.
In Supplementary Fig.~\ref{fig:KerrNoise}\textbf{b} the probe reflection $|S_{11}|$ is shown, revealing the already familiar two characteristic mirror(-quasi)-modes, generated by parametric amplification and four-wave mixing.
The lower-frequency mode -- the signal resonance -- starts as a shallow resonance absorption dip for large bias flux values and with increasing intracavity drive photon number $n_\mathrm{d}$ gets first deeper and then inverts gradually into a peak with a maximum output gain of $\sim 9\,$dB for the smallest flux values shown.
In terms of a usual linear cavity response, the cavity slowly transits from the undercoupled to the critically coupled to the overcoupled regime and finally, for bias fluxes $\Phi_\mathrm{b}/\Phi_0 \lesssim 0.5$, reaches the regime, which in a linear cavity would only be possible for a negative internal decay rate (output gain).
Simultaneously the idler mode continuously grows in height up to a net output gain of $\sim 12\,$dB.
We fit each line of Supplementary Fig.~\ref{fig:KerrNoise}\textbf{b} using Eq.~(\ref{eqn:FM_Kerr_S11}), where we use $\omega_0(\Phi_\mathrm{b})$, $\kappa_\mathrm{e}(\Phi_\mathrm{b})$ and $\mathcal{K}(\Phi_\mathrm{b})$ as fixed input parameters determined from independent measurements.
Similarly to what was described in the previous section, we obtain as fit parameters $\kappa$ and $n_\mathrm{d}$ for each flux point $\Phi_\mathrm{b}$.
Subsequently, we fit the power spectra shown in Supplementary Fig.~\ref{fig:KerrNoise}\textbf{c} with Eq.~(\ref{eqn:PSD_Kerr}) where we use all parameters except for the external and internal bath occupations $n_\mathrm{e}^\mathrm{HF}$ and $n_\mathrm{i}^\mathrm{HF}$ and the number of added photons $n_\mathrm{add}$ as used and extracted from the fits to \textbf{b}.
When fixing the added photons to $n_\mathrm{add} = 14$ and fitting $n_\mathrm{i}^\mathrm{HF}$ and $n_\mathrm{e}^\mathrm{HF}$, we obtain good agreement between theory and experiment a shown in \textbf{c}.
The values for $n_\mathrm{i}^\mathrm{HF}$ and $n_\mathrm{e}^\mathrm{HF}$ we obtain by this procedure vary in the range $0\leq n_\mathrm{i, e}^\mathrm{HF} \leq 0.08 $ without showing a clear trend with drive power, indicating that the cavity is well-thermalized to the fridge temperature and that it is not significantly heated by the drive.

\section{Supplementary Note 7: The driven Kerr quasi-modes}
In Supplementary Note~5 we derived the relevant relations for the driven HF Kerr cavity, such as reflection response and power spectral density of the noise output field.
We used these full model equations for an analysis of the total reflection reponse and output noise in Supplementary Note~6, which allows to reliably extract the "true" relevant system parameters.
In the following, we will present a second possibility to effectively describe the resulting quasi-modes when the HF Kerr cavity is strongly driven.
The goal is to describe the signal- and idler-resonances as two individual modes with each having effective linewidths, an effective suceptibility and an effective temperature.
This will turn out to be very useful for identifying the Josephson gain in the equations, and for understanding and modeling the device with photon-pressure coupling later on.
Also, we use this formalism of two individual modes in the main paper to describe the driven cavity response, in particular in the context of Fig.~2 without photon-pressure effects.

\subsection{Approximate cavity and idler susceptibilities}

In our experiment, we always have $\omega_{1/2} \neq \omega_\mathrm{d}$ and hence we can express the solutions as
\begin{eqnarray}
\tilde{\omega}_{1/2} & = & \omega_\mathrm{d} + i\frac{\kappa}{2} \pm \sqrt{\left(\Delta_\mathrm{d} - \mathcal{K} n_\mathrm{d}\right)\left(\Delta_\mathrm{d} - 3\mathcal{K} n_\mathrm{d}\right)} \nonumber \\
& = & \omega_{1/2} + i\frac{\kappa}{2}.
\end{eqnarray}
Next, we want to find a possibility to write the intracavity field as
\begin{equation}
\gamma_- = i\mathcal{G}_1 \chi_1 \sqrt{\kappa_\mathrm{e}}S_\mathrm{p} + i\mathcal{G}_2 \chi_2\sqrt{\kappa_\mathrm{e}}S_\mathrm{p}
\label{eqn:IntraSplit}
\end{equation}
with the intracavity gain of idler and cavity modes $\mathcal{G}_1$ and $\mathcal{G}_2$ and individual idler and cavity mode susceptibilities $\chi_1$ and $\chi_2$, respectively.
To find the corresponding expressions for $\mathcal{G}_1, \mathcal{G}_2$ and $\chi_1, \chi_2$ we define $A$ via
\begin{equation}
A = \frac{\chi_\mathrm{p}(0)\chi_\mathrm{p}^*(2\Omega_\mathrm{dp})}{1 - \mathcal{K}^2 n_\mathrm{d}^2 \chi_\mathrm{p}(0)\chi_\mathrm{p}^*(2\Omega_\mathrm{dp})} = \chi_1\chi_2
\end{equation}
where
\begin{equation}
\chi_n = \frac{1}{\frac{\kappa}{2} + i(\omega - \omega_n)}, ~~~ n = 1, 2
\end{equation}
and use
\begin{equation}
A = \frac{\chi_1 + \chi_2}{\kappa - 2i\Omega_\mathrm{dp}}.
\end{equation}
From here, we finally use
\begin{eqnarray}
\chi_\mathrm{g}(0) & = & \frac{A}{\chi_\mathrm{p}^*(2\Omega_\mathrm{dp})} \nonumber \\
& = & \frac{\frac{\kappa}{2} - i\left( \Delta_\mathrm{p} - 2\mathcal{K}n_\mathrm{d} + 2\Omega_\mathrm{dp} \right)}{\kappa - 2i\Omega_\mathrm{dp}}\left(\chi_1 + \chi_2\right)  \nonumber \\
& = & \mathcal{G}\left(\chi_1 + \chi_2\right).
\end{eqnarray}
If we are interested in the gain only close to the cavity and idler resonance, respectively and if $|\kappa| \ll |\Omega_{1/2}|$, we can approximate
\begin{equation}
\mathcal{G}_{1/2} \approx \frac{\Omega_{1/2} - \Delta_\mathrm{d} + 2\mathcal{K}n_\mathrm{d}}{2\Omega_{1/2}}
\end{equation}
with
\begin{equation}
\Omega_{1/2} = \omega_{1/2} - \omega_\mathrm{d} = \pm \sqrt{\left(\Delta_\mathrm{d} - \mathcal{K} n_\mathrm{d}\right)\left(\Delta_\mathrm{d} - 3\mathcal{K} n_\mathrm{d}\right)}
\end{equation}
Finally, we can express the intracavity field in the desired form Eq.~(\ref{eqn:IntraSplit}) with
\begin{equation}
\chi_\mathrm{g} \approx \mathcal{G}_1\chi_1 + \mathcal{G}_2\chi_2.
\end{equation}
Note that the idler gain $\mathcal{G}_1$ is negative in this formulation, which is equivalent to saying that this effective mode acts as if it has an inverted susceptibility.
From here on, we will use the supscripts "s" and "i" for signal and idler resonance again.

\subsection{Exact reflection and output gain at resonance}

The exact reflection of the probe tone off the driven cavity is given by
\begin{equation}
S_{11} = 1-\kappa_\mathrm{e}\chi_\mathrm{g}(0).
\end{equation}
To calculate the resonance amplitudes, we first express the susceptibility as
\begin{equation}
\chi_\mathrm{g} = \frac{\frac{\kappa}{2} - i\left(\Delta_\mathrm{d}-2\mathcal{K} n_\mathrm{d}+\Omega_\mathrm{dp}\right)}{\left[\frac{\kappa}{2} + i\left(\Delta_\mathrm{d}-2\mathcal{K} n_\mathrm{d}-\Omega_\mathrm{dp}\right)\right]\left[\frac{\kappa}{2} - i\left(\Delta_\mathrm{d}-2\mathcal{K} n_\mathrm{d}+\Omega_\mathrm{dp}\right)\right] - \mathcal{K}^2 n_\mathrm{d}^2}
\end{equation}
and use the resonance frequencies
\begin{equation}
\Omega_{\mathrm{i}/\mathrm{s}} = \pm\sqrt{\left(\Delta_\mathrm{d}-\mathcal{K} n_\mathrm{d}\right)\left(\Delta_\mathrm{d}-3\mathcal{K} n_\mathrm{d}\right)}
\end{equation}
to get
\begin{equation}
\chi_\mathrm{g}^\mathrm{res} = \frac{\frac{\kappa}{2} - i\left(\Delta_\mathrm{d}-2\mathcal{K} n_\mathrm{d}-\Omega_{\mathrm{i}/\mathrm{s}}\right)}{\frac{\kappa^2}{4} + i\kappa\Omega_{\mathrm{i}/\mathrm{s}}}.
\end{equation}
With this, the (complex and exact) output gain on resonance of the signal mode is given by
\begin{equation}
\mathcal{G}_\mathrm{s}^\mathrm{out} = 1-\frac{2\kappa_\mathrm{e}}{\kappa}\frac{\frac{\kappa}{2} - i\left(\Delta_\mathrm{d}-2\mathcal{K} n_\mathrm{d}+\Omega_0\right)}{\frac{\kappa}{2} - 2i\Omega_0}
\end{equation}
and on resonance of the idler by
\begin{equation}
\mathcal{G}_\mathrm{i}^\mathrm{out} = 1-\frac{2\kappa_\mathrm{e}}{\kappa}\frac{\frac{\kappa}{2} - i\left(\Delta_\mathrm{d}-2\mathcal{K} n_\mathrm{d}-\Omega_0\right)}{\frac{\kappa}{2} + 2i\Omega_0}
\end{equation}
where
\begin{equation}
\Omega_0 = \sqrt{\left(\Delta_\mathrm{d}-\mathcal{K} n_\mathrm{d}\right)\left(\Delta_\mathrm{d}-3\mathcal{K} n_\mathrm{d}\right)}.
\end{equation}
We can also do a similar approximation as before and then get
\begin{equation}
\mathcal{G}_\mathrm{s}^\mathrm{out} = 1-\frac{\kappa_\mathrm{e}}{\kappa}\frac{ \Delta_\mathrm{d}-2\mathcal{K} n_\mathrm{d}+\Omega_0}{\Omega_0} = 1 - \frac{2\kappa_\mathrm{e}}{\kappa}\mathcal{G}_\mathrm{s}
\label{eqn:Gout_approx}
\end{equation}
and
\begin{equation}
\mathcal{G}_\mathrm{i}^\mathrm{out} = 1+\frac{\kappa_\mathrm{e}}{\kappa}\frac{ \Delta_\mathrm{d}-2\mathcal{K} n_\mathrm{d}-\Omega_0}{\Omega_0} = 1 - \frac{2\kappa_\mathrm{e}}{\kappa}\mathcal{G}_\mathrm{i}.
\end{equation}
In this case, the difference in output gain between the cavity and the idler is given by
\begin{equation}
\mathcal{G}_\mathrm{i}^\mathrm{out} - \mathcal{G}_\mathrm{s}^\mathrm{out} = 2\frac{\kappa_\mathrm{e}}{\kappa}\frac{ 2\mathcal{K} n_\mathrm{d} - \Delta_\mathrm{d}}{\Omega_0}
\end{equation}
and for the intracavity gains we get
\begin{equation}
\mathcal{G}_\mathrm{s} = -\mathcal{G}_\mathrm{i} + 1.
\end{equation}
Note that for the full expression we get $\mathcal{G}(\Omega) = -\mathcal{G}^*(-\Omega)+1$.

\subsection{Effective external linewidths in reflection}

Using the individual susceptibilities derived above, we can also express the reflection as
\begin{equation}
S_{11} = 1-\kappa_\mathrm{e}\mathcal{G}_\mathrm{s}\chi_\mathrm{s} - \kappa_\mathrm{e}\mathcal{G}_\mathrm{i}\chi_\mathrm{i}
\end{equation}
from where we can define the apparent external linewidths $\kappa_2 = \kappa_\mathrm{e}\mathcal{G}_\mathrm{i}$ and $\kappa_1 = \kappa_\mathrm{e}\mathcal{G}_\mathrm{s}$.
Note that these effective external linewidths only describe the phenomenology of the reflection, it cannot be applied to describe the intracavity dynamics.
If we are only interested in the behaviour close to the signal mode, i.e., far detuned from the idler mode, we get
\begin{equation}
S_{11} =1 - \kappa_1 \chi_\mathrm{s}
\end{equation}
where
\begin{equation}
\kappa_1 \approx \kappa_\mathrm{e}\frac{\Omega_0 + \Delta_\mathrm{d} - 2\mathcal{K}n_\mathrm{d}}{2\Omega_0}.
\end{equation}

\subsection{Data analysis for main paper Fig.~2}

We use the equations described in the previous section for an analysis of the signal resonance as presented in main paper Fig.~2.
In particular, we use for the background-corrected reflection
\begin{equation}
S_{11} = 1-\kappa_1\chi_\mathrm{s}
\end{equation}
with 
\begin{equation}
\chi_\mathrm{s} = \frac{1}{\frac{\kappa}{2} + i\left(\omega - \omega_\mathrm{s} \right)}.	
\end{equation}
In addition, we use for the apparent internal linewidth $\kappa_\mathrm{i}^\mathrm{eff} = \kappa - \kappa_1$.

\subsection{Effective temperature of the driven HF cavity signal mode}

For the intracavity field we got above
\begin{equation}
\hat{c} = \sqrt{\kappa_\mathrm{e}}\chi_\mathrm{g}\hat{\xi}_\mathrm{e}(\Omega) + \sqrt{\kappa_\mathrm{i}}\chi_\mathrm{g}\hat{\xi}_\mathrm{i}(\Omega) + i\mathcal{K}n_\mathrm{d}\overline{\chi}_\mathrm{p}\chi_\mathrm{g}\sqrt{\kappa_\mathrm{e}}\hat{\xi}_\mathrm{e}^\dagger(-\Omega) + i\mathcal{K}n_\mathrm{d}\overline{\chi}_\mathrm{p}\chi_\mathrm{g}\sqrt{\kappa_\mathrm{i}}\hat{\xi}_\mathrm{i}^\dagger(-\Omega)
\end{equation}
with which we can get the effective photon power spectral density of the driven HF mode
\begin{eqnarray}
S_n^\mathrm{HF} & = & \langle \hat{c}^\dagger \hat{c}\rangle \nonumber \\
& = & \kappa|\chi_\mathrm{g}|^2 n_\mathrm{th}^\mathrm{HF} + \kappa\mathcal{K}^2 n_\mathrm{d}^2 |\chi_\mathrm{g}|^2|\overline{\chi}_\mathrm{p}|^2(n_\mathrm{th}^\mathrm{HF} + 1)
\end{eqnarray}
where we also introduced
\begin{equation}
n_\mathrm{th}^\mathrm{HF} = \frac{\kappa_\mathrm{e}n_\mathrm{e}^\mathrm{HF} + \kappa_\mathrm{i}n_\mathrm{i}^\mathrm{HF}}{\kappa}.
\end{equation}
For negligible thermal occupation, we obtain
\begin{eqnarray}
S_{n, \mathrm{q}}^\mathrm{HF} = \kappa\mathcal{K}^2 n_\mathrm{d}^2 |\chi_\mathrm{g}|^2|\overline{\chi}_\mathrm{p}|^2
\end{eqnarray}
or using the above approximations and only for the signal mode
\begin{eqnarray}
S_{n, \mathrm{q}}^\mathrm{s} = \kappa\mathcal{K}^2 n_\mathrm{d}^2 |\mathcal{G}_\mathrm{s}|^2 |\chi_\mathrm{s}|^2|\overline{\chi}_\mathrm{p}|^2.
\end{eqnarray}
By comparison with the result of a linear mode with susceptibility $\chi_\mathrm{c}$ and thermal occupation $n_\mathrm{c}$
\begin{eqnarray}
S_n^\mathrm{lin} = \kappa |\chi_\mathrm{c}|^2n_\mathrm{th}^\mathrm{lin}
\end{eqnarray}
we can determine the effective signal mode occupation as
\begin{equation}
\tilde{n}_\mathrm{q}^\mathrm{HF} = \mathcal{K}^2 n_\mathrm{d}^2 |\mathcal{G}_\mathrm{s}|^2 |\overline{\chi}_\mathrm{p}|^2 \approx \frac{\mathcal{K}^2 n_\mathrm{d}^2}{|\kappa + 2i\Omega_\mathrm{s}|^2}
\end{equation}
and the effective signal mode temperature as
\begin{equation}
T_\mathrm{s}^\mathrm{eff} = \frac{\hbar\omega_\mathrm{s}}{k_\mathrm{B}}\tilde{n}_\mathrm{q}^\mathrm{HF}.
\end{equation}

\section{Supplementary Note 8: Theory of a photon-pressure Kerr cavity}

\subsection{Resonance frequency and the single-photon coupling rate}
\begin{figure}[h!]
	\centerline{\includegraphics[trim = {0cm, 14cm, 0cm, 5.0cm}, clip, scale=0.6]{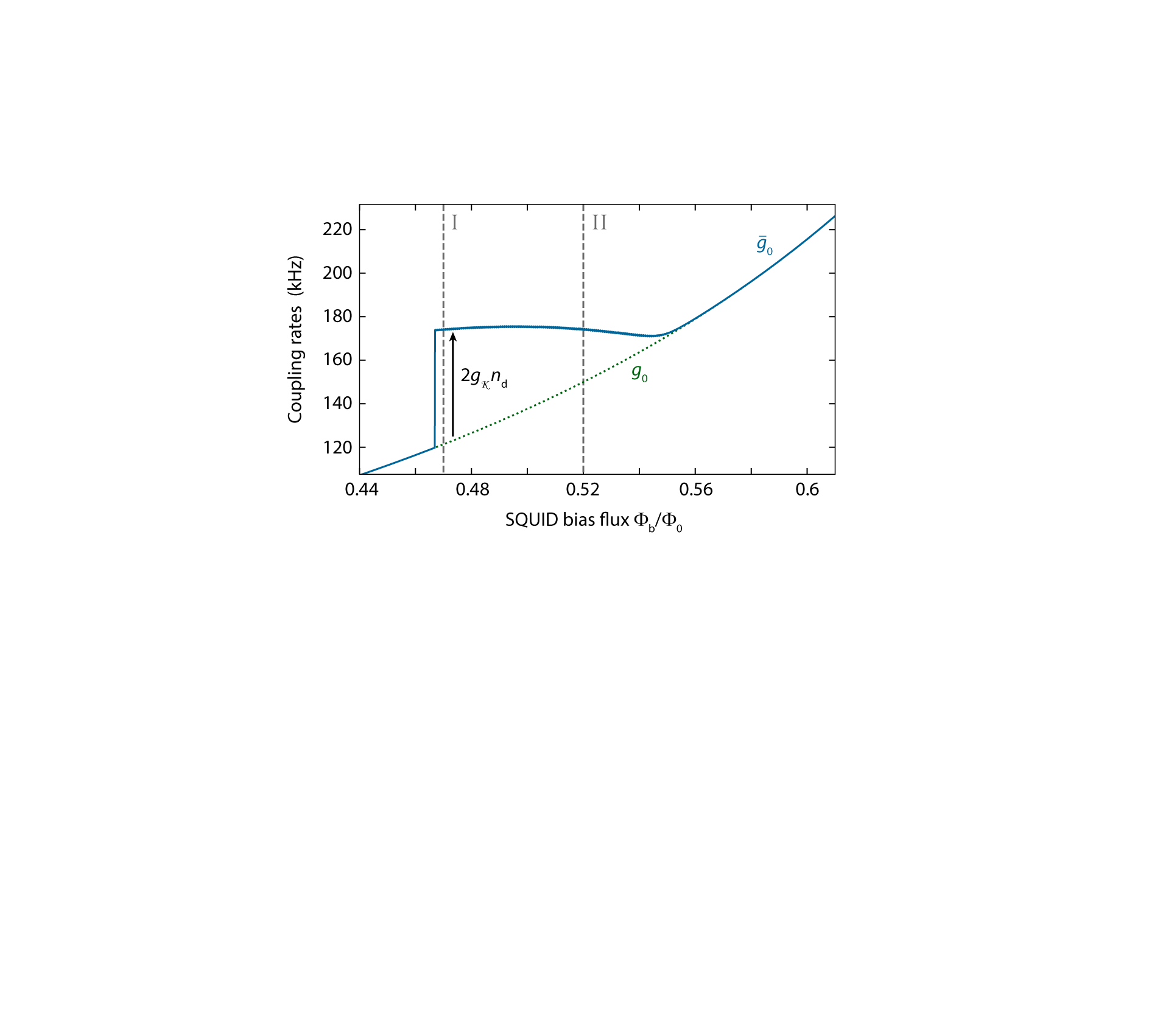}}
	\caption{\textsf{\textbf{Enhancing the single-photon coupling rate with a strong  drive.} Green dotted line shows the single-photon coupling rate $g_0$ of the undriven cavity vs magnetic bias flux in the SQUID. The blue curve shows the single-photon coupling rate $\tilde{g}_0$ including the Kerr enhancement $2g_\mathcal{K}n_\mathrm{d}$ following Eq.~(\ref{eqn:enhanced_g0}). For the calculation, we use the Kerr anharmonicity $\mathcal{K}(\Phi_\mathrm{b})$ and the drive photon number $n_\mathrm{d}$ as obtained from the modelling of the bare cavity flux dependence and the driven cavity flux dependence as shown and discussed in main paper Fig.~2. The vertical dashed lines correspond to the operation points I and II.}}
	\label{fig:g0_enhanced}
\end{figure}

The (classical) resonance frequency of a photon-pressure Kerr cavity, that depends on both, the intracavity photon number $n_\mathrm{c} = |\alpha|^2$ and the RF flux through the SQUID $\Phi$, is given by the Taylor approximation
\begin{equation}
\omega_0(n_\mathrm{c}, \Phi) = \omega_0 + \frac{\partial \omega_0}{\partial n_\mathrm{c}}n_\mathrm{c} + \frac{\partial \omega_0}{\partial \Phi}\Phi + \frac{1}{2}\frac{\partial^2 \omega_0}{\partial n_\mathrm{c}^2}n_\mathrm{c}^2 + \frac{\partial^2 \omega_0}{\partial\Phi\partial n_\mathrm{c}}n_\mathrm{c}\Phi + ...
\end{equation}
where we limited the expansion to terms linear in $\Phi$.
Using
\begin{equation}
\frac{\partial \omega_0}{\partial n_\mathrm{c}} = \mathcal{K}, ~~~~ \frac{\partial^2 \omega_0}{\partial n_\mathrm{c}^2} = 0, ~~~~ \frac{\partial \omega_0}{\partial \Phi} = G_0, ~~~~ \frac{\partial \mathcal{K}}{\partial \Phi} = G_\mathcal{K}
\end{equation}
we get
\begin{equation}
\omega_0(n_\mathrm{c}, \Phi) \approx \omega_0 +\mathcal{K}n_\mathrm{c} +  G_0\Phi + G_\mathcal{K}n_\mathrm{c}\Phi.
\end{equation}
As final version, we write
\begin{equation}
\omega_0(n_\mathrm{c}, \Phi) \approx \omega_0 +\mathcal{K}n_\mathrm{c} - G\Phi
\end{equation}
where the modified cavity pull is given by
\begin{equation}
G = -\left(G_0 +  G_\mathcal{K}n_\mathrm{c}\right).
\end{equation}
We will see below, that in the multi-tone driven and pumped system, the actual linearized pull for a photon-pressure pump on one of the sidebands in addition to the drive is enhanced by $2G_\mathcal{K}n_\mathrm{d}$ with the drive-induced intracavity photon number $n_\mathrm{d}$.
The single-photon coupling rate from this then is
\begin{equation}
\tilde{g}_0 = -\left(G_0 +  2G_\mathcal{K}n_\mathrm{d} \right)\Phi_\mathrm{zpf} = g_0 + 2g_\mathcal{K}n_\mathrm{d}
\label{eqn:enhanced_g0}
\end{equation}
with $\Phi_\mathrm{zpf} = 635\,\mu\Phi_0$ as discussed above.
Supplementary Fig.~\ref{fig:g0_enhanced} shows the effect of this higher order correction to the total single-photon coupling rate in the relevant regime for the experiments presented here.
In fact, by driving the system strongly we enhance the single-photon coupling rate by up to $\sim 45\%$ compared to the undriven system.

\subsection{Classical equations of motion}

We model the photon-pressure system with the classical equations of motion
\begin{eqnarray}
\ddot{\Phi} & = & -\Omega_0^2 \Phi - \Gamma_0 \dot{\Phi} - \frac{\hbar G_0}{C_\mathrm{RF}}|\alpha|^2 - \frac{\hbar G_\mathcal{K}}{2C_\mathrm{RF}}|\alpha|^4 \\
\dot{\alpha} & = & \left[i\left(\omega_0 + \mathcal{K}|\alpha|^2 + G_0\Phi + G_\mathcal{K}|\alpha|^2\Phi  \right) -\frac{\kappa}{2} \right]\alpha + i\sqrt{\kappa_\mathrm{e}}S_\mathrm{in}
\end{eqnarray}
where $\Omega_0$ and $\Gamma_0$ are the resonance frequency and linewidth of the RF circuit, $\Phi$ is its flux, and $G_0, G_\mathcal{K}$ as defined in the previous section are the HF cavity and Kerr pull parameters.

\subsection{Linearized three-tone solution}

For the main parts of our experiment, we send two pump tones (one parametric drive and one photon-pressure sideband pump) to the HF cavity and consider in addition either a small probe tone or input noise.
The input field in the first case is given by
\begin{equation}
S_\mathrm{in} = S_\mathrm{d}e^{i(\omega_\mathrm{d}t + \phi_\mathrm{d})} + S_\mathrm{p}e^{i\omega_\mathrm{p}t} + S_0(t)e^{i\omega_\mathrm{p}t}
\end{equation}
where $S_\mathrm{d}$ is the parametric drive, $S_\mathrm{p}$ is the photon-pressure sideband pump and $S_0$ is a time-dependent fluctuation field, e.g. a probe tone.
As solution for the RF flux and the intracavity field, we make the Ansatz
\begin{eqnarray}
\Phi & = & \Phi_\mathrm{eq} + \delta\Phi(t) \\
\alpha & = & \alpha_\mathrm{d}e^{i\omega_\mathrm{d}t} + \gamma_-e^{i\omega_\mathrm{p}t} + \gamma_+e^{i(2\omega_\mathrm{d}-\omega_\mathrm{p})t} + \delta\alpha(t)e^{i\omega_\mathrm{p}t}.
\end{eqnarray}
As writing down all terms, which show up when this Ansatz is injected into the equations of motion, would be very space-consuming without adding any real information, we just give the surviving relevant terms for the linearized equations of motion.
For the RF oscilator, we omit off-resonant drive terms (terms without $\delta\alpha$ or $\delta\alpha^*$) oscillating at $\pm\Omega_\mathrm{dp}$ or $\pm2\Omega_\mathrm{dp}$, we separate off the steady-state solution, and perform the linearization by $\alpha_\mathrm{d}^2 + |\gamma_-|^2 + |\gamma_+|^2 + |\delta\alpha|^2 \approx \alpha_\mathrm{d}^2$.
We also drop higher order terms in $\gamma_-, \gamma_+$ and $\delta\alpha$.
The resulting remaining equation of motion is
\begin{eqnarray}
\delta\ddot{\Phi} & = & -\Omega_0^2\delta\Phi - \Gamma_0\delta\dot{\Phi} + \frac{\hbar G' \alpha_\mathrm{d}}{C_\mathrm{RF}}\left[\delta\alpha^*e^{i\Omega_\mathrm{dp}t} +  \delta\alpha e^{-i\Omega_\mathrm{dp}t} \right] + \frac{\hbar \tilde{G} }{C_\mathrm{RF}}\left[\gamma_-\delta\alpha^* +  \gamma_-^*\delta\alpha \right] + \frac{\hbar \tilde{G} }{C_\mathrm{RF}}\left[\gamma_+\delta\alpha^*e^{i2\Omega_\mathrm{dp}t} +  \gamma_+^*\delta\alpha e^{-i2\Omega_\mathrm{dp}t} \right]. \nonumber \\
\end{eqnarray}
with
\begin{equation}
G' = -G_0 - G_\mathcal{K}n_\mathrm{d}, ~~~~~ \tilde{G} = -G_0 - 2G_\mathcal{K}n_\mathrm{d}.
\end{equation}
For the intracavity field, we get with the same procedure
\begin{eqnarray}
& & \delta\dot{\alpha} + \left[ \frac{\kappa}{2}+i\left(\Delta_\mathrm{p} - 2\mathcal{K}n_\mathrm{d}\right) \right]\delta\alpha \nonumber \\
& = & \left[-i\left(\Delta_\mathrm{d} - \mathcal{K}n_\mathrm{d}\right) - \frac{\kappa}{2} \right]\alpha_\mathrm{d}e^{i\Omega_\mathrm{dp}t} + \left[-i\left(\Delta_\mathrm{p} - 2\mathcal{K}n_\mathrm{d}\right) - \frac{\kappa}{2} \right]\gamma_- + \left[-i\left(\Delta_\mathrm{p} - 2\mathcal{K}n_\mathrm{d} + 2\Omega_\mathrm{dp}\right) - \frac{\kappa}{2} \right]\gamma_+e^{i2\Omega_\mathrm{dp}t}  \nonumber \\
& & -iG'\delta\Phi\alpha_\mathrm{d}e^{i\Omega_\mathrm{dp}t} -i\tilde{G}_-\delta\Phi\gamma_- -i\tilde{G}_+\delta\Phi\gamma_+ e^{i2\Omega_\mathrm{dp}t}    \nonumber \\
& & +i\mathcal{K}n_\mathrm{d}e^{i2\Omega_\mathrm{dp}t}\left[\gamma_-^* + \gamma_+^*e^{-i2\Omega_\mathrm{dp}t} +\delta\alpha^* \right]\nonumber \\
& & + i\sqrt{\kappa_\mathrm{e}}S_\mathrm{d}e^{i\Omega_\mathrm{dp}t}e^{i\phi_\mathrm{d}} + i\sqrt{\kappa_\mathrm{e}}S_\mathrm{p} + i\sqrt{\kappa_\mathrm{e}}S_0.
\end{eqnarray}
with
\begin{equation}
\tilde{G}_- = -G_0 - 2\mathcal{K}n_\mathrm{d}\left(1 + \frac{\gamma_+^*}{2\gamma_-}\right), ~~~~~ \tilde{G}_+ = -G_0 - 2\mathcal{K}n_\mathrm{d}\left(1 + \frac{\gamma_-^*}{2\gamma_+}\right).
\end{equation}
As $|\gamma_-| \gg |\gamma_+|$ in our experiment and as all contributions for our results due to $|\gamma_+|$ are essentially negligible in our pump frequency setting, we from here on use for the sake of simplicity $\tilde{G}_- = \tilde{G}_+ = \tilde{G} = -G_0 - 2\mathcal{K}n_\mathrm{d}$, which for sure is a good approximation for $\tilde{G}_-$, the only relevant parameter for this work.
We split this equation into four individual equations
\begin{eqnarray}
\left[ \frac{\kappa}{2}+i\left(\Delta_\mathrm{d} - \mathcal{K}n_\mathrm{d}\right) \right]\alpha_\mathrm{d} & = & i\sqrt{\kappa_\mathrm{e}}S_\mathrm{d}e^{i\phi_\mathrm{d}} \\
\left[ \frac{\kappa}{2}+i\left(\Delta_\mathrm{p} - 2\mathcal{K}n_\mathrm{d}\right) \right]\gamma_- - i\mathcal{K}n_\mathrm{d}\gamma_+^* & = &  i\sqrt{\kappa_\mathrm{e}}S_\mathrm{p}  \\
\left[ \frac{\kappa}{2}+i\left(\Delta_\mathrm{p} - 2\mathcal{K}n_\mathrm{d} + 2\Omega_\mathrm{dp}\right) \right]\gamma_+ - i\mathcal{K}n_\mathrm{d}\gamma_-^* & = &  0  \\
\delta\dot{\alpha} +  \left[ \frac{\kappa}{2}+i\left(\Delta_\mathrm{p} - 2\mathcal{K}n_\mathrm{d}\right) \right]\delta\alpha + iG'\delta\Phi \alpha_\mathrm{d}e^{i\Omega_\mathrm{dp}t} +  i\tilde{G}\delta\Phi\left(\gamma_- + \gamma_+e^{i2\Omega_\mathrm{dp}t}\right) - i\mathcal{K}n_\mathrm{d}\delta\alpha^*e^{i2\Omega_\mathrm{dp}t} & = & i\sqrt{\kappa_\mathrm{e}}S_0
\end{eqnarray}
of which the first three are the standard equations for the linearized two-tone driven Kerr oscillator and the fourth describes the dynamics of the fluctuation fields.
The two coupled equations of motion read after Fourier transform
\begin{eqnarray}
\frac{\delta\Phi(\Omega)}{\chi_0(\Omega)} & = & \frac{\hbar \tilde{G}}{C_\mathrm{RF}}\left[ \gamma_-\delta\alpha^*(-\Omega) + \gamma_-^*\delta\alpha(\Omega) \right] + \frac{\hbar G' \alpha_\mathrm{d}}{C_\mathrm{RF}}\left[ \delta\alpha^*(-\Omega + \Omega_\mathrm{dp}) + \delta\alpha(\Omega + \Omega_\mathrm{dp}) \right] \nonumber \\
& & +\frac{\hbar \tilde{G}}{C_\mathrm{RF}}\left[ \gamma_+\delta\alpha^*(-\Omega + 2\Omega_\mathrm{dp}) + \gamma_+^*\delta\alpha(\Omega + 2\Omega_\mathrm{dp}) \right] \\
\frac{\delta\alpha(\Omega)}{\chi_\mathrm{p}(\Omega)} & = & -i\tilde{G}\gamma_-\delta\Phi(\Omega) - iG'\alpha_\mathrm{d}\delta\Phi(\Omega - \Omega_\mathrm{dp}) - i\tilde{G}\gamma_+\delta\Phi(\Omega - 2\Omega_\mathrm{dp}) + i\mathcal{K}n_\mathrm{d}\delta\alpha^*(-\Omega + 2\Omega_\mathrm{dp}) + i\sqrt{\kappa_\mathrm{e}}S_0(\Omega).
\end{eqnarray}
As usual, we eliminate first the parametric coupling by using 
\begin{eqnarray}
\frac{\delta\alpha^*(-\Omega + 2\Omega_\mathrm{dp})}{\chi_\mathrm{p}^*(-\Omega + 2\Omega_\mathrm{dp})} & = & i\tilde{G}\gamma_-^*\delta\Phi^*(-\Omega + 2\Omega_\mathrm{dp}) + iG'\alpha_\mathrm{d}\delta\Phi^*(-\Omega + \Omega_\mathrm{dp}) + i\tilde{G}\gamma_+^*\delta\Phi^*(-\Omega) - i\mathcal{K}n_\mathrm{d}\delta\alpha(\Omega) - i\sqrt{\kappa_\mathrm{e}}S_0^*(-\Omega + 2\Omega_\mathrm{dp}). \nonumber \\
\end{eqnarray}
and obtain with $\delta\Phi(\Omega) = \delta\Phi^*(-\Omega)$
\begin{eqnarray}
\frac{\delta\alpha(\Omega)}{\chi_\mathrm{g}(\Omega)} & = & -\mathcal{K}n_\mathrm{d}\tilde{G}\gamma_-^* \chi_\mathrm{p}^*(-\Omega + 2\Omega_\mathrm{dp})\delta\Phi(\Omega - 2\Omega_\mathrm{dp}) - i\tilde{G}\gamma_+\delta\Phi(\Omega - 2\Omega_\mathrm{dp}) \nonumber \\
& & -\mathcal{K}n_\mathrm{d}G'\alpha_\mathrm{d} \chi_\mathrm{p}^*(-\Omega + 2\Omega_\mathrm{dp})\delta\Phi(\Omega - \Omega_\mathrm{dp}) - iG'\alpha_\mathrm{d}\delta\Phi(\Omega - \Omega_\mathrm{dp}) \nonumber \\
& & -\mathcal{K}n_\mathrm{d}\tilde{G}\gamma_+^* \chi_\mathrm{p}^*(-\Omega + 2\Omega_\mathrm{dp})\delta\Phi(\Omega) - i\tilde{G}\gamma_-\delta\Phi(\Omega) \nonumber \\
& & + i\sqrt{\kappa_\mathrm{e}}S_0(\Omega) + \mathcal{K}n_\mathrm{d}\chi_\mathrm{p}^*(-\Omega + 2\Omega_\mathrm{dp})\sqrt{\kappa_\mathrm{e}}S_0^*(-\Omega + 2\Omega_\mathrm{dp}).
\end{eqnarray}

\subsection{Multi-tone Kerr dynamical backaction}

To obtain a first order approximation for the dynamical backaction under multi-tone driving, we simplify the six intracavity fields that will lead to dynamical backaction and keep only the terms proportional to $\delta\Phi(\Omega)$.
In addition, we consider only a single frequency probe input field around one mechanical frequency detuned from the $\gamma_-$ field and obtain
\begin{eqnarray}
\frac{\delta\alpha(\Omega)}{\chi_\mathrm{g}(\Omega)} & = & -\mathcal{K}n_\mathrm{d}\tilde{G}\gamma_+^* \chi_\mathrm{p}^*(-\Omega + 2\Omega_\mathrm{dp})\delta\Phi(\Omega) - i\tilde{G}\gamma_-\delta\Phi(\Omega) + i\sqrt{\kappa_\mathrm{e}}S_0(\Omega) \\
\frac{\delta\alpha^*(-\Omega)}{\chi_\mathrm{g}^*(-\Omega)} & = & -\mathcal{K}n_\mathrm{d}\tilde{G}\gamma_+ \chi_\mathrm{p}(\Omega + 2\Omega_\mathrm{dp})\delta\Phi(\Omega) + i\tilde{G}\gamma_-^*\delta\Phi(\Omega) \\
\frac{\delta\alpha(\Omega + \Omega_\mathrm{dp})}{\chi_\mathrm{g}(\Omega + \Omega_\mathrm{dp})} & = & -\mathcal{K}n_\mathrm{d}G'\alpha_\mathrm{d} \chi_\mathrm{p}^*(-\Omega + \Omega_\mathrm{dp})\delta\Phi(\Omega) - iG'\alpha_\mathrm{d}\delta\Phi(\Omega) \\
\frac{\delta\alpha^*(-\Omega + \Omega_\mathrm{dp})}{\chi_\mathrm{g}^*(-\Omega + \Omega_\mathrm{dp})} & = & -\mathcal{K}n_\mathrm{d}G'\alpha_\mathrm{d} \chi_\mathrm{p}(\Omega + \Omega_\mathrm{dp})\delta\Phi(\Omega) + iG'\alpha_\mathrm{d}\delta\Phi(\Omega) \\
\frac{\delta\alpha(\Omega + 2\Omega_\mathrm{dp})}{\chi_\mathrm{g}(\Omega + 2\Omega_\mathrm{dp})} & = & -\mathcal{K}n_\mathrm{d}\tilde{G}\gamma_-^* \chi_\mathrm{p}^*(-\Omega)\delta\Phi(\Omega) - i\tilde{G}\gamma_+\delta\Phi(\Omega) \\
\frac{\delta\alpha^*(-\Omega + 2\Omega_\mathrm{dp})}{\chi_\mathrm{g}^*(-\Omega + 2\Omega_\mathrm{dp})} & = & -\mathcal{K}n_\mathrm{d}\tilde{G}\gamma_- \chi_\mathrm{p}(\Omega)\delta\Phi(\Omega) + i\tilde{G}\gamma_+^*\delta\Phi(\Omega) + \mathcal{K}n_\mathrm{d}\chi_\mathrm{p}(\Omega)\sqrt{\kappa_\mathrm{e}}S_0(\Omega).
\end{eqnarray}
Injecting all these Fourier components into the RF flux equation of motion and using $g_- = \gamma_-\tilde{G}\Phi_\mathrm{zpf}, g_+ = \gamma_+\tilde{G}\Phi_\mathrm{zpf}$ and $g_\alpha = \alpha_\mathrm{d}G'\Phi_\mathrm{zpf}$ leads to
\begin{equation}
\delta\Phi\left[\Omega_0^2 - \Omega^2 +i\Omega\Gamma_0 + i2\Omega_0\left(\Sigma_\alpha + \Sigma_- + \Sigma_+ + \Sigma_\pm \right) \right] = i2\Omega_0\Phi_\mathrm{zpf}\chi_\mathrm{g}(\Omega)\left[ g_-^* - i\mathcal{K}n_\mathrm{d}g_+\chi_\mathrm{p}^*(-\Omega + 2\Omega_\mathrm{dp}) \right]\sqrt{\kappa_\mathrm{e}}S_0(\Omega)
\end{equation}
where
\begin{eqnarray}
\Sigma_\alpha & = & g_\alpha^2\left(\chi_\mathrm{g}(\Omega + \Omega_\mathrm{dp})\left[1 - i\mathcal{K}n_\mathrm{d}\chi_\mathrm{p}^*(-\Omega + \Omega_\mathrm{dp})  \right]  -\chi_\mathrm{g}^*(-\Omega + \Omega_\mathrm{dp})\left[1 + i\mathcal{K}n_\mathrm{d}\chi_\mathrm{p}(\Omega + \Omega_\mathrm{dp})  \right] \right)
\end{eqnarray}
captures the already earlier derived backaction from the $\alpha_\mathrm{d}$ field, and
\begin{eqnarray}
\Sigma_- & = & |g_-|^2\left[\chi_\mathrm{g}(\Omega) - \chi_\mathrm{g}^*(-\Omega) \right] \\
\Sigma_+ & = & |g_+|^2\left[\chi_\mathrm{g}(\Omega + 2\Omega_\mathrm{dp}) - \chi_\mathrm{g}^*(-\Omega + 2\Omega_\mathrm{dp}) \right]
\end{eqnarray}
capture the direct backaction from the $\gamma_-$ and $\gamma_+$ fields.
The last contribution
\begin{eqnarray}
\Sigma_\pm & = & -i\mathcal{K}n_\mathrm{d}\left[g_-g_+ + g_-^* g_+^*  \right]\left[\chi_\mathrm{g}(\Omega)\chi_\mathrm{p}^*(-\Omega + 2\Omega_\mathrm{dp}) + \chi_\mathrm{g}^*(-\Omega)\chi_\mathrm{p}(\Omega + 2\Omega_\mathrm{dp})\right]
\end{eqnarray}
takes into account interference of RF mode sidebands due to parametric-drive-induced degenerate four-wave mixing processes in the cavity.
The full effective RF mode susceptibility under multi-tone driving is hence given by
\begin{equation}
\chi_0^\mathrm{full}(\Omega) = \frac{1}{\Omega_0^2 - \Omega^2 + i\Omega\Gamma_0 + i2\Omega_0\left[\Sigma_\alpha(\Omega) + \Sigma_-(\Omega) + \Sigma_+(\Omega) + \Sigma_\pm (\Omega) \right]}
\end{equation}
\subsection{HF reflection response - Version I}
One way to ontain the reflection response at the HF cavity in a limited frequency range is to consider particular drive and detuning configurations and omit non-significant terms straight away.
For an optomechanical pump on the red sideband of the cavity quasi-mode and a probe tone around the cavity mode, we can in this approximation simplify the equations of motion to
\begin{eqnarray}
\frac{\delta\Phi(\Omega)}{\chi_0^\mathrm{full}(\Omega)} & = & i2\Omega_0\Phi_\mathrm{zpf}\chi_\mathrm{g}(\Omega)\left[g_-^* - i\mathcal{K}n_\mathrm{d}g_+\chi_\mathrm{p}^*(-\Omega + 2\Omega_\mathrm{dp})\right]\sqrt{\kappa_\mathrm{e}}S_0(\Omega) \\
\frac{\delta\alpha(\Omega)}{\chi_\mathrm{g}(\Omega)} & = & -i\tilde{G}\left[ \gamma_- -i\mathcal{K}n_\mathrm{d}\tilde{G} \gamma_+^*\chi_\mathrm{p}^*(-\Omega + 2\Omega_\mathrm{dp})\right]	\delta\Phi(\Omega) + i\sqrt{\kappa_\mathrm{e}}S_0(\Omega).
\end{eqnarray}
In combination, we obtain
\begin{eqnarray}
\delta\alpha(\Omega) = i\chi_\mathrm{g}(\Omega)\left(1 -i2\Omega_0\chi_\mathrm{g}(\Omega)\chi_0^\mathrm{full}(\Omega)\left[g_-^* - i\mathcal{K}n_\mathrm{d}g_+\chi_\mathrm{p}^*(-\Omega + 2\Omega_\mathrm{dp})  \right] \left[g_- - i\mathcal{K}n_\mathrm{d}g_+^*\chi_\mathrm{p}^*(-\Omega + 2\Omega_\mathrm{dp})  \right]  \right)\sqrt{\kappa_\mathrm{e}}S_0(\Omega) \nonumber \\
\end{eqnarray}
and for the reflection
\begin{eqnarray}
S_{11} = 1-\kappa_\mathrm{e}\chi_\mathrm{g}(\Omega)\left(1 -i2\Omega_0\chi_\mathrm{g}(\Omega)\chi_0^\mathrm{full}(\Omega)\left[g_-^* - i\mathcal{K}n_\mathrm{d}g_+\chi_\mathrm{p}^*(-\Omega + 2\Omega_\mathrm{dp})  \right] \left[g_- - i\mathcal{K}n_\mathrm{d}g_+^*\chi_\mathrm{p}^*(-\Omega + 2\Omega_\mathrm{dp})  \right]  \right). \nonumber \\
\label{eqn:FullOMIT}
\end{eqnarray}
\subsection{HF reflection response - Version II}
Alternatively, we can as first step eliminate the flux contribution from the intracavity equation of motion and afterwards perform the elimination of the parametric coupling between signal and idler.
We start for this approach with the equations
\begin{eqnarray}
\frac{\delta\Phi(\Omega)}{\chi_0(\Omega)} & = & \frac{\hbar \tilde{G}}{C_\mathrm{RF}}\left[ \gamma_-\delta\alpha^*(-\Omega) + \gamma_-^*\delta\alpha(\Omega) \right] + \frac{\hbar G' \alpha_\mathrm{d}}{C_\mathrm{RF}}\left[ \delta\alpha^*(-\Omega + \Omega_\mathrm{dp}) + \delta\alpha(\Omega + \Omega_\mathrm{dp}) \right] \nonumber \\
& & +\frac{\hbar \tilde{G}}{C_\mathrm{RF}}\left[ \gamma_+\delta\alpha^*(-\Omega + 2\Omega_\mathrm{dp}) + \gamma_+^*\delta\alpha(\Omega + 2\Omega_\mathrm{dp}) \right] \\
\frac{\delta\alpha(\Omega)}{\chi_\mathrm{p}(\Omega)} & = & -i\tilde{G}\gamma_-\delta\Phi(\Omega) - iG'\alpha_\mathrm{d}\delta\Phi(\Omega - \Omega_\mathrm{dp}) - i\tilde{G}\gamma_+\delta\Phi(\Omega - 2\Omega_\mathrm{dp}) + i\mathcal{K}n_\mathrm{d}\delta\alpha^*(-\Omega + 2\Omega_\mathrm{dp}) + i\sqrt{\kappa_\mathrm{e}}S_0(\Omega).
\end{eqnarray}
and keep only intracavity field contributions in the three fluxes which are either oscillating at $\Omega$ or at $-\Omega + 2\Omega_\mathrm{dp}$, as we are deep in the sideband-resolved regime.
As result, we get
\begin{eqnarray}
\frac{\delta\alpha(\Omega)}{\chi_\mathrm{p}(\Omega)} & = & -i2\Omega_0|g_-|^2\chi_0(\Omega)\delta\alpha(\Omega) -i2\Omega_0g_- g_+\chi_0(\Omega)\delta\alpha^*(-\Omega + 2\Omega_\mathrm{dp}) \nonumber \\
& & - i2\Omega_0g_\alpha^2 \chi_0(\Omega - \Omega_\mathrm{dp})\delta\alpha(\Omega) - i2\Omega_0g_\alpha^2 \chi_0(\Omega - \Omega_\mathrm{dp}) \delta\alpha^*(-\Omega + 2\Omega_\mathrm{dp})\nonumber \\
& & - i 2\Omega_0 |g_+|^2 \chi_0(\Omega - 2\Omega_\mathrm{dp})\delta\alpha(\Omega) - i2\Omega_0g_+g_-\chi_0(\Omega - 2\Omega_\mathrm{dp})\delta\alpha^*(-\Omega + 2\Omega_\mathrm{dp}) \nonumber \\
& & + i\mathcal{K}n_\mathrm{d}\delta\alpha^*(-\Omega + 2\Omega_\mathrm{dp}) + i\sqrt{\kappa_\mathrm{e}}S_0(\Omega) \nonumber \\
& = & -i2\Omega_0\left[|g_-|^2\chi_0(\Omega) + g_\alpha^2 \chi_0(\Omega - \Omega_\mathrm{dp}) + |g_+|^2\chi_0(\Omega - 2\Omega_\mathrm{dp}) \right] \delta\alpha(\Omega) \nonumber \\
& & +i\left[ \mathcal{K}n_\mathrm{d}-2\Omega_0g_\alpha^2\chi_0(\Omega - \Omega_\mathrm{dp}) - 2\Omega_0g_+g_-\left[ \chi_0(\Omega) + \chi_0(\Omega - 2\Omega_\mathrm{dp}) \right] \right] \delta\alpha^*(-\Omega + 2\Omega_\mathrm{dp}) + i\sqrt{\kappa_\mathrm{e}}S_0(\Omega). ~~~~~
\end{eqnarray}
With the definitions
\begin{eqnarray}
\chi_\mathrm{om}(\Omega) & = & \frac{\chi_\mathrm{p}(\Omega)}{1+i2\Omega_0\left[|g_-|^2\chi_0(\Omega) + g_\alpha^2 \chi_0(\Omega - \Omega_\mathrm{dp}) + |g_+|^2\chi_0(\Omega - 2\Omega_\mathrm{dp}) \right]\chi_\mathrm{p}(\Omega)} \\
\chi_\mathrm{k}(\Omega) & = & \frac{\chi_\mathrm{om}(\Omega)}{1 - \mathcal{A}(\Omega)\mathcal{B}(\Omega) \chi_\mathrm{om}(\Omega)\chi_\mathrm{om}^*(-\Omega + 2\Omega_\mathrm{dp})} \\
\end{eqnarray}
where
\begin{eqnarray}
\mathcal{A}(\Omega) & = & \mathcal{K}n_\mathrm{d}-2\Omega_0g_\alpha^2\chi_0(\Omega - \Omega_\mathrm{dp}) - 2\Omega_0g_+g_-\left[ \chi_0(\Omega) + \chi_0(\Omega - 2\Omega_\mathrm{dp}) \right] \\
\mathcal{B}(\Omega) & = & \mathcal{K}n_\mathrm{d}-2\Omega_0g_\alpha^2\chi_0(\Omega - \Omega_\mathrm{dp}) - 2\Omega_0g_+^*g_-^*\left[ \chi_0(\Omega) + \chi_0(\Omega - 2\Omega_\mathrm{dp}) \right]
\end{eqnarray}
we can also write the intracavity field very short as
\begin{equation}
\delta\alpha(\Omega) = i\chi_\mathrm{k}(\Omega)\sqrt{\kappa_\mathrm{e}}S_0(\Omega)
\end{equation}
and the reflection as
\begin{equation}
S_{11} = 1 - \kappa_\mathrm{e}\chi_\mathrm{k}(\Omega).
\label{eqn:PPIT_full_v2}
\end{equation}
The theory lines in main paper Fig.~3 and Supplementary Fig.~\ref{fig:PPmediumgain} are plotted using this equation (\ref{eqn:PPIT_full_v2}).
\subsection{HF reflection response - Approximation}
In the experimental setting we investigate here, we can approximate the reflection using $g_\alpha = g_+ = 0$ in Eq.~(\ref{eqn:FullOMIT}) and get
\begin{equation}
S_{11} \approx 1 - \kappa_\mathrm{e}\chi_\mathrm{g}(\Omega)\left[1 - i2\Omega_0 |g_-|^2\chi_\mathrm{g}(\Omega) \chi_0^\mathrm{eff}(\Omega)  \right]
\end{equation}
where
\begin{equation}
\chi_0^\mathrm{eff}(\Omega) = \frac{1}{\Omega_0^2 - \Omega^2 + i\Omega \Gamma_0 + i2\Omega_0|g_-|^2\chi_\mathrm{g}(\Omega)}
\end{equation}
Now remembering that we can express the cavity susceptibility around the JPA signal resonance as
\begin{equation}
\chi_\mathrm{g}(\Omega) = \mathcal{G}_\mathrm{s}\chi_\mathrm{s}(\Omega)
\end{equation}
with the signal resonance gain $\mathcal{G}_\mathrm{s}$ yields
\begin{equation}
\chi_0^\mathrm{eff}(\Omega) = \frac{1}{\Omega_0^2 - \Omega^2 + i\Omega \Gamma_0 + i2\Omega_0\mathcal{G}_\mathrm{s}|g_-|^2\chi_s(\Omega)}
\end{equation}
and
\begin{equation}
S_{11} \approx 1 - \mathcal{G}_\mathrm{s}\kappa_\mathrm{e}\chi_\mathrm{s}(\Omega)\left[1 - i2\Omega_0 \mathcal{G}_\mathrm{s}|g_-|^2\chi_\mathrm{s}(\Omega) \chi_0^\mathrm{eff}(\Omega)  \right]
\end{equation}
This is fully equivalent to a linear optomechanical system with
\begin{equation}
\kappa_\mathrm{e}^\mathrm{lin} = \mathcal{G}_\mathrm{s}\kappa_\mathrm{e} = \kappa_1, ~~~~~ g_\mathrm{eff} = \sqrt{\mathcal{G}_\mathrm{s}n_-}\tilde{g}_0.
\end{equation}
To determine the effective cooperativity of main paper Fig.~3 and Supplementary Fig.~\ref{fig:PPmediumgain} datasets, we therefore use
\begin{equation}
S_{11} \approx 1 - \kappa_1\chi_\mathrm{s}(\Omega)\left[1 - i2\Omega_0 g_\mathrm{eff}^2\chi_\mathrm{s}(\Omega) \chi_0^\mathrm{eff}(\Omega)  \right]
\label{eqn:PPIT_approx}
\end{equation}
for fitting and then calculate the effective cooperativity
\begin{equation}
\mathcal{C}_\mathrm{eff} = \frac{4 g_\mathrm{eff}^2}{\kappa \Gamma_0}.
\end{equation}
Note, however, that the theory lines in Fig.~3 and Supplementary Fig.~\ref{fig:PPmediumgain} are not plotted based on this equation, as Eq.~(\ref{eqn:PPIT_approx}) does not show the slight asymmetry in the reflection induced by frequency dependent gain.
\subsection{A multi-tone driven photon-pressure Kerr system with noise}
The Fourier transformed equations of motion are given by
\begin{eqnarray}
\frac{\hat{b}_0}{\chi_{+, 0}} & = & -i\left(g_-^*\hat{c}_0 + g_-\hat{c}_0^\dagger\right) - ig_\alpha\left(\hat{c}_{1} + \hat{c}_{1}^\dagger \right) - i\left(g_+^*\hat{c}_{2} + g_+\hat{c}_{2}^\dagger \right) + \sqrt{\Gamma_\mathrm{e}}\hat{\zeta}_{\mathrm{e}, 0}  + \sqrt{\Gamma_\mathrm{i}}\hat{\zeta}_{\mathrm{i}, 0} \\
\frac{\hat{c}_0}{\chi_{\mathrm{p}, 0}} & = & - ig_-\left(\hat{b}_0 + \hat{b}_0^\dagger\right) - ig_\alpha\left(\hat{b}_{-1} + \hat{b}_{1}^\dagger\right) - ig_+\left(\hat{b}_{-2} + \hat{b}_{2}^\dagger\right) + i\mathcal{K}n_\mathrm{d}\hat{c}_{2}^\dagger + \sqrt{\kappa_\mathrm{e}}\hat{\xi}_{\mathrm{e}, 0} + \sqrt{\kappa_\mathrm{i}}\hat{\xi}_{\mathrm{i}, 0}.
\end{eqnarray}
where we introduced the short version of the Fourier components $\hat{a}_k = \hat{a}(\Omega + k\Omega_\mathrm{dp})$ and $\hat{a}_k^\dagger = \hat{a}^\dagger(-\Omega + k\Omega_\mathrm{dp})$ for $\hat{a} = \hat{b}, \hat{c}$ and $k \in \mathbb{Z}$.
We start again by calculating the approximate expressions
\begin{eqnarray}
\hat{b}_0 + \hat{b}_0^\dagger & \approx & 2\Omega_0g_-^*\chi_{0, 0}\hat{c}_0 +2\Omega_0g_+\chi_{0, 0}\hat{c}_{2}^\dagger + \chi_{+, 0}\hat{S}_0 + \overline{\chi}_{+, 0}\hat{S}_0^\dagger \\
\hat{b}_{-1} + \hat{b}_{1}^\dagger & \approx & 2\Omega_0g_\alpha\chi_{0, -1}\hat{c}_0 + 2\Omega_0g_\alpha\chi_{0, -1}\hat{c}_{2}^\dagger + \chi_{+, -1}\hat{S}_{-1} + \overline{\chi}_{+, 1}\hat{S}_{1}^\dagger \\
\hat{b}_{-2} + \hat{b}_{2}^\dagger & \approx & 2\Omega_0g_+^*\chi_{0, -2}\hat{c}_0 + 2\Omega_0g_-\chi_{0, -2}\hat{c}_{2}^\dagger + \chi_{+, -2}\hat{S}_{-2} + \overline{\chi}_{+, 2}\hat{S}_{2}^\dagger
\end{eqnarray}
where we kept only the most relevant contributions and with the short version $\hat{S}_k = \sqrt{\Gamma_\mathrm{e}}\hat{\zeta}_{\mathrm{e}, k} + \sqrt{\Gamma_\mathrm{i}}\hat{\zeta}_{\mathrm{i}, k}$.
Combining this with the equation for $\hat{c}$ we obtain
\begin{eqnarray}
\frac{\hat{c}_0}{\chi_{\mathrm{om}, 0}} & = & i\mathcal{A}(\Omega)\hat{c}_2^\dagger - ig_-\left[ \chi_{+, 0}\hat{S}_0 + \overline{\chi}_{+, 0}\hat{S}_0^\dagger \right] - ig_\alpha \left[ \chi_{+, -1}\hat{S}_{-1} + \overline{\chi}_{+, 1}\hat{S}_{1}^\dagger \right] - ig_+\left[ \chi_{+, -2}\hat{S}_{-2} + \overline{\chi}_{+, 2}\hat{S}_{2}^\dagger \right] + \sqrt{\kappa_\mathrm{e}}\hat{\xi}_{\mathrm{e}, 0} + \sqrt{\kappa_\mathrm{i}}\hat{\xi}_{\mathrm{i}, 0} \nonumber \\
\frac{\hat{c}_{2}^\dagger}{\overline{\chi}_{\mathrm{om}, 2}} & = & -i\mathcal{B}(\Omega)\hat{c}_0 + ig_-^*\left[ \chi_{+, -2}\hat{S}_{-2} + \overline{\chi}_{+, 2}\hat{S}_2^\dagger \right] + ig_\alpha \left[ \chi_{+, -1}\hat{S}_{-1} + \overline{\chi}_{+, 1}\hat{S}_{1}^\dagger \right] + ig_+^*\left[ \chi_{+, 0}\hat{S}_{0} + \overline{\chi}_{+, 0}\hat{S}_{0}^\dagger \right] + \sqrt{\kappa_\mathrm{e}}\hat{\xi}_{\mathrm{e}, 2}^\dagger + \sqrt{\kappa_\mathrm{i}}\hat{\xi}_{\mathrm{i}, 2}^\dagger. \nonumber
\end{eqnarray}
These two equations can be combined and we obtain
\begin{eqnarray}
\frac{\hat{c}_0}{\chi_{\mathrm{k}, 0}} & = & -i\left(g_- - i g_+^*\mathcal{A}(\Omega)\overline{\chi}_{\mathrm{om}, 2} \right)\left[ \chi_{+, 0}\hat{S}_0 + \overline{\chi}_{+, 0}\hat{S}_0^\dagger \right] \nonumber \\
& & -ig_\alpha\left(1 - i \mathcal{A}(\Omega)\overline{\chi}_{\mathrm{om}, 2} \right)\left[ \chi_{+, -1}\hat{S}_{-1} + \overline{\chi}_{+, 1}\hat{S}_{1}^\dagger \right] \nonumber \\
& & -i\left(g_+ - i g_-^*\mathcal{A}(\Omega)\overline{\chi}_{\mathrm{om}, 2} \right)\left[ \chi_{+, -2}\hat{S}_{-2} + \overline{\chi}_{+, 2}\hat{S}_{2}^\dagger \right] \nonumber \\
& & + \sqrt{\kappa_\mathrm{e}}\hat{\xi}_{\mathrm{e}, 0} + \sqrt{\kappa_\mathrm{i}}\hat{\xi}_{\mathrm{i}, 0} + i\mathcal{A}(\Omega)\overline{\chi}_{\mathrm{om}, 2}\sqrt{\kappa_\mathrm{e}}\hat{\xi}_{\mathrm{e}, 2}^\dagger + i\mathcal{A}(\Omega)\overline{\chi}_{\mathrm{om}, 2}\sqrt{\kappa_\mathrm{i}}\hat{\xi}_{\mathrm{i}, 2}^\dagger
\end{eqnarray}
\subsection{HF cavity output noise}
The HF cavity output field can now be calculated as
\begin{eqnarray}
\hat{c}_\mathrm{out} & = & i\sqrt{\kappa_\mathrm{e}}\chi_{\mathrm{k}, 0}\left(g_- - i g_+^*\mathcal{A}(\Omega)\overline{\chi}_{\mathrm{om}, 2} \right)\left[ \chi_{+, 0}\hat{S}_0 + \overline{\chi}_{+, 0}\hat{S}_0^\dagger \right] \nonumber \\
& & + i\sqrt{\kappa_\mathrm{e}}\chi_{\mathrm{k}, 0}g_\alpha\left(1 - i \mathcal{A}(\Omega)\overline{\chi}_{\mathrm{om}, 2} \right)\left[ \chi_{+, -1}\hat{S}_{-1} + \overline{\chi}_{+, 1}\hat{S}_{1}^\dagger \right] \nonumber \\
& & + i\sqrt{\kappa_\mathrm{e}}\chi_{\mathrm{k}, 0}\left(g_+ - i g_-^*\mathcal{A}(\Omega)\overline{\chi}_{\mathrm{om}, 2} \right)\left[ \chi_{+, -2}\hat{S}_{-2} + \overline{\chi}_{+, 2}\hat{S}_{2}^\dagger \right] \nonumber \\
& & + \left( 1 - \kappa_\mathrm{e}\chi_{\mathrm{k}, 0}\right)\hat{\xi}_{\mathrm{e}, 0} + \sqrt{\kappa_\mathrm{e}\kappa_\mathrm{i}}\chi_{\mathrm{k}, 0}\hat{\xi}_{\mathrm{i}, 0} + i\kappa_\mathrm{e}\chi_{\mathrm{k}, 0}\mathcal{A}(\Omega)\overline{\chi}_{\mathrm{om}, 2}\hat{\xi}_{\mathrm{e}, 2}^\dagger + i\sqrt{\kappa_\mathrm{e}\kappa_\mathrm{i}}\chi_{\mathrm{k}, 0}\mathcal{A}(\Omega)\overline{\chi}_{\mathrm{om}, 2}\hat{\xi}_{\mathrm{i}, 2}^\dagger
\end{eqnarray}
and from this the symmetric power spectral density of the output field in units of quanta as
\begin{eqnarray}
S_{nn} & = & \kappa_\mathrm{e}|\chi_{\mathrm{k}, 0}|^2\left|g_- - i g_+^*\mathcal{A}(\Omega)\overline{\chi}_{\mathrm{om}, 2} \right|^2 \left(|\chi_{+, 0}|^2 + |\overline{\chi}_{+, 0}|^2\right)\left[ \Gamma_\mathrm{e}\left(n_\mathrm{e, 0}^\mathrm{RF} + \frac{1}{2}\right) + \Gamma_\mathrm{i}\left(n_\mathrm{i, 0}^\mathrm{RF} + \frac{1}{2}\right) \right] \nonumber \\
& & + \kappa_\mathrm{e}|\chi_{\mathrm{k}, 0}|^2g_\alpha^2\left|1 - i \mathcal{A}(\Omega)\overline{\chi}_{\mathrm{om}, 2} \right|^2 \left( |\chi_{+, -1}|^2 + |\overline{\chi}_{+, 1}|^2 \right) \left[ \Gamma_\mathrm{e}\left(n_\mathrm{e, -1}^\mathrm{RF} + \frac{1}{2}\right) + \Gamma_\mathrm{i}\left(n_\mathrm{i, -1}^\mathrm{RF} + \frac{1}{2}\right) \right] \nonumber \\
& & + \kappa_\mathrm{e}|\chi_{\mathrm{k}, 0}|^2\left|g_+ - i g_-^*\mathcal{A}(\Omega)\overline{\chi}_{\mathrm{om}, 2} \right|^2\left( |\chi_{+, -2}|^2 + |\overline{\chi}_{+, 2}|^2 \right)\left[ \Gamma_\mathrm{e}\left(n_\mathrm{e, -2}^\mathrm{RF} + \frac{1}{2}\right) + \Gamma_\mathrm{i}\left(n_\mathrm{i, -2}^\mathrm{RF} + \frac{1}{2}\right) \right] \nonumber \\
& & + \left| 1 - \kappa_\mathrm{e}\chi_{\mathrm{k}, 0}\right|^2\left(n_\mathrm{e}^\mathrm{HF}+\frac{1}{2}\right) + \kappa_\mathrm{e}\kappa_\mathrm{i}|\chi_{\mathrm{k}, 0}|^2\left(n_\mathrm{i}^\mathrm{HF}+\frac{1}{2}\right) \nonumber \\
& & + \kappa_\mathrm{e}^2|\chi_{\mathrm{k}, 0}\mathcal{A}(\Omega)\overline{\chi}_{\mathrm{om}, 2}|^2\left(n_\mathrm{e}^\mathrm{HF}+\frac{1}{2}\right) + \kappa_\mathrm{e}\kappa_\mathrm{i}|\chi_{\mathrm{k}, 0}\mathcal{A}(\Omega)\overline{\chi}_{\mathrm{om}, 2}|^2\left(n_\mathrm{i}^\mathrm{HF}+\frac{1}{2}\right).
\end{eqnarray}
Keeping only the leading terms for the experiment described in the main paper and adding the effective HEMT amplifier added noise $n_\mathrm{add}$, we get
\begin{eqnarray}
S_{nn} & = & n_\mathrm{add} + \kappa_\mathrm{e}|\chi_{\mathrm{k}, 0}|^2\left|g_- - i g_+^*\mathcal{A}(\Omega)\overline{\chi}_{\mathrm{om}, 2} \right|^2 |\chi_{+, 0}|^2\left[ \Gamma_\mathrm{e}\left(n_\mathrm{e}^\mathrm{RF} + \frac{1}{2}\right) + \Gamma_\mathrm{i}\left(n_\mathrm{i}^\mathrm{RF} + \frac{1}{2}\right) \right] \nonumber \\
& & + \left| 1 - \kappa_\mathrm{e}\chi_{\mathrm{k}, 0}\right|^2\left(n_\mathrm{e}^\mathrm{HF}+\frac{1}{2}\right) + \kappa_\mathrm{e}\kappa_\mathrm{i}|\chi_{\mathrm{k}, 0}|^2\left(n_\mathrm{i}^\mathrm{HF}+\frac{1}{2}\right) \nonumber \\
& & + \kappa_\mathrm{e}^2|\chi_{\mathrm{k}, 0}\mathcal{A}(\Omega)\overline{\chi}_{\mathrm{om}, 2}|^2\left(n_\mathrm{e}^\mathrm{HF}+\frac{1}{2}\right) + \kappa_\mathrm{e}\kappa_\mathrm{i}|\chi_{\mathrm{k}, 0}\mathcal{A}(\Omega)\overline{\chi}_{\mathrm{om}, 2}|^2\left(n_\mathrm{i}^\mathrm{HF}+\frac{1}{2}\right).
\label{eqn:FullKerrNoise}
\end{eqnarray}
We use Eq.~(\ref{eqn:FullKerrNoise}) to fit the noise datasets shown in main paper Fig.~4, main paper Fig.~5 and Supplementary Fig.~\ref{fig:SCmediumgain}, where we set $n_\mathrm{i}^\mathrm{HF} = n_\mathrm{e}^\mathrm{HF} = 0$, $n_\mathrm{add} = 15$ and use as single fit parameter $n_\mathrm{th}^\mathrm{RF} = \left(\Gamma_\mathrm{e}n_\mathrm{e}^\mathrm{RF} + \Gamma_\mathrm{i}n_\mathrm{i}^\mathrm{RF}\right)/\Gamma_0$.
All other photon-pressure and device parameters going into this equation, we obtain from independent measurements and data analyses.
\subsection{RF residual occupation}
To calculate the residual occupation of the RF mode in the red-sideband cooling scheme, we use
\begin{eqnarray}
\frac{\hat{c}_0}{\chi_{\mathrm{g}, 0}} & = & -i\left[g_- - ig_+^*\mathcal{K}n_\mathrm{d}\overline{\chi}_\mathrm{p, 2} \right]\left(\hat{b}_0 + \hat{b}_0^\dagger \right) -ig_\alpha\left[1 - i\mathcal{K}n_\mathrm{d}\overline{\chi}_\mathrm{p, 2} \right]\left(\hat{b}_{-1} + \hat{b}_1^\dagger \right) -i\left[g_+ - ig_-^*\mathcal{K}n_\mathrm{d}\overline{\chi}_\mathrm{p, 2} \right]\left(\hat{b}_{-2} + \hat{b}_2^\dagger \right) \nonumber \\ & & + \hat{Z}_0 + i\mathcal{K}n_\mathrm{d}\overline{\chi}_{\mathrm{p}, 2}\hat{Z}_{2}^\dagger 
\end{eqnarray}
with $\hat{Z}_k = \sqrt{\kappa_\mathrm{e}}\hat{\xi}_{\mathrm{e}, k} + \sqrt{\kappa_\mathrm{i}}\hat{\xi}_{\mathrm{i}, k}$
and the corresponding equations for the other five Fourier components.
Keeping only the leading terms, we get
\begin{eqnarray}
\frac{\hat{b}_0}{\chi_{+, 0}^\mathrm{eff}} & = & -i\chi_\mathrm{g, 0}\left[g_-^* - ig_+\mathcal{K}n_\mathrm{d}\overline{\chi}_\mathrm{p, 2}  \right]\hat{Z}_0 - ig_\alpha\chi_\mathrm{g, 1}\left( 1 - i\mathcal{K}n_\mathrm{d}\overline{\chi}_\mathrm{p, 1} \right)\hat{Z}_1 - ig_\alpha\overline{\chi}_\mathrm{g, 1}\left( 1 + i\mathcal{K}n_\mathrm{d}\chi_\mathrm{p, 1} \right)\hat{Z}_1^\dagger \nonumber \\
& & - i\overline{\chi}_\mathrm{g, 2}\left[g_+ + ig_-^*\mathcal{K}n_\mathrm{d}\chi_\mathrm{p, 0} \right]\hat{Z}_2^\dagger + \sqrt{\Gamma_\mathrm{e}}\hat{\zeta}_\mathrm{e} + \sqrt{\Gamma_\mathrm{i}}\hat{\zeta}_\mathrm{i}
\end{eqnarray}
with
\begin{equation}
\chi_{+, 0}^\mathrm{eff} = \frac{1}{\frac{\Gamma_0}{2} + i(\Omega - \Omega_0) + \Sigma_\alpha + \Sigma_- + \Sigma_+ + \Sigma_\pm}
\end{equation}
From this, we can calculate the RF circuit power spectral density in units of quanta as
\begin{eqnarray}
S_n^\mathrm{RF} & = & \kappa|\chi_\mathrm{g, 0}|^2 |\chi_{+, 0}^\mathrm{eff}|^2\left| g_-^* - ig_+\mathcal{K}n_\mathrm{d}\overline{\chi}_\mathrm{p, 2}  \right|^2 n_\mathrm{th}^\mathrm{HF} + \kappa g_\alpha^2|\chi_\mathrm{g, 1}|^2 |\chi_{+, 0}^\mathrm{eff}|^2\left| 1- i\mathcal{K}n_\mathrm{d}\overline{\chi}_\mathrm{p, 1}  \right|^2 n_\mathrm{th}^\mathrm{HF} \nonumber \\
& & + \kappa g_\alpha^2|\overline{\chi}_\mathrm{g, 1}|^2 |\chi_{+, 0}^\mathrm{eff}|^2\left| 1+ i\mathcal{K}n_\mathrm{d}\chi_\mathrm{p, 1}  \right|^2 \left(n_\mathrm{th}^\mathrm{HF} + 1\right)  + \kappa |\overline{\chi}_\mathrm{g, 2}|^2 |\chi_{+, 0}^\mathrm{eff}|^2\left| g_+ + ig_-^*\mathcal{K}n_\mathrm{d}\chi_\mathrm{p, 0}  \right|^2 \left(n_\mathrm{th}^\mathrm{HF} + 1\right) \nonumber \\
& & + |\chi_{+, 0}^\mathrm{eff}|^2 \Gamma_\mathrm{e}n_\mathrm{e}^\mathrm{RF} + |\chi_{+, 0}^\mathrm{eff}|^2 \Gamma_\mathrm{i}n_\mathrm{i}^\mathrm{RF}
\end{eqnarray}
where we defined
\begin{equation}
n_\mathrm{th}^\mathrm{HF} = \frac{\kappa_\mathrm{e}n_\mathrm{e}^\mathrm{HF} + \kappa_\mathrm{i}n_\mathrm{i}^\mathrm{HF}}{\kappa}
\end{equation}
and assumed for simplicity that the HF input noise quanta do not depend on the exact frequency, as $n_\mathrm{th}^\mathrm{HF} \ll 1$ in any case.
As final step and for $n_\mathrm{th}^\mathrm{HF} \ll 1$, we can identify the final RF mode occupation as
\begin{equation}
n_\mathrm{fin}^\mathrm{RF} \approx n_\mathrm{rpsn}^\mathrm{RF} + n_\mathrm{cool}^\mathrm{RF}
\label{eqn:nRFfin}
\end{equation}
with the cooled original RF occupation
\begin{equation}
n_\mathrm{cool}^\mathrm{RF} = \int |\chi_{+, 0}^\mathrm{eff}|^2 \Gamma_\mathrm{e}n_\mathrm{e}^\mathrm{RF} \frac{d\Omega}{2\pi} + \int|\chi_{+, 0}^\mathrm{eff}|^2 \Gamma_\mathrm{i}n_\mathrm{i}^\mathrm{RF}\frac{d\Omega}{2\pi},
\label{eqn:nRFcool}
\end{equation}
and the HF cavity induced radiation pressure shot noise occupation
\begin{eqnarray}
n_\mathrm{rpsn}^\mathrm{RF} & = & \int \kappa g_\alpha^2|\overline{\chi}_\mathrm{g, 1}|^2 |\chi_{+, 0}^\mathrm{eff}|^2\left| 1+ i\mathcal{K}n_\mathrm{d}\chi_\mathrm{p, 1}  \right|^2  \frac{d\Omega}{2\pi} + \int \kappa |\overline{\chi}_\mathrm{g, 2}|^2 |\chi_{+, 0}^\mathrm{eff}|^2\left| g_+ + ig_-^*\mathcal{K}n_\mathrm{d}\chi_\mathrm{p, 0}  \right|^2  \frac{d\Omega}{2\pi}
\label{eqn:nRFrpsn}
\end{eqnarray}
that originates from the Josephson-amplified quantum fluctuations of the driven HF cavity.
To obtain the data points of the final RF mode occupation, as shown in main paper Fig.~5 and Supplementary Fig.~\ref{fig:SCmediumgain}, we calculate these integrals numerically based on the parameters we use and obtain for the fits of the output spectral densities with Eq.~(\ref{eqn:FullKerrNoise}).

\subsection{HF residual occupation}
Analogously, we can calculate the residual HF occupation in the red-sideband cooling scheme and obtain for the power spectral density in units of quanta
\begin{eqnarray}
S_n^\mathrm{HF} & = & \kappa|\chi_\mathrm{g, 0}|^2\left| 1 - \chi_\mathrm{g, 0}\chi_{+, 0}^\mathrm{eff} \left[g_- - ig_+^*\mathcal{K}n_\mathrm{d}\overline{\chi}_\mathrm{p, 2} \right]\left[g_-^* - ig_+\mathcal{K}n_\mathrm{d}\overline{\chi}_\mathrm{p, 2} \right]\right|^2n_\mathrm{th}^\mathrm{HF}  \nonumber \\ 
& & + ~\kappa|\chi_\mathrm{g, 0}|^2\left| i\mathcal{K}n_\mathrm{d}\overline{\chi}_\mathrm{p, 2} - \overline{\chi}_\mathrm{g, 2}\chi_{+, 0}^\mathrm{eff} \left[g_- - ig_+^*\mathcal{K}n_\mathrm{d}\overline{\chi}_\mathrm{p, 2} \right]\left[g_+ + ig_-^*\mathcal{K}n_\mathrm{d}\chi_\mathrm{p, 0} \right]\right|^2\left(n_\mathrm{th}^\mathrm{HF}+1\right) \nonumber \\
& & + ~\Gamma_0|\chi_\mathrm{g, 0}|^2\left| g_- - ig_+^*\mathcal{K}n_\mathrm{d}\overline{\chi}_\mathrm{p, 2}\right|^2|\chi_{+, 0}^\mathrm{eff}|^2 n_\mathrm{th}^\mathrm{RF}.
\end{eqnarray}
To obtain from this the final effective HF mode occupation, we integrate this PSD over the frequencies $\omega \leq \omega_\mathrm{d}$, as we are interested in the effective occupation of the signal mode only. 
\subsection{Approximate noise treatment with the effective signal mode}
The Fourier transformed and approximated ($g_\alpha = g_+ = 0$, $\kappa/\Omega_0 \ll 1$, $\omega_\mathrm{p} \approx \omega_\mathrm{s} - \Omega_0$) equations of motion are
\begin{eqnarray}
\frac{\hat{b}}{\chi_+} & = & -ig_-^*\hat{c} + \sqrt{\Gamma_\mathrm{e}}\hat{\zeta}_\mathrm{e}  + \sqrt{\Gamma_\mathrm{i}}\hat{\zeta}_\mathrm{i} \\
\frac{\hat{c}}{\chi_\mathrm{p}} & = & - ig_-\hat{b} + i\mathcal{K}n_\mathrm{d}\hat{c}^\dagger + \sqrt{\kappa_\mathrm{e}}\hat{\xi}_\mathrm{e} + \sqrt{\kappa_\mathrm{i}}\hat{\xi}_\mathrm{i} \\
\frac{\hat{c}^\dagger}{\overline{\chi}_\mathrm{p}} & = & - i\mathcal{K}n_\mathrm{d}\hat{c} + \sqrt{\kappa_\mathrm{e}}\hat{\xi}_\mathrm{e}^\dagger + \sqrt{\kappa_\mathrm{i}}\hat{\xi}_\mathrm{i}^\dagger.
\end{eqnarray}
Note that the annihilation operators of the cavity fields and noise here are at $\Omega$, while the annihilation operators are at $-\Omega + 2\Omega_\mathrm{dp}$. 
In the next step we obtain
\begin{eqnarray}
\frac{\hat{b}}{\chi_+} & = & -ig_-^*\hat{c} + \sqrt{\Gamma_\mathrm{e}}\hat{\zeta}_\mathrm{e}  + \sqrt{\Gamma_\mathrm{i}}\hat{\zeta}_\mathrm{i} \\
\frac{\hat{c}}{\chi_\mathrm{g}} & = & - ig_-\hat{b} + \sqrt{\kappa_\mathrm{e}}\hat{\xi}_\mathrm{e} + \sqrt{\kappa_\mathrm{i}}\hat{\xi}_\mathrm{i} + i\mathcal{K}n_\mathrm{d}\overline{\chi}_\mathrm{p}\sqrt{\kappa_\mathrm{e}}\hat{\xi}_\mathrm{e}^\dagger + i\mathcal{K}n_\mathrm{d}\overline{\chi}_\mathrm{p}\sqrt{\kappa_\mathrm{i}}\hat{\xi}_\mathrm{i}^\dagger.
\end{eqnarray}
or
\begin{eqnarray}
\frac{\hat{b}}{\chi_+} & = & -ig_-^*\hat{c} + \sqrt{\Gamma_\mathrm{e}}\hat{\zeta}_\mathrm{e}  + \sqrt{\Gamma_\mathrm{i}}\hat{\zeta}_\mathrm{i} \\
\frac{\hat{c}}{\chi_\mathrm{s}} & = & - ig_-\mathcal{G}_\mathrm{s}\hat{b} + \sqrt{\kappa_\mathrm{e}}\mathcal{G}_\mathrm{s}\hat{\xi}_\mathrm{e} + \sqrt{\kappa_\mathrm{i}}\mathcal{G}_\mathrm{s}\hat{\xi}_\mathrm{i} + i\mathcal{K}n_\mathrm{d}\overline{\chi}_\mathrm{p}\sqrt{\kappa_\mathrm{e}}\mathcal{G}_\mathrm{s}\hat{\xi}_\mathrm{e}^\dagger + i\mathcal{K}n_\mathrm{d}\overline{\chi}_\mathrm{p}\sqrt{\kappa_\mathrm{i}}\mathcal{G}_\mathrm{s}\hat{\xi}_\mathrm{i}^\dagger.
\end{eqnarray}
which we can use to eliminate the other mode from each equation
\begin{eqnarray}
\frac{\hat{b}}{\chi_+^\mathrm{eff}} & = & -ig_-^*\mathcal{G}_\mathrm{s}\chi_\mathrm{s}\sqrt{\kappa_\mathrm{e}}\hat{\xi}_\mathrm{e} -ig_-^*\mathcal{G}_\mathrm{s}\chi_\mathrm{s} \sqrt{\kappa_\mathrm{i}}\hat{\xi}_\mathrm{i} + g_-^*\mathcal{G}_\mathrm{s}\chi_\mathrm{s} \mathcal{K}n_\mathrm{d}\overline{\chi}_\mathrm{p}\sqrt{\kappa_\mathrm{e}}\hat{\xi}_\mathrm{e}^\dagger + g_-^*\mathcal{G}_\mathrm{s}\chi_\mathrm{s}\mathcal{K}n_\mathrm{d}\overline{\chi}_\mathrm{p}\sqrt{\kappa_\mathrm{i}}\hat{\xi}_\mathrm{i}^\dagger + \sqrt{\Gamma_\mathrm{e}}\hat{\zeta}_\mathrm{e}  + \sqrt{\Gamma_\mathrm{i}}\hat{\zeta}_\mathrm{i} \\
\frac{\hat{c}}{\chi_\mathrm{s}^\mathrm{eff}} & = & \sqrt{\kappa_\mathrm{e}}\mathcal{G}_\mathrm{s}\hat{\xi}_\mathrm{e} + \sqrt{\kappa_\mathrm{i}}\mathcal{G}_\mathrm{s}\hat{\xi}_\mathrm{i} + i\mathcal{K}n_\mathrm{d}\overline{\chi}_\mathrm{p}\sqrt{\kappa_\mathrm{e}}\mathcal{G}_\mathrm{s}\hat{\xi}_\mathrm{e}^\dagger + i\mathcal{K}n_\mathrm{d}\overline{\chi}_\mathrm{p}\sqrt{\kappa_\mathrm{i}}\mathcal{G}_\mathrm{s}\hat{\xi}_\mathrm{i}^\dagger -ig_-\chi_+ \sqrt{\Gamma_\mathrm{e}}\mathcal{G}_\mathrm{s}\hat{\zeta}_\mathrm{e}  -ig_-\chi_+ \sqrt{\Gamma_\mathrm{i}}\mathcal{G}_\mathrm{s}\hat{\zeta}_\mathrm{i}
\end{eqnarray}
with
\begin{equation}
\chi_+^\mathrm{eff} = \frac{\chi_+}{1 + g_\mathrm{eff}^2\chi_+\chi_\mathrm{s}}, ~~~~~ \chi_\mathrm{s}^\mathrm{eff} = \frac{\chi_\mathrm{s}}{1 + g_\mathrm{eff}^2\chi_+\chi_\mathrm{s}}.
\end{equation}
For the photon spectral density in the HF mode, we get with this approximately
\begin{eqnarray}
S_{n}^\mathrm{HF} & = & \langle\hat{c}^\dagger\hat{c}\rangle \\
& = & \kappa|\chi_\mathrm{s}^\mathrm{eff}|^2\mathcal{G}_\mathrm{s}^2 n_\mathrm{th}^\mathrm{HF} + \kappa|\chi_\mathrm{s}^\mathrm{eff}|^2 \frac{\mathcal{K}^2 n_\mathrm{d}^2}{|\kappa + 2i\Omega_\mathrm{s}|^2}  (n_\mathrm{th}^\mathrm{HF} + 1) + g_\mathrm{eff}^2|\chi_+|^2|\chi_\mathrm{s}^\mathrm{eff}|^2\Gamma_0 \mathcal{G}_\mathrm{s}n_\mathrm{th}^\mathrm{RF}.
\end{eqnarray}
For the HF mode, this is fully equivalent to a standard photon-pressure system with
\begin{equation}
\tilde{n}_\mathrm{th}^\mathrm{HF} = \mathcal{G}_\mathrm{s}^2 n_\mathrm{th}^\mathrm{HF} + \frac{\mathcal{K}^2 n_\mathrm{d}^2}{|\kappa + 2i\Omega_\mathrm{s}|^2}  (n_\mathrm{th}^\mathrm{HF} + 1), ~~~~~ \tilde{n}_\mathrm{th}^\mathrm{RF} = \mathcal{G}_\mathrm{s}n_\mathrm{th}^\mathrm{RF}
\end{equation}
or
\begin{equation}
\tilde{n}_\mathrm{q}^\mathrm{HF} = \frac{\mathcal{K}^2 n_\mathrm{d}^2}{|\kappa + 2i\Omega_\mathrm{s}|^2}, ~~~~~ \tilde{n}_\mathrm{th}^\mathrm{RF} = \mathcal{G}_\mathrm{s}n_\mathrm{th}^\mathrm{RF}
\label{eqn:nHFtilde}
\end{equation}
for $n_\mathrm{th}^\mathrm{HF} = 0$.
Note that this also means $\tilde{n}_\mathrm{q}^\mathrm{HF} \approx \mathcal{G}_\mathrm{s}\left( \mathcal{G}_\mathrm{s} - 1 \right)$.
For the RF mode, we find
\begin{eqnarray}
S_{n}^\mathrm{RF} & = & \langle\hat{b}^\dagger\hat{b}\rangle \\
& = & \Gamma_0|\chi_+^\mathrm{eff}|^2 n_\mathrm{th}^\mathrm{RF}  + g_\mathrm{eff}^2|\chi_\mathrm{s}|^2|\chi_+^\mathrm{eff}|^2\kappa \mathcal{G}_\mathrm{s}n_\mathrm{th}^\mathrm{HF} + \kappa g_\mathrm{eff}^2|\chi_+^\mathrm{eff}|^2|\chi_\mathrm{s}|^2 \frac{\mathcal{K}^2 n_\mathrm{d}^2}{\mathcal{G}_\mathrm{s}|\kappa + 2i\Omega_\mathrm{s}|^2} (n_\mathrm{th}^\mathrm{HF} + 1).
\end{eqnarray}
which corresponds to a usual photon-pressure device with
\begin{equation}
\bar{n}_\mathrm{th}^\mathrm{HF} = \mathcal{G}_\mathrm{s} n_\mathrm{th}^\mathrm{HF} + \frac{\mathcal{K}^2 n_\mathrm{d}^2}{\mathcal{G}_\mathrm{s}|\kappa + 2i\Omega_\mathrm{s}|^2}  (n_\mathrm{th}^\mathrm{HF} + 1), ~~~~~ \bar{n}_\mathrm{th}^\mathrm{RF} = n_\mathrm{th}^\mathrm{RF}
\end{equation}
or
\begin{equation}
\bar{n}_\mathrm{q}^\mathrm{HF} = \frac{\mathcal{K}^2 n_\mathrm{d}^2}{\mathcal{G}_\mathrm{s}|\kappa + 2i\Omega_\mathrm{s}|^2}, ~~~~~ \bar{n}_\mathrm{th}^\mathrm{RF} = n_\mathrm{th}^\mathrm{RF}
\label{eqn:nHFbar}
\end{equation}
for $n_\mathrm{th}^\mathrm{HF} = 0$.
The latter results reveal an interesting asymmetry.
From the viewpoint of the RF mode, the thermal occupation of both modes is redced by $1/\mathcal{G}_\mathrm{s}$ compared to the viewpoint of the HF mode
\begin{equation}
\bar{n}_\mathrm{th}^\mathrm{HF} = \frac{	\tilde{n}_\mathrm{th}^\mathrm{HF}}{\mathcal{G}_\mathrm{s}}, ~~~~~ \bar{n}_\mathrm{q}^\mathrm{HF} = \frac{	\tilde{n}_\mathrm{q}^\mathrm{HF}}{\mathcal{G}_\mathrm{s}}, ~~~~~ \bar{n}_\mathrm{th}^\mathrm{RF} = \frac{	\tilde{n}_\mathrm{th}^\mathrm{RF}}{\mathcal{G}_\mathrm{s}}
\end{equation}
which can be interpreted as half of the power gain in the HF mode occuring before the energy exchange with the RF mode and half after.
We can also use these results to write an analytical expression for the final cooled photon number in the HF and RF modes, respectively.
For the RF mode we obtain
\begin{equation}
n_\mathrm{fin}^\mathrm{RF} = \frac{\Gamma_0}{\kappa + \Gamma_0}\frac{4g_\mathrm{eff}^2 + \kappa(\kappa + \Gamma_0)}{4g_\mathrm{eff}^2 + \kappa\Gamma_0}n_\mathrm{th}^\mathrm{RF} + \frac{\kappa}{\kappa + \Gamma_0}\frac{4g_\mathrm{eff}^2 }{4g_\mathrm{eff}^2 + \kappa\Gamma_0}\bar{n}_\mathrm{q}^\mathrm{HF}
\label{eqn:nRFfinanal}
\end{equation}
where $\bar{n}_\mathrm{q}^\mathrm{HF}$ is given by Eq.~(\ref{eqn:nHFbar}) and is approximately $\bar{n}_\mathrm{q}^\mathrm{HF} \approx \mathcal{G}_\mathrm{s}-1$.
For the HF mode we get
\begin{equation}
n_\mathrm{fin}^\mathrm{HF} = \frac{\kappa}{\kappa + \Gamma_0}\frac{4g_\mathrm{eff}^2 + \Gamma_0(\kappa + \Gamma_0)}{4g_\mathrm{eff}^2 + \kappa\Gamma_0}\tilde{n}_\mathrm{q}^\mathrm{HF} + \frac{\Gamma_0}{\kappa + \Gamma_0}\frac{4g_\mathrm{eff}^2 }{4g_\mathrm{eff}^2 + \kappa\Gamma_0}\tilde{n}_\mathrm{th}^\mathrm{RF}
\label{eqn:nHFfinanal}
\end{equation}
where $\tilde{n}_\mathrm{q}^\mathrm{HF}$ and $\tilde{n}_\mathrm{th}^\mathrm{RF}$ are given by Eq.~(\ref{eqn:nHFtilde}) and $\tilde{n}_\mathrm{q}^\mathrm{HF} \approx \mathcal{G}_\mathrm{s}(\mathcal{G}_\mathrm{s}-1)$.
Finally, we will formulate the HF cavity output field (which is what we observe in our experiment) in terms of these effective occupations.
We get first 
\begin{eqnarray}
\hat{c}_\mathrm{out} & = & \left(1-\kappa_\mathrm{e}\mathcal{G}_\mathrm{s}\chi_\mathrm{s}^\mathrm{eff}\right)\hat{\xi}_\mathrm{e} - \sqrt{\kappa_\mathrm{e}\kappa_\mathrm{i}}\chi_\mathrm{s}^\mathrm{eff}\mathcal{G}_\mathrm{s}\hat{\xi}_\mathrm{i} - i\mathcal{K}n_\mathrm{d}\overline{\chi}_\mathrm{p}\kappa_\mathrm{e}\chi_\mathrm{s}^\mathrm{eff}\mathcal{G}_\mathrm{s}\hat{\xi}_\mathrm{e}^\dagger - i\mathcal{K}n_\mathrm{d}\overline{\chi}_\mathrm{p}\sqrt{\kappa_\mathrm{e}\kappa_\mathrm{i}}\chi_\mathrm{s}^\mathrm{eff}\mathcal{G}_\mathrm{s}\hat{\xi}_\mathrm{i}^\dagger \nonumber \\
& &  + ig_-\chi_+ \sqrt{\kappa_\mathrm{e}\Gamma_\mathrm{e}}\chi_\mathrm{s}^\mathrm{eff}\mathcal{G}_\mathrm{s}\hat{\zeta}_\mathrm{e}  + ig_-\chi_+ \sqrt{\kappa_\mathrm{e}\Gamma_\mathrm{i}}\chi_\mathrm{s}^\mathrm{eff}\mathcal{G}_\mathrm{s}\hat{\zeta}_\mathrm{i}
\end{eqnarray}
and with this the symmetrized PSD
\begin{eqnarray}
S_{nn} & = & \left|1-\kappa_\mathrm{e}\mathcal{G}_\mathrm{s}\chi_\mathrm{s}^\mathrm{eff}\right|^2\left(n_\mathrm{e}^\mathrm{HF} + \frac{1}{2}\right) + \kappa_\mathrm{e}\kappa_\mathrm{i}|\chi_\mathrm{s}^\mathrm{eff}|^2\mathcal{G}_\mathrm{s}^2\left(n_\mathrm{i}^\mathrm{HF} + \frac{1}{2}\right) \nonumber \\
& & + \mathcal{K}^2n_\mathrm{d}^2|\overline{\chi}_\mathrm{p}|^2\kappa_\mathrm{e}^2|\chi_\mathrm{s}^\mathrm{eff}|^2\mathcal{G}_\mathrm{s}^2\left(n_\mathrm{e}^\mathrm{HF} + \frac{1}{2}\right) + \mathcal{K}^2n_\mathrm{d}^2|\overline{\chi}_\mathrm{p}|^2\kappa_\mathrm{e}\kappa_\mathrm{i}|\chi_\mathrm{s}^\mathrm{eff}|^2\mathcal{G}_\mathrm{s}^2\left(n_\mathrm{i}^\mathrm{HF} + \frac{1}{2}\right) \nonumber \\
& &  + g_\mathrm{eff}^2|\chi_+|^2 \kappa_\mathrm{e}\Gamma_\mathrm{e}|\chi_\mathrm{s}^\mathrm{eff}|^2\mathcal{G}_\mathrm{s}\left(n_\mathrm{e}^\mathrm{RF} + \frac{1}{2}\right)  + g_\mathrm{eff}^2|\chi_+|^2 \kappa_\mathrm{e}\Gamma_\mathrm{i}|\chi_\mathrm{s}^\mathrm{eff}|^2\mathcal{G}_\mathrm{s}\left(n_\mathrm{i}^\mathrm{RF} + \frac{1}{2}\right) \nonumber \\
& = & \frac{1}{2} + n_\mathrm{e}^\mathrm{HF} + \kappa_1\kappa_\mathrm{i} |\chi_\mathrm{s}^\mathrm{eff}|^2 \mathcal{G}_\mathrm{s}\left(n_\mathrm{i}^\mathrm{HF} - n_\mathrm{e}^\mathrm{HF} \right)   \nonumber \\
& & + \kappa_1\kappa_\mathrm{e}\mathcal{K}^2n_\mathrm{d}^2|\overline{\chi}_\mathrm{p}|^2|\chi_\mathrm{s}^\mathrm{eff}|^2\mathcal{G}_\mathrm{s}\left(2n_\mathrm{e}^\mathrm{HF} + 1\right) + \kappa_1\kappa_\mathrm{i}\mathcal{K}^2n_\mathrm{d}^2|\overline{\chi}_\mathrm{p}|^2|\chi_\mathrm{s}^\mathrm{eff}|^2\mathcal{G}_\mathrm{s}\left(n_\mathrm{i}^\mathrm{HF} + n_\mathrm{e}^\mathrm{HF} + 1\right)    \nonumber \\
& & + \kappa_1 g_\mathrm{eff}^2 |\chi_+|^2 |\chi_\mathrm{s}^\mathrm{eff}|^2 \Gamma_0\left(n_\mathrm{th}^\mathrm{RF} - n_\mathrm{e}^\mathrm{HF}   \right)
\end{eqnarray}
For vanishing thermal occupation of the HF bath and including the added photons of the HEMT amplifier we get
\begin{eqnarray}
S_{nn} & = & \frac{1}{2} + n_\mathrm{add} + \kappa_1\kappa|\chi_\mathrm{s}^\mathrm{eff}|^2\bar{n}_\mathrm{q}^\mathrm{HF} + \kappa_1 g_\mathrm{eff}^2 |\chi_+|^2 |\chi_\mathrm{s}^\mathrm{eff}|^2 \Gamma_0\bar{n}_\mathrm{th}^\mathrm{RF}
\end{eqnarray}

\subsection{Imprecision noise}
One important quantity for the detection of RF signals through upconversion is the total detection imprecision noise.
The HF intracavity field with only the most dominant terms is given by
\begin{equation}
\hat{c} = - ig_-\chi_\mathrm{s}\mathcal{G}_\mathrm{s}\frac{\hat{\Phi}}{\Phi_\mathrm{zpf}} + \sqrt{\kappa_\mathrm{e}}\chi_\mathrm{s}\mathcal{G}_\mathrm{s}\hat{\xi}_\mathrm{e} + \sqrt{\kappa_\mathrm{i}}\chi_\mathrm{s}\mathcal{G}_\mathrm{s}\hat{\xi}_\mathrm{i} + i\mathcal{K}n_\mathrm{d}\overline{\chi}_\mathrm{p}\sqrt{\kappa_\mathrm{e}}\chi_\mathrm{s}\mathcal{G}_\mathrm{s}\hat{\xi}_\mathrm{e}^\dagger + i\mathcal{K}n_\mathrm{d}\overline{\chi}_\mathrm{p}\sqrt{\kappa_\mathrm{i}}\chi_\mathrm{s}\mathcal{G}_\mathrm{s}\hat{\xi}_\mathrm{i}^\dagger
\end{equation}
and the related output field by
\begin{equation}
\hat{c}_\mathrm{out} =  ig_-\chi_\mathrm{s}\sqrt{\kappa_\mathrm{e}}\mathcal{G}_\mathrm{s}\frac{\hat{\Phi}}{\Phi_\mathrm{zpf}} + \left(1- \kappa_\mathrm{e}\chi_\mathrm{s}\mathcal{G}_\mathrm{s}\right)\hat{\xi}_\mathrm{e} - \sqrt{\kappa_\mathrm{i}\kappa_\mathrm{e}}\chi_\mathrm{s}\mathcal{G}_\mathrm{s}\hat{\xi}_\mathrm{i} - i\mathcal{K}n_\mathrm{d}\overline{\chi}_\mathrm{p}\kappa_\mathrm{e}\chi_\mathrm{s}\mathcal{G}_\mathrm{s}\hat{\xi}_\mathrm{e}^\dagger - i\mathcal{K}n_\mathrm{d}\overline{\chi}_\mathrm{p}\sqrt{\kappa_\mathrm{i}\kappa_\mathrm{e}}\chi_\mathrm{s}\mathcal{G}_\mathrm{s}\hat{\xi}_\mathrm{i}^\dagger
\end{equation}
For low cooperativity (i.e. no correlations between the HF input noise and the RF flux) and including the added noise by the HEMT detection chain, we obtain for the symmetric HF output power spectral density
\begin{eqnarray}
S_{nn} & = & n_\mathrm{add} + \kappa_\mathrm{e}|g_-|^2|\chi_\mathrm{s}|^2\mathcal{G}_\mathrm{s}^2\frac{S_\Phi}{2\Phi_\mathrm{zpf}^2} + |1- \kappa_\mathrm{e}\chi_\mathrm{s}\mathcal{G}_\mathrm{s}|^2\left( n_\mathrm{e}^\mathrm{HF} + \frac{1}{2} \right) + \kappa_\mathrm{i}\kappa_\mathrm{e}|\chi_\mathrm{s}|^2\mathcal{G}_\mathrm{s}^2\left( n_\mathrm{i}^\mathrm{HF} + \frac{1}{2} \right) \nonumber \\
& & + \kappa_\mathrm{e}^2\mathcal{K}^2n_\mathrm{d}^2|\overline{\chi}_\mathrm{p}|^2|\chi_\mathrm{s}|^2\mathcal{G}_\mathrm{s}^2\left( n_\mathrm{e}^\mathrm{HF} + \frac{1}{2} \right) + \kappa_\mathrm{i}\kappa_\mathrm{e}\mathcal{K}^2n_\mathrm{d}^2|\overline{\chi}_\mathrm{p}|^2|\chi_\mathrm{s}|^2\mathcal{G}_\mathrm{s}^2\left( n_\mathrm{i}^\mathrm{HF} + \frac{1}{2} \right)
\end{eqnarray}
where we introduced the symmetric, one-sided RF flux spectral density
\begin{equation}
S_\Phi(\Omega) = \langle \hat{\Phi}^\dagger(\Omega)\hat{\Phi}(\Omega) \rangle + \langle \hat{\Phi}(\Omega)\hat{\Phi}^\dagger(\Omega) \rangle .
\end{equation}
For negligible HF thermal noise $n_\mathrm{i}^\mathrm{HF}, n_\mathrm{e}^\mathrm{HF} \ll 0.5$ we can write this as
\begin{eqnarray}
S_{{nn}, \mathrm{q}}^\mathrm{tot} & = & \frac{1}{2} + n_\mathrm{add} + \kappa_\mathrm{e}|g_-|^2|\chi_\mathrm{s}|^2\mathcal{G}_\mathrm{s}^2\frac{S_\Phi}{2\Phi_\mathrm{zpf}^2} + \kappa_\mathrm{e}\kappa\mathcal{K}^2n_\mathrm{d}^2|\overline{\chi}_\mathrm{p}|^2|\chi_\mathrm{s}|^2\mathcal{G}_\mathrm{s}^2 \nonumber \\
& = & \frac{1}{2} + n_\mathrm{add} + \kappa_\mathrm{e}|g_-|^2|\chi_\mathrm{s}|^2\mathcal{G}_\mathrm{s}^2\frac{S_\Phi}{2\Phi_\mathrm{zpf}^2} + \kappa_\mathrm{e}\kappa|\chi_\mathrm{s}|^2 \tilde{n}_\mathrm{q}^\mathrm{HF}.
\end{eqnarray}
We can calculate the apparent RF flux noise now assuming the conversion of an ideal, noiseless detector and obtain
\begin{eqnarray}
S_\Phi^\mathrm{det} & = & \frac{2\Phi_\mathrm{zpf}^2}{\kappa_\mathrm{e}|g_-|^2 |\chi_\mathrm{s}|^2 \mathcal{G}_\mathrm{s}^2}S_{nn, \mathrm{q}}^\mathrm{tot} \nonumber\\
& = & S_\Phi + \frac{2\Phi_\mathrm{zpf}^2}{\kappa_\mathrm{e}|g_-|^2 |\chi_\mathrm{s}|^2 \mathcal{G}_\mathrm{s}^2}\left( \frac{1}{2} + n_\mathrm{add} + \kappa_\mathrm{e}\kappa|\chi_\mathrm{s}|^2 \tilde{n}_\mathrm{q}^\mathrm{HF} \right) \nonumber \\
& = &  S_\Phi + S_\mathrm{imp}
\end{eqnarray}
with the imprecision noise
\begin{eqnarray}
S_\mathrm{imp} & = & \frac{2\Phi_\mathrm{zpf}^2}{\kappa_\mathrm{e}|g_-|^2 |\chi_\mathrm{s}|^2 \mathcal{G}_\mathrm{s}^2}\left( \frac{1}{2} + n_\mathrm{add} + \kappa_\mathrm{e}\kappa|\chi_\mathrm{s}|^2 \tilde{n}_\mathrm{q}^\mathrm{HF} \right) \nonumber \\
& \approx & \frac{2\Phi_\mathrm{zpf}^2 }{\kappa_\mathrm{e}  g_\mathrm{eff}^2 |\chi_\mathrm{s}|^2 }\frac{n_\mathrm{add}}{\mathcal{G}_\mathrm{s}} + \frac{2\Phi_\mathrm{zpf}^2\kappa}{g_\mathrm{eff}^2 }\frac{\tilde{n}_\mathrm{q}^\mathrm{HF}}{\mathcal{G}_\mathrm{s}}
\label{eqn:imprecision}
\end{eqnarray}
where the last approximation is valid for $n_\mathrm{add} \gg 1/2$.
The first term, interestingly, is completely equivalent to usual photon-pressure or optomechanical devices, but with a Josephson-reduced added noise $n_\mathrm{add}' = n_\mathrm{add}/\mathcal{G}_\mathrm{s}$ for a given effective cooperativity.
The second term, however, is appearing due to the quantum noise Josephson heating of the HF mode and is equivalent to the imprecision noise of the JPA with
\begin{equation}
n_\mathrm{JPA} = \kappa_\mathrm{e}\kappa|\chi_\mathrm{s}|^2\tilde{n}_\mathrm{q}^\mathrm{HF}.
\end{equation}

\subsection{Discussion of imprecision noise}

The minimum in imprecision noise we achieve here of $S_\mathrm{imp}^\mathrm{min} \approx 80\,\mu\Phi_0^2\,$Hz$^{-1}$ (presented in main paper Fig.~4) seems rather large compared to other reports in similar systems\cite{Hatridge11S, LevensonFalk16S, Foroughi18S} and is also not very close to the standard quantum limit $S_\mathrm{imp}^\mathrm{min}/S_\mathrm{\Phi}^\mathrm{SQL} \approx 50$ where $S_\mathrm{\Phi}^\mathrm{SQL} = 2\Phi_\mathrm{zpf}^2/\Gamma_\mathrm{eff} \approx 1.6\,\mu\Phi_0^2\,$Hz$^{-1}$.
It is important to note several things in this context though.
In our red-sideband configuration, the upconverted RF field is amplified only by $\mathcal{G}_\mathrm{s}$ and the second $\mathcal{G}_\mathrm{s}$ factor increases the cooperativity and the dynamical backaction as presented in main paper Fig.~3.
Contrastingly, in the backaction-free, resonantly pumped case, the upconverted noise would experience a gain of $\mathcal{G}_\mathrm{s}^2$.
At the same time, the amplified quantum noise of the HF mode is $\propto \mathcal{G}_\mathrm{s}(\mathcal{G}_\mathrm{s}-1)$ and therefore dominates the output spectrum for large $\mathcal{G}_\mathrm{s}$.
Hence, there is an optimal working point for each $n_\mathrm{add}$ and $\kappa_\mathrm{e}/\kappa$ in the red-sideband case and beyond this point a larger gain will increase the imprecision noise once again.
In Supplementary Fig.\ref{fig:Imprecision} we plot the minimum imprecision noise at ($\omega_\mathrm{s}$) normalized to $S_\Phi^\mathrm{SQL}$ (equivalent to main paper Eq.~(14))
\begin{equation}
\frac{S_\mathrm{imp}^\mathrm{min}}{S_\Phi^\mathrm{SQL}} = \frac{1 + \mathcal{C}_\mathrm{eff}}{\mathcal{C}_\mathrm{eff}}\frac{\kappa}{\mathcal{G}_\mathrm{s}\kappa_\mathrm{e}}\left[\frac{1}{2} + n_\mathrm{add} + n_\mathrm{JPA} \right]
\end{equation}	
for four different numbers of added photons attributed to the detection chain outside the photon-pressure device and vs. $\kappa_\mathrm{e}/\kappa$ and $\mathcal{G}_\mathrm{s}$.
It becomes clear from these plots that the larger $n_\mathrm{add}$ and the more undercoupled the cavity is, the more we can profit from the internal amplification.
Furthermore, as we are pumping on the red sideband, the standard quantum limit can only be approached for $\mathcal{C}_\mathrm{eff} \to \infty$ but due to the similarity of the HF and RF linewidths we cannot operate at a much higher cooperativity without touching the strong coupling regime.

\begin{figure}[h!]
	\centerline{\includegraphics[trim = {0cm, 1cm, 0cm, 0cm}, clip, scale=0.7]{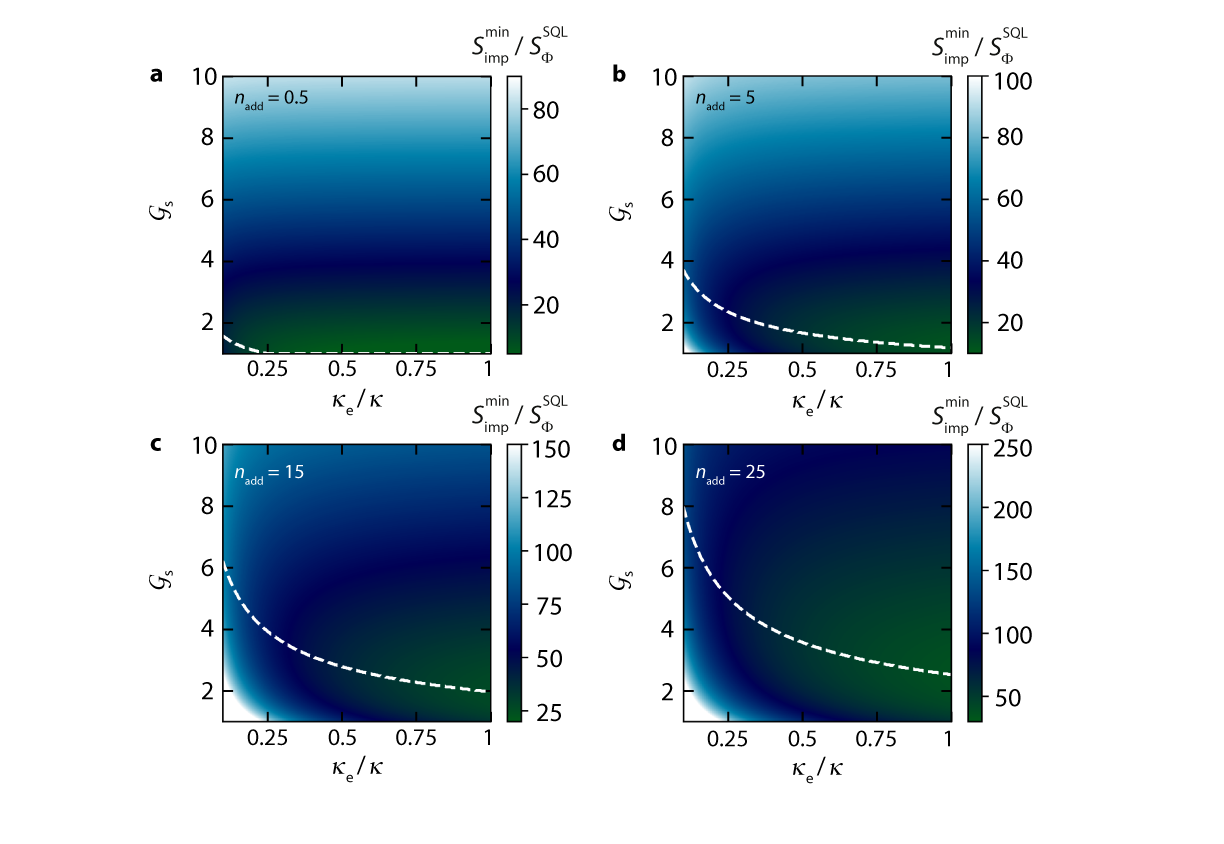}}
	\caption{\textsf{\textbf{Imprecision noise reduction with a photon-pressure Kerr amplifier depending on the device and setup parameters.} The four panels \textbf{a-d} display the imprecision noise normalized to the standard quantum limit SQL $S_\mathrm{imp}^\mathrm{min}/S_\Phi^\mathrm{SQL}$ versus intracavity Josephson gain on resonance with the signal mode $\mathcal{G}_\textrm{s}$ and the ratio of external and total linewidths $\kappa_\textrm{e}/\kappa$, for four different values of added photons by the amplification chain ($n_\textrm{add}$). The value of $n_\textrm{add}$ utilized in each panel is indicated in the respective colormap and the minimum imprecision noise is represented by the white dashed lines. While for $n_\textrm{add}=0.5$, we reach the minimum imprecision noise for considerably low values of $\mathcal{G}_\textrm{s}$, this threshold seems to increase for higher values of added photons in the amplification chain, thereby indicating a higher profit from parametric amplification. In addition, the tendency for this value to decrease with the linewidth ratio of the cavity happens consistently for every value of $n_\textrm{add}$, suggesting the highest benefit of the strong parametric drive in the case of using undercoupled cavities.}}
	\label{fig:Imprecision}
\end{figure}

One great advantage of the sideband pump detection scheme we implement here, however, is that the detection frequency is essentially limited only by the HF cavity frequency, i.e., in principle we could detect anything between kHz and GHz oscillator noise, while other experiments with resonant detection drives are restricted to the linewidth of the JPA\cite{Hatridge11S, LevensonFalk16S, Foroughi18S}.
With optimized parameters regarding the linewidths of the circuits, much higher cooperativities could be utilized as well, which would then allow for a further reduction of the imprecision noise.
We believe that the intrinsic amplification will also be extremely valuable when dynamical backaction is avoided, like for example in backaction evading measurement protocols with one red detuned and one blue detuned pump\cite{Hertzberg10S}.

\section{Supplementary Note 9: Additional data on photon-pressure induced transparency and RF noise detection}

\subsection{Experiments without parametric cavity drive}

\begin{figure}[h!]
	\centerline{\includegraphics[trim = {0cm, 6cm, 0cm, 2cm}, clip, scale=0.52]{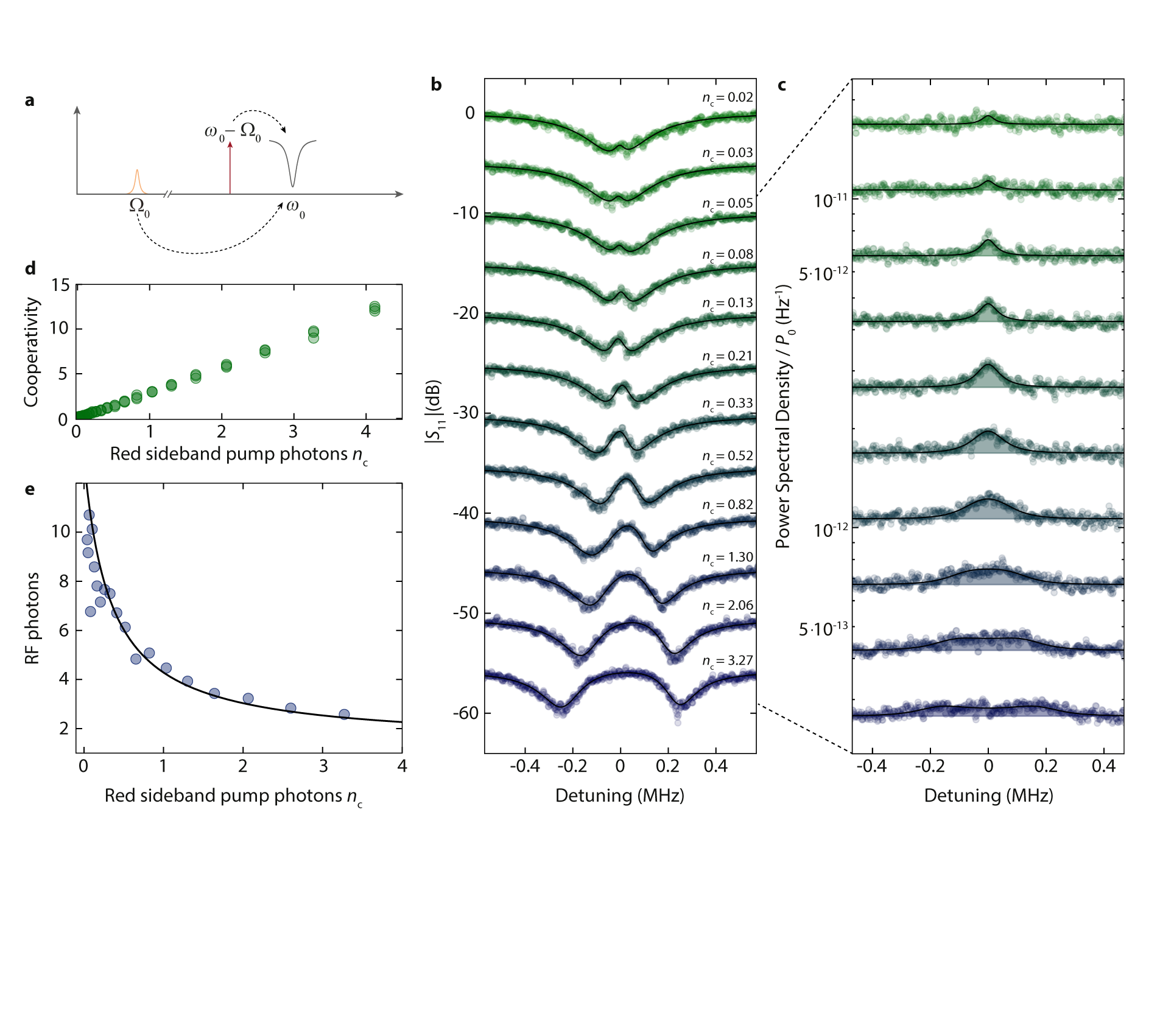}}
	\caption{\textsf{\textbf{Photon-pressure characterization of the device without a parametric drive.} Panel \textbf{a} displays a schematic of the experiment. A photon-pressure pump tone at frequency $\omega_\mathrm{p} = \omega_0 - \Omega_0$ and with variable power is sent to the red sideband of the HF cavity. For each pump power, we detect first the reflection response of the HF cavity using a VNA and then the output noise power spectrum in the same frequency window $\sim \omega_0$ with a signal analyzer, but with the VNA switched off. In \textbf{b} a series of reflection data $|S_{11}|$ is shown. From top to bottom, the red-sideband power is increased. The top curve is unshifted and with increasing power, the curves are offset by $-5\,$dB for better visibility. Small numbers next to the curves represent the average red-sideband intracavity photon numbers $n_\mathrm{c}$ at each pump power. Circles are data, lines are theory curves. In panel \textbf{c} the corresponding output power spectra are shown, normalized to the red sideband input power $P_0$. Note that for the two lowest-power $|S_{11}|$ datasets, we do not show the power spectra as they are essentially flat lines on the scale shown here. From the modelling of each linescan of $S_{11}$ with the theoretical expressions, we can then determine the cooperativity $\mathcal{C} = 4n_\mathrm{c}g_0^2\kappa^{-1}\Gamma_0^{-1}$, the result is shown in panel \textbf{d}. Note that we used more pump powers than shown in $\textbf{b}$ for the data in \textbf{d}. Also we used three linescans for each power. The cooperativity $\mathcal{C}$ grows linearly with the drive photon number as expected and for a single sideband pump photon we reach $\mathcal{C} \sim 2.8$. From the output spectra, we calculate the sideband-cooled thermal occupation of the RF mode in units of quanta, the result is shown in panel \textbf{e} vs pump power. Circles are data, line is the theoretical curve. For very small pump powers, the extraction is not very precise as the signal is very small compared to the background noise. From the theoretical modelling, we get a residual RF mode occupation of $n_\mathrm{th}^\mathrm{RF} \sim 12$, which is cooled to about $n_\mathrm{cool}^\mathrm{RF} \sim 2.5$ for the largest powers shown. }}
	\label{fig:NormalOM}
\end{figure}

The simplest photon-pressure experiment we conducted, also as part of the device calibration, was to investigate the photon-pressure interaction in absence of a parametric drive.
For this, we bias the cavity at operation point I, i.e., with $\Phi_\mathrm{b}/\Phi_0 \approx 0.48$, where we get $\tilde{g}_0 = g_0 = 2\pi\cdot 120\,$kHz and $\omega_0 = 2\pi\cdot 7.221\,$GHz.
The cavity linewidths at this bias point are approximately given by $\kappa = 2\pi\cdot 450\,$kHz and $\kappa_\mathrm{e} = 2\pi\cdot 85\,$kHz.
The RF mode parameters are $\Omega_0 = 2\pi\cdot 452\,$MHz and $\Gamma_0 = 2\pi\cdot 45\,$kHz.
For the basic experiments of photon-pressure induced transparency and RF noise detection, we apply a photon-pressure pump tone at $\omega_\mathrm{p} = \omega_0 - \Omega_0$, i.e., on the red sideband of the HF cavity, and detect both the probe tone response and the output noise around the resonance frequency of the HF cavity.
We sweep the power of the red-sideband pump and for each power record the reflection $S_{11}$ using a VNA.
In addition, we measure the HF cavity output spectrum for each pump power using a spectrum analyzer, but with the VNA switched off.
A summary of this experiment is shown and discussed in Supplementary Fig.~\ref{fig:NormalOM}.
In the reflection response $S_{11}$ we observe the characteristic small transparency peak for low powers, resembling the RF susceptibility including the dynamical backaction from the HF sideband pump field.
With increasing sideband pump power, the transparency window grows in amplitude and width and for the highest power shown here, the system enters the strong coupling regime, where the RF mode and the HF cavity hybridize, and the response develops from one mode with a transparency window into two distinct normal modes.
The response can be modelled with a single set of device parameters for the frequencies and linewidths of the two modes and using Eq.~(\ref{eqn:FullOMIT}) with $n_\mathrm{d} = 0$ and $\mathcal{K} = 0$, i.e.,
\begin{equation}
S_{11} = 1 - \kappa_\mathrm{e}\chi_\mathrm{c}(\Omega)\left[1 - i2g^2\Omega_0 \chi_\mathrm{c}(\Omega)\chi_0^\mathrm{eff}(\Omega) \right]
\end{equation}
where
\begin{equation}
\chi_\mathrm{c} = \frac{1}{\frac{\kappa}{2} + i\left( \Omega -\Omega_0 + \delta \right)}, ~~~~~ \chi_0^\mathrm{eff} = \frac{1}{\frac{\Gamma_0}{2} + i\left( \Omega -\Omega_0 \right) + ig^2\chi_\mathrm{c}(\Omega)}, ~~~~~ g = \sqrt{n_\mathrm{c}} g_0.
\end{equation}
The intracavity photon number by the sideband pump $n_\mathrm{c}$ is given by $n_\mathrm{c}\hbar\omega_\mathrm{p} = \kappa_\mathrm{e}|\chi_\mathrm{c}(0)|^2 P_0$ with the on-chip input power $P_0$.
Note that for increasing pump power a small shift of the HF cavity is visible due to pump-induced Kerr shift.
We take this into account by a small shift $\delta$ in the reflection equation without explicitly relating it to $\mathcal{K}$ for simplicity.
To model the output noise, we use Eq.~(\ref{eqn:FullKerrNoise}) for $n_\mathrm{d} = 0$ and $\mathcal{K} = 0$, and in addition take the amplifier and detection chain added noise $n_\mathrm{add}$ into account.
The equation for the power spectral density in units of quanta then is
\begin{eqnarray}
S_{nn} & = &n_\mathrm{add} + \kappa_\mathrm{e}g^2\frac{|\chi_\mathrm{c}|^2|\chi_+|^2}{|1+i2\Omega_0g^2\chi_0\chi_\mathrm{c}|^2}\Gamma_0\left(n_\mathrm{th}^\mathrm{RF} + \frac{1}{2} \right) \nonumber \\
& & + \bigg|1 - \frac{\kappa_\mathrm{e}\chi_\mathrm{c}}{1+i2\Omega_0g^2\chi_0\chi_\mathrm{c}}\bigg|^2\left( n_\mathrm{e}^\mathrm{HF} + \frac{1}{2} \right)+ \kappa_\mathrm{e}\kappa_\mathrm{i}\frac{|\chi_\mathrm{c}|^2}{|1+i2\Omega_0g^2\chi_0\chi_\mathrm{c}|^2}\left( n_\mathrm{i}^\mathrm{HF} + \frac{1}{2} \right) \nonumber \\
& \approx & \frac{1}{2} + n_\mathrm{add} + n_\mathrm{e}^\mathrm{HF} + \kappa_\mathrm{e}\kappa_\mathrm{i}\frac{|\chi_\mathrm{c}|^2}{|1+g^2\chi_+\chi_\mathrm{c}|^2}\left( n_\mathrm{i}^\mathrm{HF} - n_\mathrm{e}^\mathrm{HF} \right)  + \kappa_\mathrm{e}g^2\frac{|\chi_\mathrm{c}|^2|\chi_+|^2}{|1+g^2\chi_+\chi_\mathrm{c}|^2}\Gamma_0\left(n_\mathrm{th}^\mathrm{RF} - n_\mathrm{e}^\mathrm{HF} \right)
\end{eqnarray}
where we used for the last approximation $i2\Omega_0\chi_0 \approx \chi_+$ in the relevant frequency regime $\Omega \approx +\Omega_0$ and
\begin{equation}
n_\mathrm{th}^\mathrm{RF} = \frac{\Gamma_\mathrm{i}n_\mathrm{i}^\mathrm{RF} + \Gamma_\mathrm{e}n_\mathrm{e}^\mathrm{RF}}{\Gamma_0}
\end{equation}
the average thermal occupation of the RF mode.
Experimentally, we find the residual thermal occupation of the RF mode to be $n_\mathrm{th}^\mathrm{RF} \sim 13$ quanta from the modelling of the spectrum data and a sideband-cooled final occupation of about $n_\mathrm{cool}^\mathrm{RF} \sim 2.5$ for the largest pump powers shown here.
For the added number of photons by the amplifier chain we use $n_\mathrm{add} = 15$ here, and assume the HF cavity to be in its quantum groundstate $n_\mathrm{i}^\mathrm{HF} = n_\mathrm{e}^\mathrm{HF} = 0$.

\subsection{Experiments with medium Josephson gain}
In main paper Fig.~3, we present and discuss enhanced photon-pressure coupling between the RF mode and the strongly driven HF cavity in a large gain JPA state.
For the same drive power and drive frequency $\omega_\mathrm{d}$, but a different flux bias point $\Phi_\mathrm{b}/\Phi_0 \sim 0.52$, the intracavity drive photon number and the impact of the parametric drive regarding cavity susceptibility and Josephson gain are considerably smaller.
The signal resonance at this operation point (II) is still given by an absorption dip similar to the undriven case, but resembles the reflection response of an overcoupled cavity.
In Supplementary Fig.~\ref{fig:PPmediumgain} we present the photon-pressure response and the enhancement of the photon-pressure interaction by parametric gain in this regime, in comparison with the bare cavity and the large gain regime discussed in the main paper.
Due to the parametric drive and the related saturation of TLSs, the linewidth is reduced compared to the undriven case to $\kappa\sim 2\pi\cdot\,250$kHz.
The difference to the other two datasets (undriven cavity and large gain state) is that with the increased flux bias value also the Kerr nonlinearity $\mathcal{K}$ and the single-photon coupling rate $g_0$ are slightly different.
Therefore, the $\tilde{g}_0$ is larger than in both other datasets with $\tilde{g}_0\sim 2\pi\cdot 160\,$kHz where we used $\mathcal{K} = -2\pi\cdot 4\,$kHz and $n_\mathrm{d} = 1650$.
The intracavity Josephson gain on resonance of the signal mode is $\mathcal{G}_\mathrm{s} \approx 2.3$.
\begin{figure}[h!]
	\centerline{\includegraphics[trim = {0cm, 9cm, 0cm, 3cm}, clip, scale=0.52]{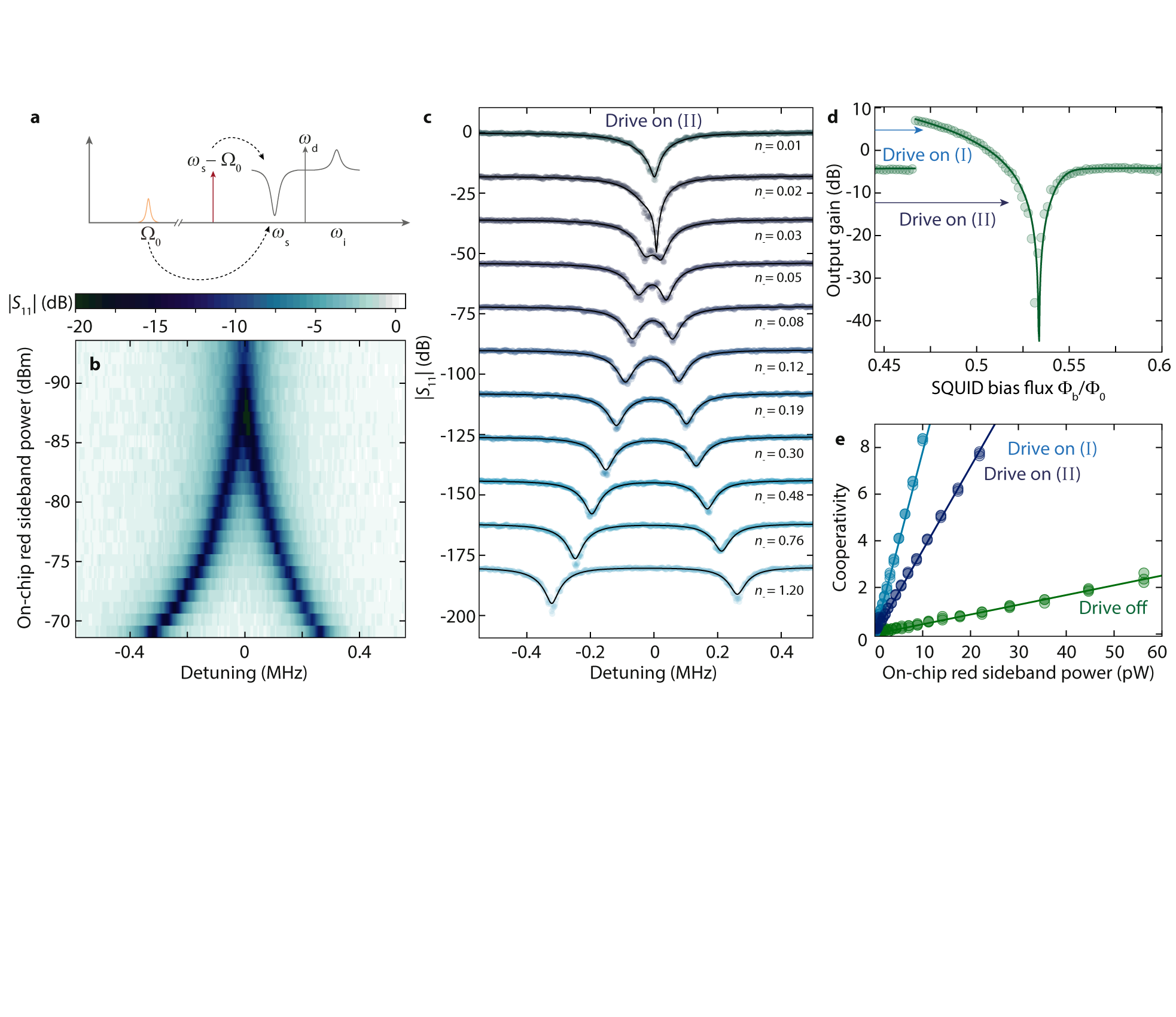}}
	\caption{\textsf{\textbf{Enhanced photon-pressure with a medium amplitude parametric drive.} \textbf{a} Schematic of the experiment. The parametric state of the cavity is activated with a strong drive tone at $\omega_\mathrm{d}$. A photon-pressure pump tone at frequency $\omega_\mathrm{p} = \omega_\mathrm{s} - \Omega_0$ and with variable power is sent to the red sideband of the HF cavity signal resonance. For each pump power, we detect the reflection response of the system using a VNA, the color-coded result is shown in panel \textbf{b}. We show selected linescans from \textbf{b} in panel \textbf{c}, circles are data, lines are theory curves using Eq.~(\ref{eqn:PPIT_full_v2}). Detuning in \textbf{b} and \textbf{c} is given with respect to the zero-pump signal mode resonance frequency $\omega_\mathrm{s}$. As in the medium gain regime the signal mode response is that of an overcoupled cavity, we observe that the photon-pressure signature in the cavity at low powers is a dip, i.e., we observe photon-presuure induced absorption. For larger powers the device enters the strong-coupling regime, where, due to the frequency dependent Josephson gain, the normal mode closer to the drive tone (right) shows less absorption than the mode far away from the drive (left). The overcoupled regime at operation point II is apparent in panel \textbf{d}, where we plot the output gain on resonance vs flux bias and point with arrows to the two operation points relevant for this work. Circles are extracted from data, line is the theoretical prediction based n Eq.~(\ref{eqn:Gout_approx}). From fits to the $S_{11}$ data with Eq.~(\ref{eqn:PPIT_approx}) (fits are not shown), we determine the effective cooperativity and plot it as function of photon-pressure red sideband on-chip pump power together with the equivalent data without drive and at operation point I. Even though the intracavity gain is moderate here, the cooperativity is significantly increased compared to the undriven case. Note that about a factor of $\sim 1.6$ of the increase in $\mathcal{C}$ is due to a slightly larger $g_0$ as the flux responsivity of the mode is larger at operation point II than at point I.}}
	\label{fig:PPmediumgain}
\end{figure}
At this operation point and gain, the signal mode resembles an overcoupled cavity, which leads to a different phenomenology than for the undercoupled case as represented by the undriven cavity, cf. Supplementary Fig.~\ref{fig:NormalOM}.
For an overcoupled cavity in a reflection measurement, the photon-pressure signature under weak red sideband pumping leads to an aborption window inside the cavity response, cf. Supplementary Fig.~\ref{fig:PPmediumgain}.
Only for larger pump powers, the dip turns into a peak and then develops into normal-mode splitting for the largest pump powers.
This tunability of the coupling is an interesting feature for signal applications, as it allows for switchable and largely tunable microwave signal propagation control.
Regarding the enhanced cooperativity we report and discuss in main paper Fig.~3, we observe also a large enhancement in this medium gain regime, as compared to the undriven case, cf. Fig.~\ref{fig:PPmediumgain}\textbf{e}.
The cooperativity is not as high as for the high gain regime (factor of 2 smaller), but still a factor of $\sim 7$ larger than in the undriven case.
Note, however, that the direct comparison shown here ($\mathcal{C}_\mathrm{eff}$ plotted versus on-chip red sideband power) is not completely fair, as at the operation point II $g_0$ is enhanced by $\sim 1.25$ due to the increased $\partial\omega_0/\partial\Phi$ of the HF mode.
In addition to characterizing the interaction in the medium gain regime by a three-tone response measurement, we also investigated sideband-cooling by taking noise traces with a spectrum analyzer for each of the powers and operation conditions of Fig.~\ref{fig:PPmediumgain}.
\begin{figure}[h!]
	\centerline{\includegraphics[trim = {0cm, 11.0cm, 0cm, 0cm}, clip, scale=0.52]{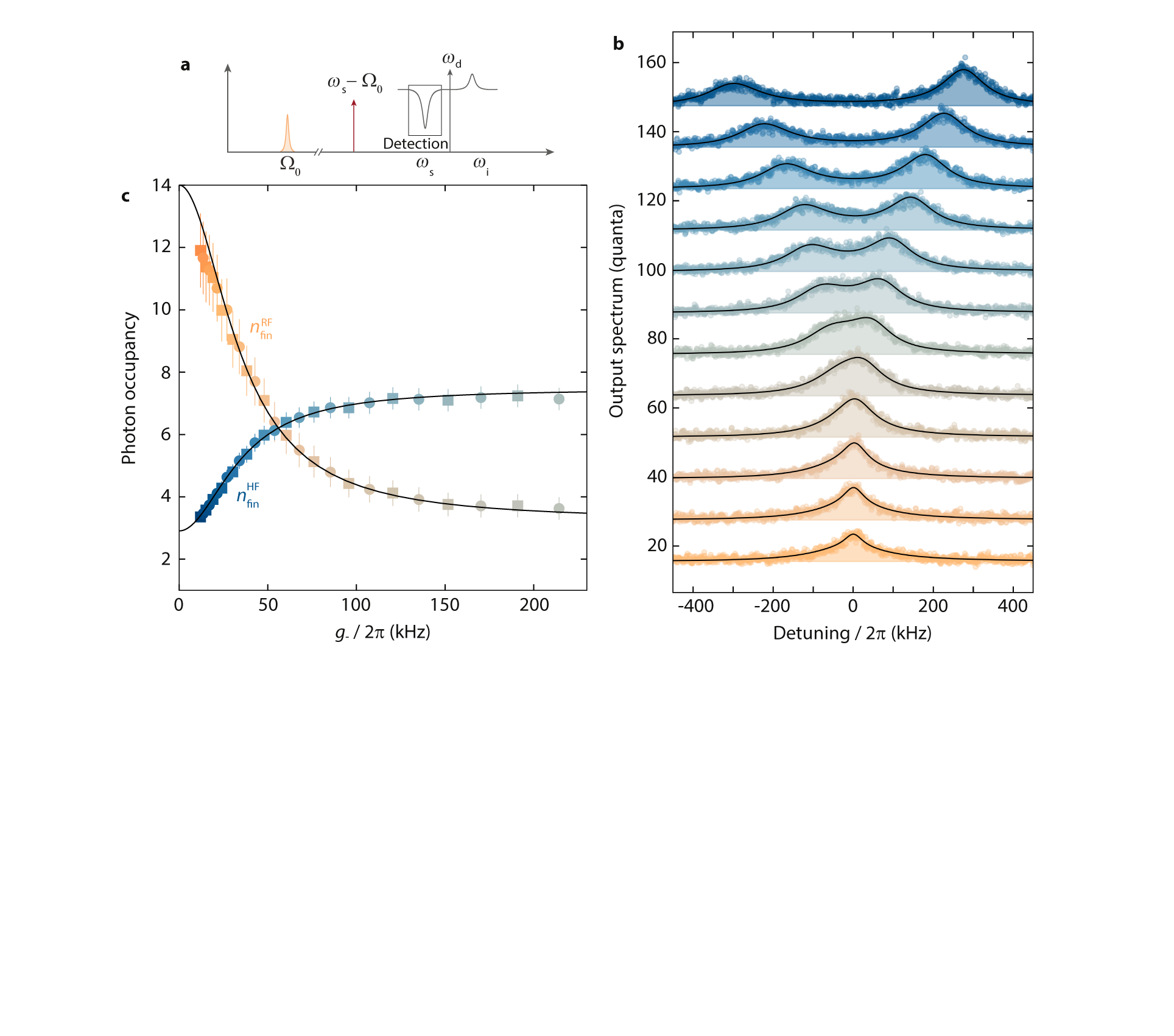}}
	\caption{\textsf{\textbf{Non-equilibrium sideband-cooling with a medium gain JPA.} \textbf{a} Schematic of the experiment. The driven HF cavity signal resonance is photon-pressure pumped on its red sideband with $\omega_\mathrm{p} = \omega_\mathrm{s}-\Omega_0$ and variable pump power. The output power spectrum around $\omega \sim \omega_\mathrm{s}$ is detected using a spectrum analyzer. The flux operation point is $\Phi_\mathrm{b}/\Phi_0 = 0.52$, the intracavity drive photon number is $n_\mathrm{d}\approx 1650$ and the intracavity gain at the signal resonance $\mathcal{G}_\mathrm{s} = 2.3$. \textbf{b} Output spectra of the signal mode in units of quanta, for increasing red-sideband pump power (bottom curve: lowest power, top curve: highest power). Circles are data, lines and shaded areas are fits using Eq.~(\ref{eqn:FullKerrNoise}) with only $n_\mathrm{th}^\mathrm{RF}$ as free parameter. Subsequent datasets and fits are offset by $+12$ quanta each for clarity, detuning is given with respect to the zero-pump signal mode resonance frequency $\omega_\mathrm{s}$. For the lowest pump powers, the output spectrum shows a narrow Lorentzian (RF noise) on top of a wider one (amplified HF quantum noise), with increasing power the two merge and develop into a pronounced normal-mode splitting. The higher frequency normal-mode acquires a larger amplitude as the Josephson gain increases with decreasing distance to the drive tone. From the fits to each linescan (and to additional, intermediate linescans not shown in \textbf{b}), we determine the sideband-cooled RF mode occupation and the resulting HF occupation, which are plotted vs photon-pressure coupling rate $g_- = \sqrt{n_-}\tilde{g}_0$ in panel \textbf{c}. Squares correspond to data, for which the spectrum is shown in \textbf{b}, circles are the intermediate data, lines are theoretical curves using Eqs.~(\ref{eqn:nRFfinanal}) and (\ref{eqn:nHFfinanal}). The residual RF mode occupation without photon-pressure pump is around $n_\mathrm{th}^\mathrm{RF} \sim 14$ and the sideband cooling reduces this thermal occupation to about $n_\mathrm{fin}^\mathrm{RF} \approx 3.5$ for the largest pump power used here $g_- \approx 2\pi\cdot 220\,$kHz. The effective HF mode occupation, arising from amplified quantum noise, is about $\tilde{n}_\mathrm{th}^\mathrm{HF} \sim 3$ without any photon-pressure pump, and increases to $n_\mathrm{fin}^\mathrm{HF} \sim 7.3$ at the highest pump powers. So similar to main paper Fig.~5, also here in the medium gain regime we get $n_\mathrm{fin}^\mathrm{HF} > n_\mathrm{fin}^\mathrm{RF}$. Error bars in \textbf{c} correspond to an estimated $10\%$ uncertainty in $n_\mathrm{fin}^\mathrm{RF}$ and $5\%$ uncertainty in $n_\mathrm{fin}^\mathrm{HF}$.}}
	\label{fig:SCmediumgain}
\end{figure}
The results are presented and discussed in Supplementary Fig.~\ref{fig:SCmediumgain}.
We observe a situation indeed intermediate between the undriven case presented in Supplementary Fig.~\ref{fig:NormalOM} and the high-gain results presented in main paper Fig.~5.

\section{Supplementary References}

\end{document}